\documentclass[aps, prr,reprint,10pt,superscriptaddress,amssymb,amsfonts,longbibliography]{revtex4-1}

\usepackage{amsmath,amssymb,amsthm,amscd,latexsym,epsfig,bm}

\usepackage[colorlinks,bookmarks=false,citecolor=blue,linkcolor=red,urlcolor=blue]{hyperref}
\usepackage{verbatim}

\usepackage[normalem]{ulem}
\usepackage[dvipsnames]{xcolor}
\usepackage{dsfont}

\usepackage{bbold}
\usepackage{booktabs}
\usepackage{longtable}
\usepackage{amstext}
\usepackage{array}
\newcolumntype{L}{>{$}l<{$}} 
\newcolumntype{C}{>{$}c<{$}} 

\def\beq{\begin{equation}}
\def\eeq{\end{equation}}
\def\be{\begin{equation}}
\def\ee{\end{equation}}

\def\Tr{\hbox{$\mathrm{Tr}$}\,}
\newcommand{\zz}{\mathbb{Z}_2}
\newcommand{\z}{\mathbb{Z}}

\def\f2{{\mathbb F}_2}

\theoremstyle{plain}
\theoremstyle{plain}

\providecommand{\theoremname}{Theorem}
\providecommand{\theoremtextname}{Theorem}

\theoremstyle{plain}
\providecommand{\propositionname}{Proposition}

\newcommand{\bsl}[1]{\boldsymbol{#1}}

\newcommand{\ii}{\mathrm{i}}

\newcommand{\eqnref}[1]{Eq.\,\eqref{#1}}
\newcommand{\figref}[1]{Fig.\,\ref{#1}}

\newcommand{\refcite}[1]{Ref.\,[\onlinecite{#1}]}

\newcommand{\eq}[1]{\begin{equation} #1 \end{equation}}

\newcommand{\eqa}[1]{\begin{align}\begin{split} #1 \end{split}\end{align}}

\usepackage{slashed}

\let\oldAA\AA
\renewcommand{\AA}{\text{\normalfont\oldAA}}

\begin{document}

\title{Effective field theories of topological crystalline insulators and topological crystals
}

\author{Sheng-Jie Huang}

\affiliation{Condensed Matter Theory Center, Department of Physics, University of Maryland, College Park, Maryland 20742-4111, USA}
\affiliation{Joint Quantum Institute, Department of Physics, University of Maryland, College Park, Maryland 20742-4111, USA}

\author{Chang-Tse Hsieh}

\affiliation{Quantum Matter Theory Research Team, RIKEN CEMS,
Wako, Saitama 351-0198, Japan}

\author{Jiabin Yu}

\affiliation{Condensed Matter Theory Center, Department of Physics, University of Maryland, College Park, Maryland 20742-4111, USA}

\date{\today}

\begin{abstract}
We present a general approach to obtain effective field theories for topological crystalline insulators whose low-energy theories are described by massive Dirac fermions. We show that these phases are characterized by the responses to spatially dependent mass parameters with interfaces. These mass interfaces implement the dimensional reduction procedure such that the state of interest is smoothly deformed into a \emph{topological crystal}, which serves as a representative state of a phase in the general classification. Effective field theories are obtained by integrating out the massive Dirac fermions, and various quantized topological terms are uncovered. Our approach can be generalized to other crystalline symmetry protected topological phases and provides a general strategy to derive effective field theories for such crystalline topological phases. 
\end{abstract}

\maketitle

\section{Introduction}
Topological phases of matter are gapped phases which are characterized by the patterns of quantum entanglement in the ground state\cite{Wen2019review}. A ground state with a short-range quantum entanglement is considered trivial since the state can be smoothly deformed into a product state of microscopic degrees of freedom without closing the energy gap. Such deformation might not be possible in the presence of symmetry, allowing us to define the so-called symmetry protected topological (SPT) phases~\cite{Schnyder2008, Kitaev2009, ryu2010, gu2009, pollmann2010, fidkowski2011, turner2011, chen2011_1dspt, chen2011_1dcomplete, cirac2011, chen2013cohomology, levin2012, kapustin2014symmetry, Else2014spt}. Explicitly, a SPT phase is a gapped phase of matter with a unique ground state that cannot be adiabatically deformed to a trivial state as long as the symmetry is preserved; the adiabatic trivialization is possible only if the symmetry is allowed to be broken.

For internal symmetries, general understanding of SPT phases has been largely achieved~\cite{freed2014shortrange, freed2019reflection,Xiong_2018sptmin, Gaiotto2019}. Besides the development of SPT phases with internal symmetries, there has been a great interest on topological phases with crystalline point group and space group symmetries~\cite{Teo2013,Shiozaki2014,Isobe2015, Shiozaki2017, Po2017indicators, Bradlyn2017, Kruthoff2017}. Typical examples include so-called topological crystalline insulators (TCIs), which are electronic insulators protected, in part, by point group or space group symmetries~\cite{Fu2011, Ando2015review, Chiu2016review}. As a consequence of crystalline symmetry, higher-order topology may occur in crystalline SPT (cSPT) phases, where anomalous gapless modes could show up on the $(d-n)$-dimensional boundary of a $d$-dimensional bulk (with $n > 1$) \cite{Schindler2018HOTI, neupert2018rev}.

There have been two general frameworks which are conjectured to give a general classification of cSPT phases. The first approach is dubbed as the ``topological crystal" approach \cite{Song2017dr, Huang2017building, rasmussen2018intrinsically, cheng2018rotation, shiozaki2018generalized, Xiong2018glide, Song2018indicator, Song2019tc, Song2019real, Shiozaki2019AHSS, Song2020, Alex2020bhospt, Huang2020tsc, Zhang2020, Geier2021defect}, in which the key idea is to deform a generic cSPT state into a real-space stacking of ``building blocks", which are lower-dimensional SPT states with effective internal symmetry. These special kinds of states are referred to as topological crystals. The classification of cSPT phases is then given by the deformation classes of the topological crystals. This approach gives a simple physical picture for understanding cSPT phases. Interesting physical signatures on the boundaries \cite{Huang2018surface, Huang20204d} or crystalline defects\cite{Geier2021defect} are usually easy to obtain in this framework. 

The second framework is based on the ``smooth states"\cite{Else2018}.
In this framework, a hypothetical lattice of ancillas (hypothetical degrees of freedom) is introduced, and the lattice of ancillas has a much smaller lattice constant than the physical lattice. A smooth state is very smooth in the lattice of ancillas. However, the radius of the spatial variation of a smooth state is on the order of the unit cell size of the physical lattice. An important consequence in this framework is that the classification of cSPT phases with a crystalline symmetry group G is the same as the classification of SPT phases with internal symmetry group G, which is known as the ``Crystalline Equivalence Principle" (See also Ref.~\onlinecite{Jiang2017, freed2019invertiblecrystal, debray2021invertible}). The topological crystal and the smooth state approach are actually equivalent as shown in Ref.~\onlinecite{Else2019defect}.

In principle, SPT phases should be characterized by their response to background gauge fields. The concept of crystalline gauge fields has been mathematically defined \cite{Else2018}; but the result is rather formal, which limits the practical usage. Further attempts have been made in order to turn it into a concrete theoretical tool either in the field theory framework\cite{Song2021polarization, gioia2021unquantized,Naren2021} or on the lattice\cite{Naren2021}. There has been an alternative proposal which characterizes cSPT phases through their response to elastic deformations\cite{Nissinen2018tetrads, Nissinen20193dqh, nissinen2020field, Nissinen2021, Else2021qc}. This approach is very physical and is well-defined in a continuous field theory. However, it has only been worked out in some simple examples, and the relation to the general classification can only be obtained formally. It's also unclear how to derive such effective field theories microscopically. 

In this paper, we propose a continuous field theoretical description of the cSPT phases. Focusing on TCIs described by massive Dirac theories, the main idea is to characterize these phases by the responses to spatially dependent mass parameters (and to the background gauge field of internal symmetry). These spatially dependent mass terms implement the dimensional reduction procedure such that the states trapped at the mass interfaces are precisely the building blocks in the corresponding topological crystal picture. The effective field theory is then obtained by integrating out the gapped fermions. Our approach not only provides well-defined effective field theories for the cSPT phases, but also gives an explicit connection between the topological crystal picture and the effective field theories. 

The rest of this paper is organized as follows: in Sec.~\ref{sec:overview}, we give a review of the topological crystal approach and summarize our main results. Cellular cohomology are used throughout this paper, which we review in Appendix~\ref{sec:cell-cohomology}. 

In Sec.~\ref{sec:1datom}, we discuss the effective field theory for 1d atomic insulators with charge conservation and the lattice translation symmetry as a simple example to illustrate our approach. Generalizations to 2d and 3d atomic insulators are also discussed and more details are given in Appendix~\ref{sec:2datomic}. We point out that the responses described by these topological terms are generalized Thouless pumps. 

In Sec.~\ref{sec:TCIs}, we apply our approach to various TCIs with point group symmetry and derive the effective field theories. Sec.~\ref{sec:1dr} consider 1d TCIs with reflection symmetry. Sec.~\ref{sec:2dTCI} generalize the discussion to 2d TCIs with $C_{N}$ rotational symmetry. Since the form of the topological term for this case might not be familiar to the readers, we provide a perturbative derivation in Appendix~\ref{sec:2drotation}. Sec.~\ref{sec:3dTCI} is devoted to 3d 2nd order TCIs protected by $C_{nv}$ symmetry, where there are gapless hinge modes for appropriate boundary conditions. The field theories we obtained are essentially the same as the axion field theory~\cite{Qi2008TFT}, where the non-trivial information of the classification is encoded in the theta angle. We also briefly discuss the physical responses as the a result of the topological terms in the effective field theories. 

Finally, we conclude in Sec.~\ref{sec:summary} with a discussion of our results and possible directions for future work. Other mathematical details are given in Appendix~\ref{sec:FD}, including general discussions on the fundamental domain for crystalline symmetry, and the construction of the map $f: M \rightarrow BG_{s}$, where $M$ is the real space manifold and the $BG_{s}$ is the classifying space for the space group $G_{s}$. Appendix~\ref{sec:BG} gives a construction of the classifying space $BG_{s}$ for a space group.


\section{General perspective and summary of the results}
\label{sec:overview}

\subsection{Review of the topological crystal approach}
Topological crystal approach is a general framework of describing and classifying cSPT phases. The main idea is that any cSPT phase is adiabatically connected to a stacking of $d_{b}$-dimensional  topological states with effective internal symmetry arranged in some crystalline pattern in $d$-dimensional space, where $d_{b}$ ranges from $0$ to $d$. This procedure is called dimensional reduction and these special kinds of states are referred to as the topological crystals in Ref.~\onlinecite{Song2019tc}. In order for the argument to go through, an important assumption is that the correlation length $\xi$ can be tuned to be arbitrarily small, and, in the presence of translation symmetry, much smaller than the size of the unit cell, which requires adding a fine mesh of trivial degrees of freedom.

A systematic way to describe a topological crystal is as follows. We define a fundamental domain (FD) to be a smallest simply connected closed part of space, subject to the condition that no two points in the region are related by a crystalline symmetry. The FD is then copied throughout space using the crystalline symmetry such that the whole of space is filled. A more formal definition of the FD is given in Appendix~\ref{sec:FD}. This construction gives a $d$-dimensional space a cell complex structure, where the $d$-cells are copies of FD. The $(d-1)$-cells lie on faces where two $d$-cells meet, with the property that no two points in the same $(d-1)$-cell are related by symmetry. This procedure continues to $0$-cells. As shown in Appendix~\ref{sec:FD}, an important property of this cell complex is that there exists a map $f:M \rightarrow BG_{s}$, where $BG_{s}$ is the classifying space of the space group $G_{s}$. A construction of the classifying space $BG_{s}$ is given in Appendix~\ref{sec:BG}. As a result, each $d$-cell can be labeled by a group element in the space group, and each path connecting a point $\boldsymbol{r}$ in a FD to $g\boldsymbol{r}$ is also labeled by an group element $g \in G_{s}$.

With this cell-complex structure, one can understand a topological crystal state by associating $d_{b}$-dimensional topological phase with each $d_{b}$-cell. These $d_{b}$-dimensional states are referred to as the ``building blocks" of the topological crystal. When the building blocks intersect in the bulk, the building blocks must be glued together so as to eliminate any gapless modes at the junctions while preserving symmetry. An ordinary crystal is a simple examples of a topological crystal state, which is formed by periodically arranged atoms as $d_{b}=0$ building blocks.

\subsection{Characterizing topological crystalline phases by responses to mass parameters}

In this section, we summarize general aspects of our approach on deriving the effective field theories for cSPT phases. The main idea of this work is to characterize cSPT phases by its response to spatially dependent mass parameters. These spatially dependent mass terms have interfaces that implement the dimensional reduction procedure such that the states trapped at the mass interfaces are precisely the building blocks in the corresponding topological crystal picture. The effective field theory is then obtained by integrating out the gapped fermions. This provides a way to connect the topological crystal picture to the effective field theories.

One of the simplest example is given in a $1$-dimensional crystal with a lattice translation symmetry and a $U(1)$ charge conservation symmetry. There is an integer topological invariant $\nu$ representing the charge per unit cell. We will show that the topological response of such a system to a spatially dependent mass parameter background and to a background $U(1)$ gauge field $A_{\mu}$ is characterized by a quantized topological term
\begin{equation}
    \frac{\nu}{2 \pi} \int  \epsilon^{\mu \nu} A_{\mu} \partial_{\nu} \phi(x) d^{2}x,
\end{equation}
where $\phi(x)$ is a phase variable that parametrizes the winding of the mass interface. The charge-$\nu$ bound state trapped at the mass interface with a non-trivial $2\pi \nu$ winding is precisely the $d_{b}=0$ building block in the topological crystal picture of this phase. Detailed arguments and the derivation are given in Sec.~\ref{sec:1datom}.

In order for this approach to make sense, it's important to consider the ``topological limit", where one tunes the correlation length $\xi$ to be arbitrarily small, especially $\xi \ll a$ (where $a$ is the size of a unit cell) in the presence of lattice translation symmetry.  One can imagine that the system is defined on a much finer lattice with lattice spacing $l \ll a$. It's still possible to describe a system in the topological limit by a continuous field theory---One consider fields that are coarse-grained with respect to the length scale $l$ so that the fields are smooth on the scale $l$ and could vary on the scale $R$ with $R \gg l$ and $R < a$. Working in the topological limit, the translation symmetry can not be viewed as an effective internal symmetry of the field theory---This property emerges only when one goes to the true ``IR limit", where the fields are smooth even on the scale $a$. In the topological limit, it thus makes sense to consider a mass interface with a characteristic length scale $w<a$ so that the dimensional reduction procedure goes through. In this paper, we will first derive the topological terms in the topological limit. Once a topological term is obtained, we are free to deform the field configurations of the mass parameters to be smooth on the scale $a$ while staying in the same classification class. The resulting theory will be validated in the IR limit. This is the general perspective that we take in this work.

\subsection{Summary of the results}
Ideally we would like to apply this method to obtain the quantized topological terms for any cSPT phase. However, doing this in full generality is still a difficult task. In this work, we instead illustrate our approach in physically relavant systems. We consider a wide range of TCIs in one, two, and three dimensions with a $U(1)$ charge conservation and a $G_{c}$ crystalline symmetry. For simplicity, we consider TCIs that can be built from building blocks with only charge $U(1)$ response, \textit{i.e.}, the building blocks are $d_{b}$-dimensional topological phases with $U(1)$ symmetry, which transform trivially under $G_{c}$. These building blocks are characterized by the  Chern-Simons term
\begin{equation}
    \mathcal{L}_{\text{CS}}^{2s+1}[A] = \frac{1}{(s+1)!}A \wedge (\frac{dA}{2\pi})^{s},
\end{equation}
where $A$ is the $U(1)$ gauge field. For these examples, we find the quantized topological terms take the following general form
\begin{equation}
    S = \int \mathcal{L}_{\text{CS}}^{d-k+1}[A] \wedge \Omega_{k},
\label{eqn:Seff-general}
\end{equation}
where $d$ is spatial dimensions and $\Omega_{k}$ is a $k$-form that corresponds to the codimension-k mass interfaces at which the building blocks are decorated. We will show that the mass interfaces that implement the dimensional reduction procedure are classified by $H^{k}(BG_{c},\mathbb{Z})$ with a twisting coefficient when $G_{c}$ contains orientation reversing elements. The $k$-form $\Omega_{k}$ in the topological term Eq.~(\ref{eqn:Seff-general}) is given by $\Omega_{k} = f^{*}\alpha$, where $f$ is the map $f:M \rightarrow BG_{c}$ given by the FDs in crystallography and $\alpha \in H^{k}(BG_{c},\mathbb{Z})$. In general, the $k$-form $\Omega_{k}$ is determined by a set of integral relations, which take the following form
\begin{equation}
    \int_{C_{\{g\}}} \Omega_{k} = N_{\{g\}},
    \label{eqn:int-cond}
\end{equation}
where $C_{\{g\}}$ is a $k$-cycle labeled by a set of group elements $\{g\}$ in $G_{c}$, and $N_{\{g\}}$ is given by
\begin{equation}
    N_{\{g\}} = \int_{C_{\{g\}}} f^{*}\alpha
\end{equation}
with $\alpha \in H^{k}(BG_{c},\mathbb{Z})$. Here $N_{\{g\}}$ is in general an integer or a $\mathbb{Z}_{n}$ number depending the cohomology group $H^{k}(BG_{c},\mathbb{Z})$. We note that, in all the examples considered in this work, it's enough to use a single group element $g \in G_{c}$ to label the $k$-cycle. Representative of $\Omega_{k}$ can be obtained by solving these equations. Table~\ref{tab:1} summarizes the examples and the results in this work. From the topological term Eq.~(\ref{eqn:Seff-general}), one can obtain electromagnetic responses of TCIs, which will be briefly discussed in the sections of each case.

\begin{table*}
\begin{center}
\setlength{\tabcolsep}{0.5em} 
{\renewcommand{\arraystretch}{1.6}
 \begin{tabular}{ |l|l|l|l|l| } 
 \hline
 Spacetime & Symmetry group $G$ & $\Omega_{k} $ in Eq.~(\ref{eqn:Seff-general}) & Integral conditions & Section \\ 
 dimensions & & & & \\
 \hline
 $1+1$D & $U(1) \times \Gamma$ & $\Omega_{1} = E$ &   $\int_{C_{t_{1}}} E = \int_{C_{t_{1}}} f^{*}\tau \in \mathbb{Z}, \tau \in H^{1}(B\Gamma,\mathbb{Z})$ & Sec.~\ref{sec:1datom} \\
 \hline
 $1+1$D & $U(1) \times D_{1}$ & $\Omega_{1}=dP$ & $\int_{C_{g_{r}}} dP = \int_{C_{g_{r}}} f^{*}r \in \mathbb{Z}_{2}, r \in H^{1}(BD_{1},\mathbb{Z}^{r})$ & Sec.~\ref{sec:1dr} \\ 
 \hline
 $2+1$D & $U(1) \times \Gamma$ & $\Omega_{2} = \frac{1}{2}\epsilon_{IJ} E^{I} \wedge E^{J}$ & $\int_{C_{t_{I}}} E^{I} = \int_{C_{t_{I}}} f^{*}\tau^{I} \in \mathbb{Z},  \tau^{I} \in H^{1}(BT_{I},\mathbb{Z})$ & App.~\ref{sec:2datomic} \\ 
 \hline 
 $2+1$D & $U(1) \times C_{N}$ & $\Omega_{2} = d\omega_{1}$ & $\int_{D_{u}} d\omega_{1} = \int_{D_{u}} f^{*}b \in \mathbb{Z}_{N}, b \in H^{2}(BC_{N},\mathbb{Z}) $ & Sec.~\ref{sec:2dTCI} \\ 
 \hline
 $3+1$D & $U(1) \times \Gamma$ & $\Omega_{3} = \frac{1}{6}\epsilon_{IJK} E^{I} \wedge E^{J} \wedge E^{K}$ & $\int_{C_{t_{I}}} E^{I} = \int_{C_{t_{I}}} f^{*}\tau^{I} \in \mathbb{Z}, \tau^{I} \in H^{1}(BT_{I},\mathbb{Z})$ & App.~\ref{sec:2datomic} \\ 
 \hline
 $3+1$D & $U(1) \times C_{nv}$ & $\Omega_{1} = dP^{(n)}$ & $\int_{C_{g_{r}}} dP^{(n)} = \int_{C_{g_{r}}} f^{*}r \in \mathbb{Z}_{2}, r \in H^{1}(BD_{1},\mathbb{Z}^{r})$ & Sec.~\ref{sec:3dTCI} \\ 
  & & &  $\int_{C_{u}} dP^{(n)} = \int_{C_{u}} f^{*}a = 0,
  a \in H^{1}(BC_{N},\mathbb{Z})$ & \\
 \hline
\end{tabular}
}
\end{center}
\caption{Summary of the results. The second column specified the symmetry group of the system. $\Gamma$, $D_{1}$, $C_{N}$, $C_{nv}$ denote lattice translation, reflection, $N$-fold rotation, and $C_{nv}$ point group, respectively. The third column gives the $k$-form $\Omega_{k}$ that appears in the topological term Eq.~(\ref{eqn:Seff-general}). The fourth column lists the integral conditions that $\Omega_{k}$ needs to be satisfied. Here we use $t_{I}$, $g_{r}$, and $u$ to denote the generator of the lattice translation in the $I$-th direction, reflection, and the $N$-fold rotations, respectively. A $1$-cycle labeled by $g$ is denoted by $C_{g}$, and the $2$-cycle labeled by the generator of the $N$-fold rotation $u$ is denoted by $D_{u}$, of which the boundary is given by $N$ $1$-cycles $C_{u}$. The last column shows the section of the paper where each case is discussed.
}
\label{tab:1}
\end{table*}


\section{Warm up: effective field theories of 1d atomic insulators}
\label{sec:1datom}

To illustrate the basic idea, we begin with a simple example: 1d atomic insulators. The relevant symmetry group is $U(1) \times \Gamma$, where $U(1)$ is the charge conservation symmetry and $\Gamma \cong \mathbb{Z}$ is a 1d discrete translation symmetry. Fermion parity is the $\zz$ subgroup of $U(1)$. We will focus on such an 1d atomic insulator whose building block picture has a charge-1 atom per unit cell. 

\begin{figure}
\center
\includegraphics[width=1\columnwidth]{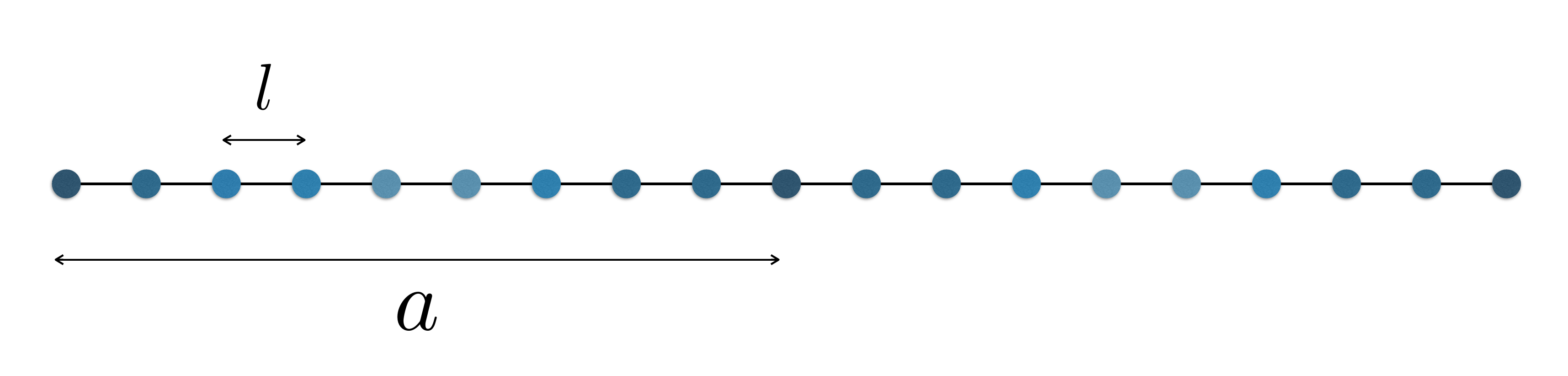}
\caption{A 1d lattice of ancillas with lattice spacing $l$ much smaller than the unit cell size $a$. The correlation length $\xi$ is assumed to be much less than the unit cell $a$. After coarse-graining with respective to the lattice space $l$, local observables have spatial variation within the unit cell. The original atoms are represented as the charge density wave (gradiant blue colors) in the lattice of ancillas.
}
\label{fig:1d-lattice}
\end{figure}

A model of a 1d atomic insulator consisting of spinless fermions on a 1d lattice with unit cell size $a$ so that there is a unit charge per unit cell. We then consider a much finer lattice by adding degrees of freedom as ancillas within each unit cell. Note that these ancillas are not the physical atoms. This setup is shown in Fig.~\ref{fig:1d-lattice}. The new model, which has a lattice spacing $l \ll a$ \footnote{To simplify the analysis, here we choose $a/l = 2m$ with $m \gg 1$ without loss of generality. One is free to choose $a/l$ to be an odd integers, which will not effect the results.}, is described by the Hamiltonian:
\begin{equation}
H = - t \sum_{x}^{L} (c_{x} c_{x+l}^{\dagger} + h.c.) - \mu \sum_{x}^{L} c_{x}^{\dagger}c_{x} + \cdots \ ,
\end{equation}
where $x$ labels the ancillas.
The ellipsis represents various perturbations consistent with the $U(1)$ and the translation symmetries $\Gamma$ while preserving the average charge per unit cell. The translation symmetry acts on the fermions by
\begin{equation}
t:  c_{x} \rightarrow c_{x+a}.
\end{equation}
By construction, the new model also has the charge $U(1)$ and translation symmetry $\Gamma$.

Now we want to derive the continuum IR limit of the theory where the fields are coarse-grained with respect to the lattice spacing $l$. To proceed, we expand the microscopic fermion operator $c_{x}$ in terms of the slowly varying low-energy fields $\psi_{R/L}$ as
\begin{equation}
c_{x} \sim \psi_{R}(x) e^{ik_{F} x} + \psi_{L}(x)  e^{- i k_{F} x},
\end{equation}
where $k_{F} = \pi/l$. Define $\psi(x) = (\psi_{R}, \psi_{L})^{T}$ as the low-energy fermion field, the continuum IR limit of the theory takes the following general form
\begin{equation}
\mathcal{L} = -i \bar{\psi} \gamma^{\mu} \partial_{\mu} \psi + i m \bar{\psi} \psi.
\label{eqn:IR_1dtrans}
\end{equation}
where $\gamma^{\mu}$ satisfies $\{\gamma^{\mu}, \gamma^{\nu}  \} = 2 g^{\mu \nu}$, and $g^{\mu \nu} =$ diag $(-1,1)$ is the Minkowski metric. We choose $\gamma^{0} = i\sigma_{y}$, $\gamma^{1} = \sigma_{x}$, and $\gamma^{01} = \sigma_{z}$.

The translation symmetry acts on the low-energy fields by
\begin{equation}
t: \psi(x) \rightarrow e^{ik_{F}a \sigma_{z}} \psi(x+a) = \psi(x+a),
\end{equation}
where we have used the fact that $k_{F}a \in 2\pi \mathbb{Z}$ for integer filling. Note that, while $\psi(x)$ varies very slowly on the scale of $l$: $\psi(x) \approx \psi(x+l)$, it's not the case on the scale of $a$: $\psi(x) \not\approx \psi(x+a)$, hence that $\psi(x)$ could vary on the scale $R \gg l$ and $R<a$.

Although this kind of model might not seem like the system one would normally consider, it has been argued that classifications and the topological properties of such systems are same as the crystalline phases in general \cite{Else2018}. In other words, all other states belonging to the same topological crystalline phase are smoothly connected to the ground state of such models (dubbed as the smooth state in Ref.~\onlinecite{Else2018}), and the smooth state can serve as a representative of the whole phase. In the next section, we are going to obtain the effective field theory in this special limit. 

\subsection{The topological term for 1d atomic insulators}
\begin{figure}
\center
\includegraphics[width=0.8\columnwidth]{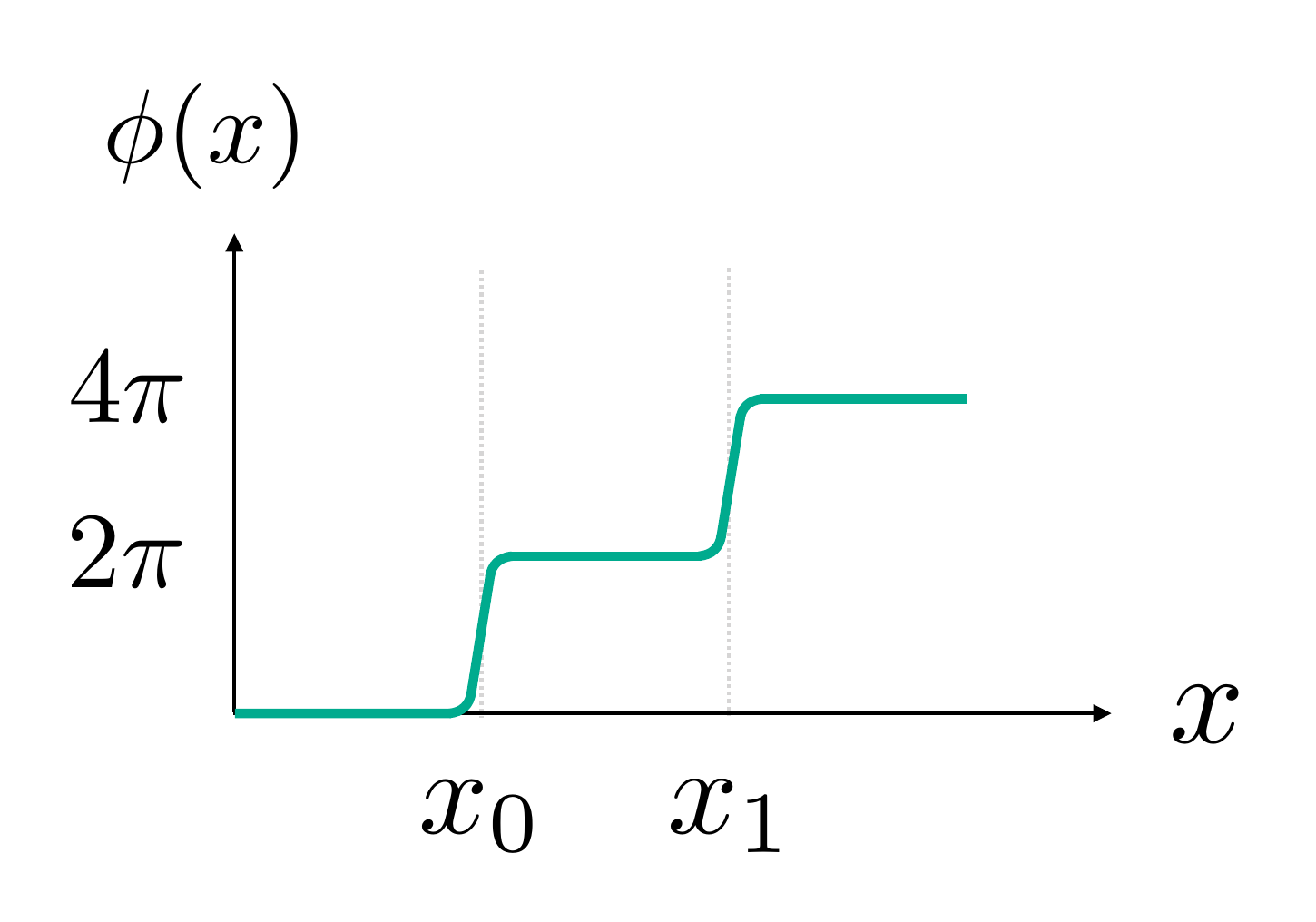}
\caption{A typical configuration of the phase $\phi(x)$ of the mass term. $\phi(x)$ jump by $2\pi$ at the location of an atom.
}
\label{fig:phi}
\end{figure}

As discussed above, the effective theory of an atomic insulator is a $1+1$d single massive Dirac theory \eqref{eqn:IR_1dtrans}. To make contact with the topological crystal picture, we add a spatially dependent mass term:
\begin{equation}
\mathcal{L}_{m} = i m_{0} \bar{\psi} e^{i \phi(x) \gamma^{01}} \psi,
\end{equation}
where $\gamma^{01} = \gamma^{0} \gamma^{1}$, $m_{0} > 0$, and we assume $m \gg m_{0}$. Here the spatial dependence of the mass term is encoded in the function $\phi(x)$. Similar to the fermion fields, $\phi(x)$ could varies on the scale $R$. Here we focus on a special configuration of $\phi(x)$ such that it's a monotonic function whose value changes abruptly by $2 \pi$ at the location of atoms as shown in Fig.~\ref{fig:phi}. It can be shown that this kind of spatially dependent mass terms trap a charge-1 bound state with a finite energy at the interfaces of $\phi(x)$. These charge-1 bound states are precisely the building blocks in the topological crystal approach and corresponds to the physical atoms \footnote{More formally, we note that the function $\phi(x)$ defines a map $\phi: X \rightarrow \mathfrak{M}_{1}$, where $\mathfrak{M}_{1}$ is the space of $1+1$d fermionic short range entangled states with $U(1)$ symmetry. The configuration of $\phi(x)$ we choose gives a noncontractible loop in $\mathfrak{M}_{1}$ every time we go through a unit cell and the bound state is associated to the winding number $\pi_{1}(\mathfrak{M}_{1}) = \z$.}.

We now discuss the classification of such mass interfaces parameterized by $\phi(x)$. We will show that the mass interfaces is classified by $H^{1}(B\Gamma,\mathbb{Z})$. To systematically discuss the configurations of $\phi(x)$, we consider the cell decomposition of $\mathbb{R}$ by the FDs. Here each one cell $\Sigma_{1,(i)}$ is simply a unit cell and is labeled by a group element $g$ in the group of discrete translation $\Gamma \cong \mathbb{Z}$. Two neighboring $1$-cells $\Sigma_{1,(i)}$, $\Sigma_{1,(j)}$ meet at a single $0$-cell $\Sigma_{0,(ij)}$, which is labeled by the generator $t$ of the translation group $\Gamma$. This kind of cell decompositions by the FDs applies to any crystalline symmetry. 

With this cell decomposition, we can now discuss the configurations of $\phi(x)$ more systematically. We will first focus on the discontinuous configurations of $\phi(x)$ since the general structure can be seen more clearly in this limit. A discontinuous configurations of $\phi(x)$ is modeled by having a constant function in each $1$-cell $\Sigma_{1,(i)}$. At the intersecting $0$-cell $\Sigma_{0,(ij)}$, where $\Sigma_{1,(i)}$ and $\Sigma_{1,(j)}$ meet, we have the relation
\begin{equation}
\phi(\Sigma_{1,(j)}) = \phi(\Sigma_{1,(i)}) + 2 \pi \tau(\Sigma_{0,(ij)}), \ \tau(\Sigma_{0,(ij)}) \in \mathbb{Z}.
\label{eqn:phi_1dtrans}
\end{equation}
The integer-valued function $\tau$ satisfies
\begin{equation}
\tau(\Sigma_{0,(ij)}) + \tau(\Sigma_{0,(jk)}) = \tau(\Sigma_{0,(ik)}).
\end{equation}
Moreover, there is a redundancy since, if we modify the configuration of $\phi$ as
\begin{eqnarray}
\phi(\Sigma_{1,(i)}) &\rightarrow& \phi(\Sigma_{1,(i)}) + 2\pi h(\Sigma_{1,(i)}),
\nonumber\\
\tau(\Sigma_{0,(ij)}) &\rightarrow& \tau(\Sigma_{0,(ij)}) + h(\Sigma_{1,(j)}) - h(\Sigma_{1,(i)})
\label{eqn:phi_trans_cobdy}
\end{eqnarray}
with $h(\Sigma_{1,(i)}) \in \mathbb{Z}$, we obtain the same configuration of $\phi$, which means that $\tau$ is a $\mathbb{Z}$-valued cocycle in $H^{1}(M,\mathbb{Z})$. This, however, does not mean that $\phi(x)$ is classified by the cellular cohomology $H^{1}(M,\mathbb{Z})$ of the manifold $M$, and we are not considering arbitrary interfaces. What we are interested in is the \emph{symmetric deformation classes} of the interface configurations of $\phi$ with a charge-1 particle at each interface. Taking the typical configuration of $\phi(x)$ shown in Fig.~\ref{fig:phi} as an example, one can see that all the deformations, which respects the translation symmetry, will not change the $2 \pi$ jumps at the locations of atoms. This structure can be captured by identifying the integer-valued transition function $\tau$ as the pullback $\tau = f ^{*}\alpha$, where $\alpha \in H^{1}(B\Gamma,\mathbb{Z})$ by the map $f : M \rightarrow B\Gamma$. The general construction of the map $f$ is given in Appendix~\ref{sec:FD}. In our case the map $f$ is constructed as follows. We note that for each $1$-cell $\Sigma_{1,(i)}$, there is a dual $0$-cell $\Sigma^{\vee}_{0,(i)}$. Similarly, for each $0$-cell $\Sigma_{0,(ij)}$, there is a dual $1$-cell $\Sigma^{\vee}_{1,(ij)}$. The map $f$ is constructed such that every dual $0$-cell $\Sigma^{\vee}_{0,(i)}$ is mapped to the based point $*$ in $B\Gamma$, and a dual $1$-cells $\Sigma^{\vee}_{1,(ij)}$ is mapped to the non-trivial loop labeled by the generator $t \in \pi_{1}(B\Gamma) = \mathbb{Z}$:
\begin{eqnarray}
f : M &\rightarrow& B\Gamma
\nonumber\\
\Sigma^{\vee}_{0,(i)} &\mapsto& *
\nonumber\\
\Sigma^{\vee}_{1,(ij)} &\mapsto& t \in \pi_{1}(B\Gamma) = \mathbb{Z}.
\end{eqnarray}
As a result, dual $1$-cells (or the original $0$-cells) are in one-to-one correspondence with the generator $t$ of the translation group $\Gamma$. This has a following implication: let $x_{0}$ be a point inside a $1$-cell, a path connecting $x_{0}$ to $gx_{0}$ is labeled by a group element $g \in \Gamma$. This property is generally true for any space group $G_{s}$---any such path can be labeled by a group element $g \in G_{s}$.

Since we have $\tau = f ^{*}\alpha$, the redundancy of the mass interfaces Eq.~(\ref{eqn:phi_trans_cobdy}) is restricted as we now discussed. We recall that $H^{1}(B\Gamma,\mathbb{Z}) \cong H^{1}(\Gamma,\mathbb{Z})$ is the quotient of 1-cocycles by 1-coboundaries. The cocycle condition reads
\begin{equation}
\alpha(g_{1}) + \alpha(g_{2}) = \alpha(g_{1}g_{2}).
\end{equation}
In other words, $\alpha(g)$ is a group homomorphism of $\Gamma$. The coboundaries in this case are all trivial:
\begin{eqnarray}
\delta \mu: \Gamma &\rightarrow& \mathbb{Z} \nonumber
\\
g &\mapsto& 0.
\label{eqn:gamma_cobdy}
\end{eqnarray}
One can easily see that $H^{1}(B\Gamma,\mathbb{Z}) = \mathbb{Z}$ with the generator given by $\alpha(t) = 1$ where $t$ is the generator of the translation group $\Gamma$. Using our construction of the map $f$, we have $ h(\Sigma_{1,j}) = h(\Sigma_{1,i})$ due to Eq.~(\ref{eqn:gamma_cobdy}). The only remaining redundancy is given an overall shift by $2 \pi$, which will not affect the value of $\tau$, and hence the $2\pi$ jumps of $\phi(x)$ at the locations of atoms. This is similar to the redundancy of changing the integer labeling of the atoms. Therefore, the symmetric deformation classes of the interfaces are classified by $H^{1}(B\Gamma,\mathbb{Z}) = \mathbb{Z}$.

To obtain an effective theory, we couple the fermion to an external background $U(1)$ gauge field $A_{\mu}$. By integrate out the massive fermion, the effective action contains the following topological term:
\begin{eqnarray}
S_{\text{eff}} &=& \frac{1}{4 \pi} \int \epsilon^{\mu \nu} \phi(x) F_{\mu \nu} d^{2}x
\nonumber\\
&=& \frac{1}{2 \pi} \int  \epsilon^{\mu \nu} A_{\mu} \partial_{\nu} \phi(x) d^{2}x.
\label{eqn:Seff1d}
\end{eqnarray}
By requiring the effective action to be gauge invariant under $A_{\mu} \rightarrow A_{\mu} + \partial_{\mu} \alpha$, we find the current
\begin{equation}
J^{\mu} = \frac{1}{2 \pi} \epsilon^{\mu \nu} \partial_{\nu} \phi
\end{equation}
is conserved. 

In the limit where $\phi(x)$ is discontinuous, the density is given by
\begin{equation}
\rho = \frac{1}{2 \pi} \partial_{x} \phi =  \sum_{i} \delta(x - x_{i}),
\end{equation}
where the discrete nature of the density of an atomic insulator is recovered. Note that, if we define the theory on a finite system with size $L$, the total charge of the system is
\begin{equation}
Q_{\text{tot}} = \int_{0}^{L} \frac{1}{2 \pi} \partial_{x} \phi dx = \frac{1}{2 \pi} \left (\phi(L)-\phi(0)) \right) = N.
\end{equation}
This topological term also reveals the ``Thouless pump" response for the atomic insulators\cite{Thouless1983}---There is a net charge flows through the system when the phase field $\phi$ winds $2\pi$ in time. 

In general, there is a coefficient $\nu$ in front of the topological term Eq.~(\ref{eqn:Seff1d}). To show this coefficient is quantized, we proceed with the following argument. Consider $\phi$ is time-independent and the spatial dependence is given by Eq.~(\ref{eqn:phi_1dtrans}) with $\tau = 1$. We then integrate along the $x$-direction and the topological term becomes
\begin{eqnarray}
S_{\text{eff}} &=& \frac{\nu}{2\pi} \int  \epsilon^{\mu \nu} A_{\mu} \partial_{\nu}\phi d^{2}x
\nonumber\\
&\sim& \nu \sum_{i} \int A_{0} \delta(x-x_{i}) d^{2}x
\nonumber\\
&=& \sum_{i} \nu \int A_{0}(x_i) dt\ ,
\end{eqnarray}
where the gauge transformations of $A_{0}(x_i)$ can be different for different $i$.
We obtain a sum of $0+1$d effective actions, each of which is the effective action of an atom couple to the $U(1)$ gauge field. By the gauge invariance, we see the $\nu$ has to be quantized to integers.

\subsection{Taking the smooth limit}
In the previous section, we take the limit where $\phi(x)$ is discontinuous in order to make a clear connection to the building block picture. We also see that these discontinuous configurations of $\phi(x)$ are classified by $\alpha \in H^{1}(B\Gamma,\mathbb{Z})$. Here we show that it's possible to take a limit where $\phi(x)$ is a smooth function such that $\phi(x)$ is still classified by $H^{1}(B\Gamma,\mathbb{Z})$. Moreover, we are going to take the limit where $\phi(x)$ is as smooth as possible such that $\partial_{x} \phi(x)$ is uniform. After taking such smooth limit, we will obtain an effective field theory which works in the usual IR limit, where the correlation length $\xi$ doesn't have to be much smaller than the unit cell $a$.

Let's recall that the defining property of an element $\alpha \in H^{1}(B\Gamma,\mathbb{Z})$ is that the pairing satisfies
\begin{equation}
\int_{C_{1}} \alpha \in \mathbb{Z},
\end{equation}
where $C_{1}$ is the non-trivial $1$-cycle of $B\Gamma$. We pull this back by using the map $f : M \rightarrow B\Gamma$, and the corresponding $1$-cocycle $e_{1} \in H^{1}(M,\mathbb{Z})$ satisfies
\begin{equation}
\int_{\Sigma_{1}^{\vee}} e_{1} = \int_{\Sigma_{1}^{\vee}} f ^{*}\alpha  \in \mathbb{Z},
\label{eqn:e-cond}
\end{equation}
where $\Sigma_{1}^{\vee}$ is a dual $1$-cell (unit cell) in the real space. This integral essentially counts the number of atoms in a unit cell. 

We would like to obtain a low-energy effective field theory where all the fields are smooth. At the same time, we want to preserve the classification of $\phi(x)$ given by $H^{1}(B\Gamma,\mathbb{Z})$. The way to achieve this is to consider closed $1$-forms with integral periods \footnote{Recall that a closed $k$-form $\omega$ on M has integral periods if, for every smooth $k$-cycle $C$ in M, the integral $\int_{C} \omega$ is an integer. Moreover, a closed $k$-form $\omega$ has integral periods if and only if the de Rham class of $\omega$ lies in the image of the change-of-coefficients map
\begin{equation}
H^{k}(M,\mathbb{Z}) \rightarrow H^{k}(M,\mathbb{R}) \cong H^{k}_{dR}(M),
\end{equation}
where $H^{k}_{dR}(M)$ denotes the de Rham cohomology of $M$ \cite{Simons2007}. Loosely speaking, a closed $k$-form with an integral period serves as a differential form representative of an element in $H^{k}(M,\mathbb{Z})$.}. Now we replace the cocycle on the left-hand side of Eq.~(\ref{eqn:e-cond}) by an smooth differential $1$-form $E_{1}$ with integral periods:
\begin{equation}
    \int_{\Sigma_{1}^{\vee}} E_{1} = N_{t} \in \mathbb{Z},
    \label{eqn:e1-int-cond}
\end{equation}
where $E_{1} = d \phi /2\pi$, and $\phi$ is a smooth function. Then we identify the integer $N_{t}$ with the right-hand side of Eq.~(\ref{eqn:e-cond}):
\begin{equation}
    N_{t} = \int_{\Sigma_{1}^{\vee}} f ^{*}\alpha  \in \mathbb{Z}.
\end{equation}

Representatives of the smooth $1$-form $E_{1}$ can be obtained by solving Eq.~(\ref{eqn:e1-int-cond}). We consider a smooth function $\phi(x)$ satisfying
\begin{equation}
\int_{x_{0}}^{x_{0}+a} d \phi = (\phi(x_{0}+a) - \phi(x_{0})) = 2\pi \tau,
\label{eqn:1d-phi-cond}
\end{equation}
where we have restored the unit cell size $a$ for the sake of clarity. An example of such function, which is as smooth as possible and satisfying Eq.~(\ref{eqn:1d-phi-cond}), is given by
\begin{equation}
\phi(x) = \frac{2 \pi}{a} \tau x = b_{1} \tau x,
\end{equation}
where $b_{1} = 2\pi/a$ is the reciprocal lattice vector (This is essentially the ``labelling" field introduced by Haldane \cite{Haldane1981}). We can then define a smooth $1$-form
\begin{equation}
E_{1} = \frac{1}{2\pi} \partial_{x} \phi(x) dx =  \frac{1}{2\pi} \tau b_{1}dx.
\end{equation}
This is a closed $1$-form with integral period since, if we integrate over a unit cell, we have
\begin{equation}
\int_{\Sigma_{1}^{\vee}} E_{1} = \int_{x_{0}}^{x_{0}+a} \frac{1}{2\pi} \tau b_{1} dx = \tau \in \mathbb{Z},
\end{equation}
which is the property that we want.

In general, there will be time-dependence in $\phi$ so that we can define the time-component of the $1$-form $E_{0} = \partial_{t} \phi dt / 2 \pi$. Written in terms of these smooth $1$-form $E$, we have the following topological term:
\begin{eqnarray}
\int  A \wedge E = \frac{1}{2\pi} \int  \epsilon^{\mu \nu} A_{\mu} b_{\nu} d^{2}x.
\end{eqnarray}
We have thus reproduced the effective field theory of an atomic insulator in Ref.~\onlinecite{Nissinen2018tetrads, Nissinen20193dqh, nissinen2020field, Nissinen2021, Else2021qc}. As one can see from the above discussion, the $1$-form $E = d\phi/2\pi$ basically tells us where to decorate the 0d building blocks. Therefore, our approach gives a direct correspondence between the topological terms and the topological crystal picture.

This discussion can be generalized to atomic insulators in higher dimensions. For example, in 2d and 3d, we expect there are topological terms of the form:
\begin{eqnarray}
S_{\text{eff}} &=& \frac{\nu}{2} \int  \epsilon_{IJ} A \wedge   E^{I} \wedge E^{J}.
\label{eqn:2d-atomic-topo}
\nonumber\\
S_{\text{eff}} &=& \frac{\nu}{6} \int \epsilon_{IJK} A \wedge E^{I} \wedge E^{J} \wedge E^{K},
\label{eqn:3d-atomic-topo}
\end{eqnarray}
where, for a translation in the $I$th direction, $E^{I}$ is the differential form representative of the cocycle in $H^{1}(BT_{I},\mathbb{Z})$. In Appendix~\ref{sec:2datomic}, we give a detailed derivation of the topological term Eq.~(\ref{eqn:2d-atomic-topo}) for the 2d atomic insulators. Generalizing to 3d atomic insulators is straightforward. 

The physical meaning of the topological terms Eq.~(\ref{eqn:2d-atomic-topo}) is that there is a charge-$\nu$ per unit cell as one can see from the effective action of the mass interface obtained by integrating out the spatial directions. Those terms also describe higher dimensional analogs of the Thouless pump\cite{Kapustin2020thouless,hsin2020berry}. Such kind of topological terms for atomic insulators are discussed in Ref.~\onlinecite{Else2021qc} in the context of topological elasticity theory (see also Ref.~\onlinecite{Nissinen2018tetrads,Nissinen20193dqh,nissinen2020field,Nissinen2021}), where $\theta^{I}$ fields are interpreted as the phonon fields. Since the topological terms we obtained above take essentially the same form, this suggests that there could be an elasticity interpretation for the spatially dependent mass terms. Indeed, in the case where the spatially dependent mass terms are generated from coupling to lattice deformations, our topological terms are the topological terms in the elasticity theory.


\section{Topological terms of topological crystalline insulators with point group symmetry}
\label{sec:TCIs}

In this section, we are going to discuss various effective field theories for TCIs with point group symmetries. We will facus on the reflection, $C_{N}$ rotation, and $C_{nv}$ symmetries for 1d, 2d, and 3d TCIs. The 1d and 2d TCIs that we are going to discuss are built by placing 0d charges at the high symmetry points, which do not support protected boundary gapless modes. For 3d TCIs, we will consider the 2nd-order topological phases with gapless chiral hinge models, whose building block picture is given by placing 2d IQH states on high symmetry planes.

\subsection{1d insulators with reflection symmetry}
\label{sec:1dr}
We now move on to discuss topological crystalline insulators with reflection symmetry in 1d. The symmetry group we focus on is $G = U(1) \times D_{1}$, where the reflection group $D_{1} \cong \mathbb{Z}_{2}$. We are going to focus on the phase of which the building block picture is given by placing a 0d state carrying a unit $U(1)$ charge and a trivial irreducible representation of $D_{1}$ at the reflection center. 

\begin{figure}
\center
\includegraphics[width=0.8\columnwidth]{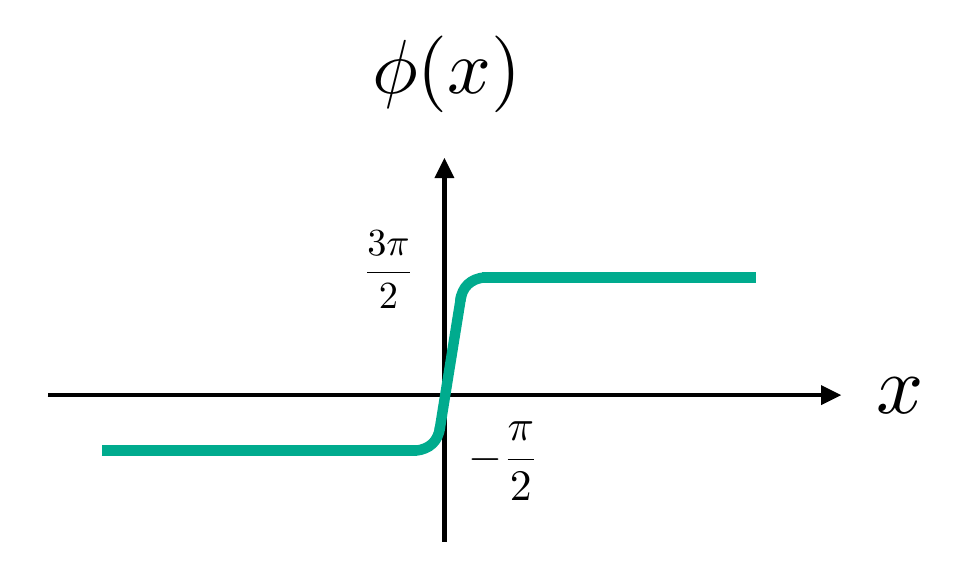}
\caption{A typical configuration of $\phi(x)$ which supports a charge-1 bound state at the reflection center.
}
\label{fig:phi-r}
\end{figure}

The low energy theory of this kind of insulators is given by the following massive Dirac theory
\begin{equation}
\mathcal{L} = -i \bar{\psi} \gamma^{\mu} \partial_{\mu} \psi -   m_{1} \bar{\psi} \gamma^{01}  \psi
\label{eqn:dirac-r}
\end{equation}
with the reflection symmetry acting on the fermions by
\begin{equation}
g_{r}: \psi(x,t) \rightarrow \gamma_{1}\psi(-x,t).
\end{equation}
We are going to show that this Dirac theory indeed describes the TCI we are interested in. We proceed with the dimensional reduction procedure by adding a spatially varying mass term:
\begin{equation}
\mathcal{L}_{m} = - i m_{2} \bar{\psi} e^{i \phi(x)\gamma^{01}}  \psi,
\end{equation}
where we require that $m_{1} \gg m_{2} > 0$. 
The reflection symmetry requires that $\phi(x) = \pi - \phi(-x)$ mod $2\pi$. The phase $\phi(x)$ could wind non-trivially in space subjecting to the constraint given by the reflection symmetry. If we consider a configuration of $\phi(x)$ such that it winds $2\pi$ along a path passing through the reflection center (as shown in Fig.~\ref{fig:phi-r}), by solving the bound state directly for this mass interface, one can confirm that there is a charge-1 bound state sitting at the reflection center. This is precisely the building block picture for this phase, which justifies the claim that Eq.~(\ref{eqn:dirac-r}) describes the TCI we are interested in.

\begin{figure}
\center
\includegraphics[width=0.8\columnwidth]{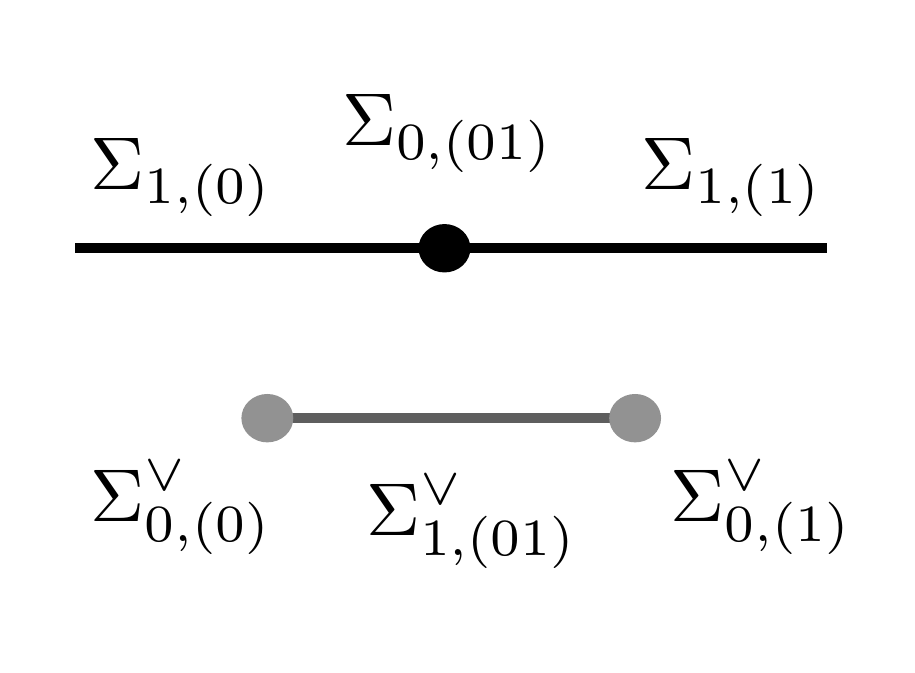}
\caption{Reflection symmetric cell-decomposition (above) and the dual-cell-decomposition (below).
}
\label{fig:Dualcell-r}
\end{figure}

We are going to show that the mass interfaces parameterized by $\phi(x)$ are classified by the cellular cohomology $H^{1}(B D_{1}, \mathbb{Z}^{r})$ with a twisting coefficient $\mathbb{Z}^{r}$, which will be defined below. To proceed, we first discuss the cell decomposition given by the FDs. As shown in Fig.~\ref{fig:Dualcell-r}, there are two 1-cells $\Sigma_{1,(0)}$, $\Sigma_{1,(1)}$ and a single 0-cell $\Sigma_{0,(01)}$ at the reflection center. The dual cell structure is also shown in Fig.~\ref{fig:Dualcell-r}, which contains two dual 0-cells $\Sigma^{\vee}_{0,(0)}$, $\Sigma^{\vee}_{0,(1)}$ and one dual 1-cell $\Sigma^{\vee}_{1,(01)}$. With this choice of cell-decomposition, there is a map $f : M \rightarrow BD_{1}$. Written explicitly, the map $f : \Sigma^{\vee}_{0,(0)} \mapsto *$, $ \Sigma^{\vee}_{0,(1)} \mapsto *$,  $ \Sigma^{\vee}_{1,(01)} \mapsto g_{r}$, where $g_{r} \in \pi_{1}(BD_{1}) \cong D_{1}$ denotes the non-trivial group element in $D_{1}$.

We consider the configurations of $\phi(x)$ such that it's a constant function within the two 1-cells $\Sigma_{1,(0)}$, $\Sigma_{1,(1)}$. At the intersecting 0-cell $\Sigma_{0,(01)}$, the constant functions satisfy the relation:
\begin{equation}
\phi(\Sigma_{1,(1)}) = \phi(\Sigma_{1,(0)}) + 2\pi r(\Sigma_{0,(01)}),
\label{eqn:r}
\end{equation}
where $r(\Sigma_{0,(01)}) \in \mathbb{Z}$. The reflection symmetry gives the following condition
\begin{equation}
\phi(\Sigma_{1,(0)}) + \phi(\Sigma_{1,(1)}) = \pi \ \text{mod} \ 2\pi
\label{eqn:prcond-1d}
\end{equation}
as well as a non-trivial action on the integer-valued function $r$
\begin{equation}
g_{r} \cdot r = -r.
\label{eqn:rtwist}
\end{equation}
However, there is a redundancy since, if we modify the configuration of $\phi$ as
\begin{eqnarray}
\phi(\Sigma_{1,(i)}) &\rightarrow& \phi(\Sigma_{1,(i)}) + 2\pi h(\Sigma_{1,(i)}), 
\nonumber\\
r(\Sigma_{0,(01)}) &\rightarrow& r(\Sigma_{0,(01)}) + h(\Sigma_{1,(1)}) - h(\Sigma_{1,(0)}),
\end{eqnarray}
we obtain the same configuration of $\phi(x)$. These conditions tell us that $r$ is a $\mathbb{Z}$-valued cocycle in $H^{1}(M,\mathbb{Z}^{r})$ with a twisting coefficient given by Eq.~(\ref{eqn:rtwist}).

We consider the mass interface such that $r$ is given by the pullback $r = f ^{*}\upsilon $ of $\upsilon  \in H^{1}(BD_{1},\mathbb{Z}^{r}) \cong \mathbb{Z}_{2}$. To understand this $\mathbb{Z}_{2}$ classification, we note that the coboundaries in $H^{1}(BD_{1},\mathbb{Z}^{r})$ are of the form
\begin{eqnarray}
\delta \nu^{0}:D_{1} & \rightarrow & \mathbb{Z} \nonumber \\
g_{r} & \mapsto & 2m
\label{r-coboundaries}
\end{eqnarray}
for some $m \in \mathbb{Z}$, where $\nu^{0}$ is a function taking the based point $* \in BD_{1}$ to an integer $\nu^{0}(*) \in \mathbb{Z}$. By using the map $f:M \rightarrow BD_{1}$, we have
\begin{eqnarray}
  h(\Sigma_{1,(0)}) &=& \nu^{0}(f(\Sigma_{1,(0)})) = \nu^{0}(*),
  \nonumber\\
  h(\Sigma_{1,(1)}) &=& \nu^{0}(f(\Sigma_{1,(1)})) = \nu^{0}(g_{r} \cdot *),
\end{eqnarray}
where $g \cdot * \sim *$ in the classifying space $BD_{1}$. Using Eq.~(\ref{r-coboundaries}), we have $h(\Sigma_{1,(1)}) - h(\Sigma_{1,(0)}) = \delta \nu^{0} = 2m$. As a result, there is an equivalence relation $r \sim r+2m$, and the non-trivial winding of the phase $\phi(x)$ has a $\mathbb{Z}_{2}$ classification. Pictorially, it's easy to see that a interface with a $4\pi$ jump can be deformed into a configuration with no interface while preserving the reflection symmetry.

After coupling to the $U(1)$ gauge fields and integrated out the massive fermions, we obtain an topological term which takes the same form as Eq.~(\ref{eqn:Seff1d}). The difference is that $\phi(x)$ has to satisfy Eq.~(\ref{eqn:prcond-1d}) due to the reflection symmetry. We thus define a 1-form
\begin{equation}
    dP = \frac{d\phi}{2\pi}.
\end{equation}
The topological term becomes
\begin{eqnarray}
S_{\text{eff}} &=& k\int A \wedge dP,
\label{eqn:Seff1dr}
\end{eqnarray}
where the coefficient $k=1$ in our example.
Note that the coefficient $k$ is $\mathbb{Z}_{2}$ valued since the mass interfaces have $\mathbb{Z}_{2}$ classification.
In the next section, we are going to show how to connect the topological term Eq.~(\ref{eqn:Seff1dr}) to the known result in Ref.~\onlinecite{Ramamurthy2015, nissinen2020field} by taking the smooth limit for $\phi(x)$.

\subsubsection{Smooth limit}
Now we would like to consider smooth configurations of $\phi(x)$ such that the mass interfaces are still classified by $H^{1}(BD_{1},\mathbb{Z}^{r})$. To proceed, we first rewrite Eq.~(\ref{eqn:r}) in terms of the cellular cohomology:
\begin{equation}
 \int_{\Sigma_{1}^{\vee}} dP'_{1} = \int_{\partial \Sigma_{1}^{\vee}}P'_{1} = r,
\label{eqn:dP-cell-cond}
\end{equation}
where, for the sake of convenience, we have defined $P'_{1} = \phi/2\pi-1/4$ to subtract the constant $\pi/2$ contribution in $\phi$. $P'_{1}$ is now simply odd under refection: $g_{r}: P'_{1} \rightarrow -P'_{1}$. 

What we are looking for is the smooth version of Eq.~(\ref{eqn:dP-cell-cond}) such that the $1$-form $dP_{1}'$ is constructed by smooth functions. This is achieved by considering the following smooth 1-form with integral periods:
\begin{eqnarray}
\int_{-x_{0}}^{g_{r}\cdot (-x_{0})} d\tilde{P}_{1} &=&  \int_{-x_{0}}^{x_{0}} d\tilde{P}_{1} 
\nonumber\\
&=& \tilde{P}_{1}(x_{0}) - \tilde{P}_{1}(-x_{0}) 
\nonumber\\
&=& r,
\label{eqn:ptcond}
\end{eqnarray}
where $|x_{0}| \gg \xi$ and $\tilde{P}$ is smooth. We can thus write the low-energy effective action in terms of the smooth $1$-form $\tilde{P}$:
\begin{eqnarray}
S_{\text{eff}} &=& k \int A \wedge d\tilde{P}.
\label{eqn:Seff1dpt}
\end{eqnarray}
Under the reflection symmetry $\tilde{P} \rightarrow - \tilde{P}$. After an integration by parts, we recognized that Eq.~(\ref{eqn:Seff1dpt}) is essentially the effective action obtained in Ref.~\onlinecite{Ramamurthy2015, nissinen2020field}, and $\tilde{P}$ can be interpreted as the spatially dependent electric polarization.

\subsection{2d insulators with rotational symmetry}
\label{sec:2dTCI}
Here we discuss the topological term of fermionic TCIs with $U(1) \times C_{N}$ symmetry
\footnote{More precisely, the symmetry is $(U(1) \times C_{N})/\mathbb{Z}_2$, as the $C_N$ rotation $U$ defined in Eq.~(\ref{eqn:2d-cn}) obeys $U^N = (-1)^{N_f}$, where $N_f$ is the total fermion number under $U(1)$.
}.
The classification of these systems has been computed in Ref.~\onlinecite{Meng2018rot} by using the topological crystal approach. For the sake of simplicity, here we focus on the states with no charge and thermal Hall conductivity, and with a building block picture given by placing a $0$d state carrying a unit $U(1)$ charge and a trivial $C_{N}$ charge at the rotational center.

The low energy field theory of this state is given by a 2+1D massive Dirac theory:
\begin{equation}
\mathcal{L} = -i \bar{\Psi} \gamma^{\mu} \partial_{\mu} \Psi + i m_{0} \bar{\Psi} \sigma_{3} \Psi,
\label{eqn:2d-dirac-cn}
\end{equation}
where $\Psi = (\psi_{1},\psi_{2})$, and $\sigma_{i}$ are Pauli matrices in the flavor space. The mass term here guarantees that the Chern number is zero. The $C_{N}$ rotation acts on the fermions by
\begin{equation}
u : \Psi(\boldsymbol{r}) \rightarrow \exp{\left(\frac{i}{2}\frac{ 2\pi}{N} \gamma^{0}\sigma_{3}\right)}\Psi(R\boldsymbol{r}).
\label{eqn:2d-cn}
\end{equation} 

One can recover the building block state by adding the following spatially varying mass term:
\begin{equation}
\mathcal{L}_{m} = i m \bar{\Psi} \left(n^{1}(\boldsymbol{r})\sigma_{1} + n^{2}(\boldsymbol{r})\sigma_{2} \right) \Psi,
\end{equation}
where we require that $m_{0} \gg m > 0$
and consider the configuration of $n^{1}$ and $n^{2}$ such that there is a bound state carrying a unit $U(1)$ charge at the origin. An example of such configuration would be a ``hedgehog" with a singularity at the origin. Usually a hedgehog configuration is invariant under a continuous rotational symmetry, here we only require that it's invariant under a discrete $C_{N}$ rotation. We note that Shiozaki shows there there is an isomorphism between the group of 0d building blocks and the K group of the Dirac Hamiltonians with the hedgehog-mass potential with a unit winding number \cite{Shiozaki2019dirac}. Starting from a 0d Hamiltonian describing the 0d building state, one can obtain the massive Dirac theory Eq.~(\ref{eqn:2d-dirac-cn}) by using his general construction.

Since the system is rotational invariant, it's convenient to parametrize the mass term by
\begin{equation}
n^{1} = n_{r} \cos{n_\theta}, \ n^{2} = n_{r} \sin{n_\theta}.
\label{eqn:n-parameters}
\end{equation}
where $n_{r}$ can be taken to be a constant almost everywhere and $n_{r} \rightarrow 0$ as $r \rightarrow 0$. Following from the $C_{N}$ transformation on the fermions Eq.~(\ref{eqn:2d-cn}), $n_{\theta}$ must satisfy
\begin{equation}
    n_{\theta}(\theta + \frac{2\pi}{N}) = n_{\theta}(\theta)+ \frac{2\pi}{N} \ \text{mod} \ 2\pi.
\label{eqn:ntheta-cond}
\end{equation}
In general, the singularity at the origin is described by a one-form $\omega_{1}$ such that $d\omega_{1} = (\partial_{i}\partial_{j} - \partial_{j}\partial_{i})n_{\theta} \neq 0$. As we will discuss in detail below, this one-form $\omega_{1}$ is classified by $H^{1}(BC_{N}, \mathbb{Z}_N)$.

After coupling the fermion to a $U(1)$ gauge field and integrate out the massive fermions, the effective theory contains the following topological term:
\begin{eqnarray}
S_{\text{eff}} = \int A \wedge \omega_{2}
\label{eqn:Seff-cn}
\end{eqnarray}
where the 2-form $\omega_{2}$ is the Euler class in $H^{2}(M,\mathbb{Z})$. A perturbative derivation of this topological term is given in Appendix~\ref{sec:2drotation}. In general, we have $\omega_{2} = n^{*}\tau_{2}$, where $\tau_{2}$ is $2$-form in the space of the mass parameters. However, we are going to show that the topological crystal picture tells us that the $2$-form $\omega_{2}$ should be identified with the pullback $f^{*} \alpha_{2}$, where $\alpha_{2} \in H^{2}(BC_{N},\mathbb{Z})$ and the map $f:M \rightarrow BC_{N}$ is the map from the manifold $M$ to the classifying space $BC_{N}$.

The appearance of the Euler class $H^{2}(M,\mathbb{Z})$ is very natural (see Ref.~\onlinecite{Else2019defect, Hason2020smith} for encountering the Euler class in similar situations). Note that, since $n^{1}$ and $n^{2}$ transform as the regular representation of the $C_{N}$ rotational symmetry, it makes sense to view them as a vector field $\boldsymbol{n} = (n^{1},n^{2})$. Formally, we have a $\mathbb{R}^{2}$ real vector bundle $V$ over the manifold $M$ (We will always assume the manifold $M$ to be a euclidean space $\mathbb{E}^{2}$), and the vector field $\boldsymbol{n}$ is the section of this bundle: $n: M \rightarrow V$ such that $\pi \circ n = id$, where $\pi$ is the projection $\pi: V \rightarrow M$. The singular configurations of $\boldsymbol{n}$ corresponds to the zero sections of the vector bundle $V$, and it's well-known that the Euler class $e(V) \in H^{2}({M,\mathbb{Z}})$ counts the number of zero sections\cite{Bott1982}. 

\begin{figure}
\center
\includegraphics[width=1\columnwidth]{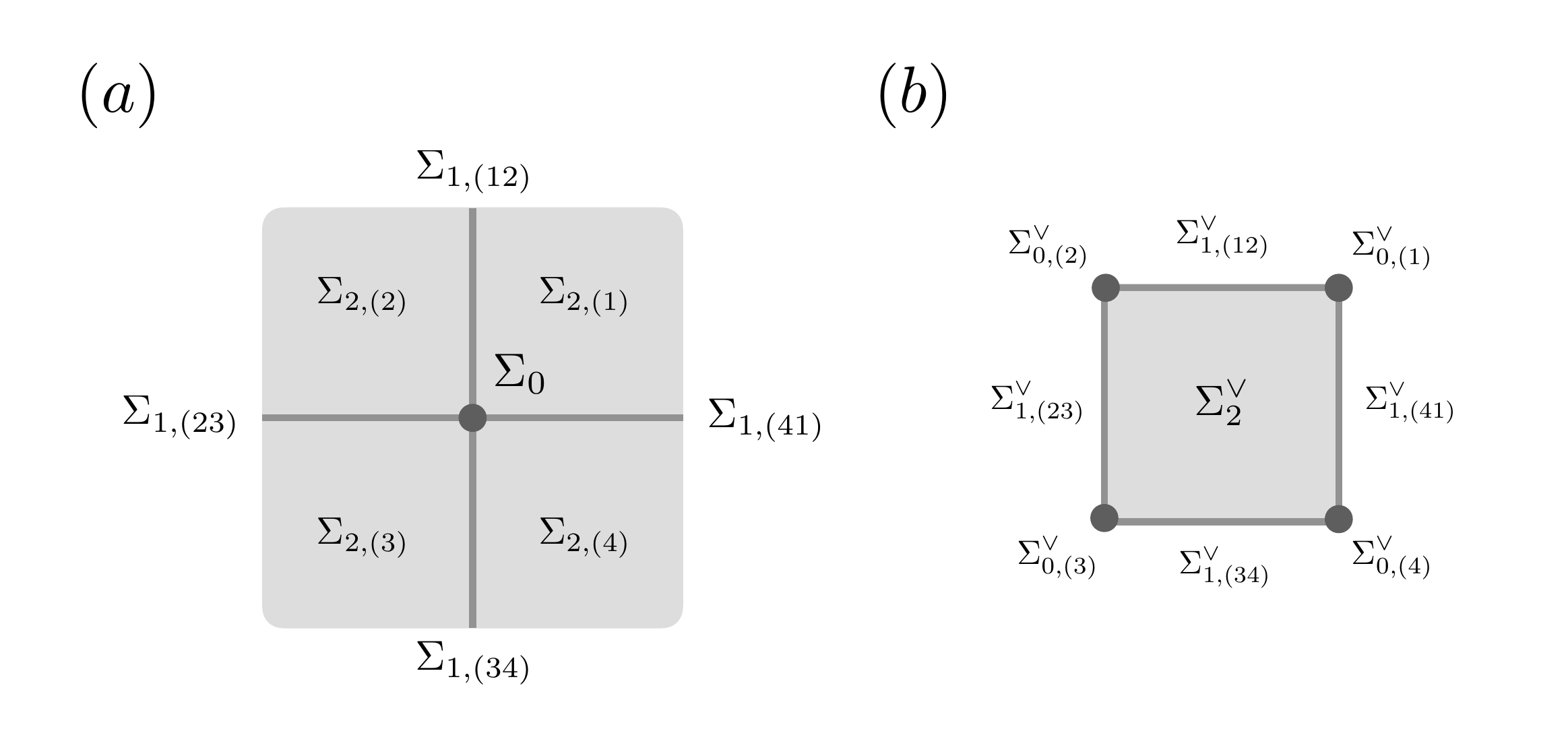}
\caption{(a) The cell decomposition for a system with $C_{4}$ symmetry, and (b) the corresponding dual-cell decomposition.
}
\label{fig:c4_cell}
\end{figure}

Now we discuss how to construct the $2$-form $\omega_{2}$ explicitly. We begin with the the cell decomposition of $\mathbb{R}^{2}$ given by the FDs. Each $2$-cell $\Sigma_{2,(i)}$ is labeled by a group element $g \in C_{N}$. Two neighboring two cells $\Sigma_{2,(i)}$, $\Sigma_{2,(j)}$ meet at a single $1$-cell $\Sigma_{1,(ij)}$, which is labeled by the generator $u \in C_{N}$. There is a unique $0$-cell $\Sigma_{0}$ sitting at the rotational center. An example of the $C_{4}$ symmetric cell-decomposition is shown in Fig.~\ref{fig:c4_cell}(a), and the corresponding dual-cell decomposition is shown in Fig.~\ref{fig:c4_cell}(b).

\begin{figure}
\center
\includegraphics[width=0.8\columnwidth]{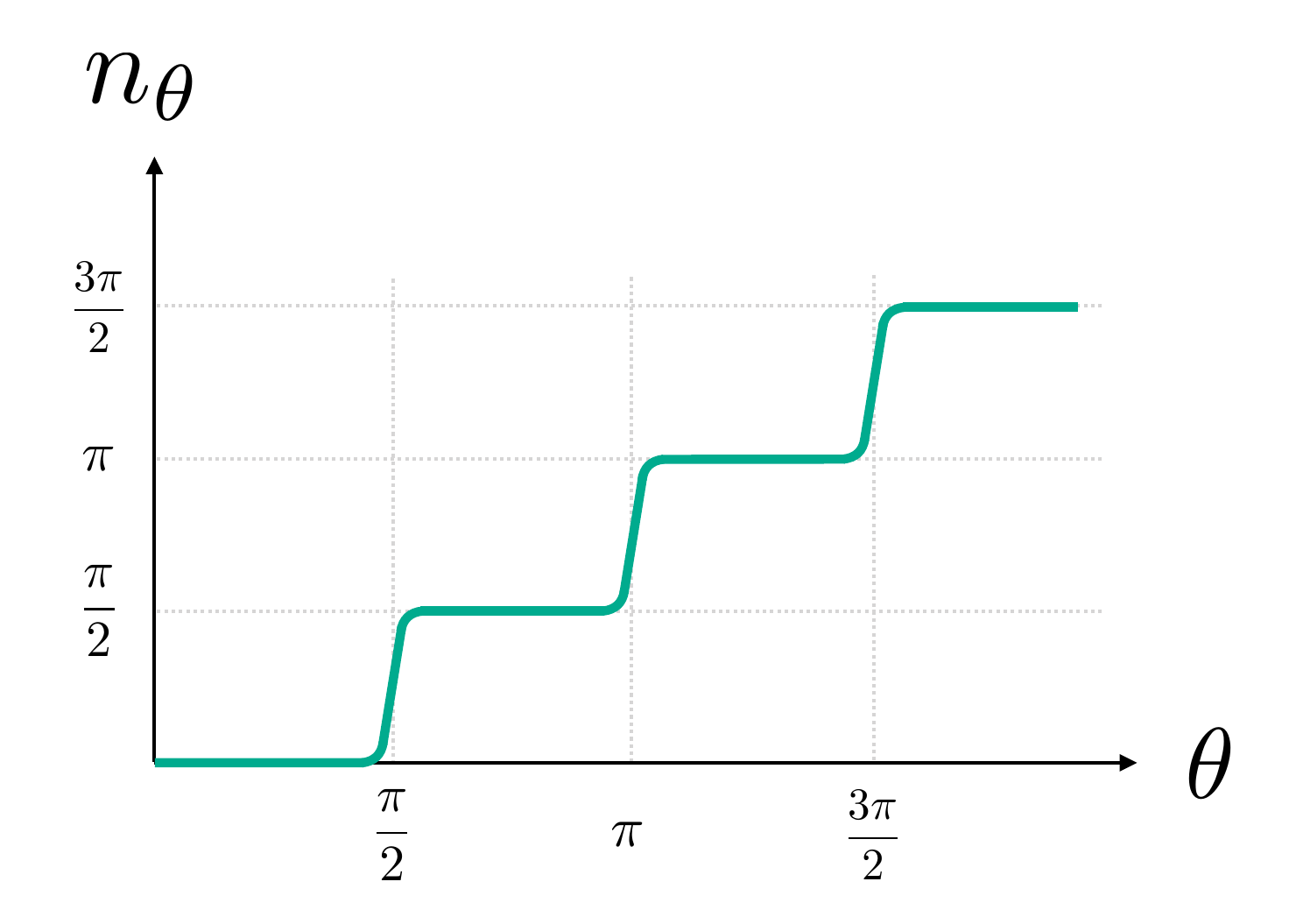}
\caption{A typical configuration of $n_{\theta}$ for systems with $C_{4}$ symmetry. 
}
\label{fig:nt}
\end{figure}

We now discuss the configurations of the mass term parameterized by Eq.~(\ref{eqn:n-parameters}) from a more general perspective. Since we only require that the mass term to be invariant under a discrete $C_{N}$ rotation, $n_{\theta}$ can be chosen to be a constant in each $2$-cell $\Sigma_{2,(i)}$. At the intersections of $2$-cells, $n_{\theta}$ could jump abruptly. We proceed with the following systematically discussion. At the $1$-cell $\Sigma_{1,(ij)}$ where $\Sigma_{2,(i)}$ and $\Sigma_{2,(j)}$ meet, we have the relation
\begin{equation}
n_{\theta}(\Sigma_{2,(j)}) = n_{\theta}(\Sigma_{2,(i)}) + \frac{2 \pi}{N} c(\Sigma_{1,(ij)}), \ c(\Sigma_{1,(ij)}) \in \mathbb{Z}.
\label{eqn:nt}
\end{equation}
A typical configuration of $n_{\theta}$ for $C_{4}$ symmetric system is shown in Fig.~\ref{fig:nt}. The integer-valued function $c$ satisfies
\begin{equation}
c(\Sigma_{1,(ij)}) + c(\Sigma_{1,(jk)}) = c(\Sigma_{1,(ik)}).
\end{equation}
There is a redundancy of the function $c(\Sigma_{1,(ij)})$ since we obtain the same configuration of $n_{\theta}$ after we modify the configuration as
\begin{eqnarray}
n_{\theta}(\Sigma_{2,(i)}) &\rightarrow& n_{\theta}(\Sigma_{2,(i)}) + 2\pi h(\Sigma_{2,(i)}), 
\nonumber\\
c(\Sigma_{1,(ij)}) &\rightarrow& c(\Sigma_{1,(ij)}) + h(\Sigma_{2,(j)}) - h(\Sigma_{2,(i)}),
\end{eqnarray}
, where $h(\Sigma_{2,(i)}) \in \mathbb{Z}$. Formally, $c$ is a $\mathbb{Z}$-valued cocycle in $H^{1}(M,\mathbb{Z})$. Here we consider the interface such that $c = f ^{*}\alpha_{1}$, where $\alpha_{1} \in H^{1}(BC_{N},\mathbb{Z}_{N})$, which is lifted to $C^{1}(BC_{N},\mathbb{Z})$ by the embedding $\mathbb{Z}_{N} \sim [0,N-1) \subset \mathbb{Z}$. The deformation classes of the $n_{\theta}$ interfaces are classified by $H^{1}(BC_{N},\mathbb{Z}_{N}) = \mathbb{Z}_{N}$. If we choose $c(\Sigma_{1,(ij)}) = 1$, we obtain a class of interfaces of $n_{\theta}$ satisfying Eq.~(\ref{eqn:ntheta-cond}). Such choice can always be made since we can always choose $\alpha_{1}(u) = 1$, where $u$ is the generator of the $C_{N}$ rotation and $\alpha(u) \in H^{1}(BC_{N},\mathbb{Z}_{N})$. We then have 
\begin{equation}
    c(\Sigma_{1,(ij)}) = \alpha_{1}(f(\Sigma_{1,(ij)}))
    = \alpha_{1}(u) = 1,
\label{eqn:alpha-g}
\end{equation}
where we have used the fact that $f: \Sigma_{1,(ij)} \mapsto u$ for every $1$-cell $\Sigma_{1,(ij)}$.

We now use the cocycle $c \in H^{1}(M,\mathbb{Z})$ to construct the desire $2$-form in the topological term. At the level of cellular cohomology, we would like to have some 2-cocycle $w = \delta c $ satisfying the following property:
\begin{eqnarray}
\frac{1}{N}\int_{\Sigma_{2}^{\vee}} w = \frac{1}{N}\int_{\Sigma_{2}^{\vee}} \delta c = \frac{1}{N}\int_{\partial \Sigma_{2}^{\vee}} c = 1.
\label{eqn:int-u}
\end{eqnarray}
Physically, this means that $n_{\theta}$ rotates by $2\pi$ as it goes around the origin along the $1$-cells, which, intuitively, corresponds to having a vortex configuration of $\boldsymbol{n}$ with strength $1$. There will be a localized charge-1 bound state trapped at the vortex core, which reproduces the building block picture. We claim the property Eq.~(\ref{eqn:int-u}) is satisfied if $\delta c = f^{*} \delta \alpha_{1}$. Using Eq.~(\ref{eqn:alpha-g}), we can explicitly check Eq,~(\ref{eqn:int-u}) is satisfied:
\begin{eqnarray}
\frac{1}{N}\int_{\Sigma_{2}^{\vee}} w &=& \frac{1}{N}\int_{\partial \Sigma_{2}^{\vee}} c
\nonumber\\
&=& \frac{1}{N} \sum_{i=1}^{N} c(\Sigma_{1,(i,i+1)}) 
\nonumber\\
&=& \frac{1}{N} \sum_{i=1}^{N} \alpha_{1}(g)
\nonumber\\
&=& 1.
\label{eqn:int-u-2}
\end{eqnarray}

The two form $\omega_{2}$ we are looking for is the differential form representative of the $2$-cocycle $w$, which is a $2$-form with integral periods. Now we discuss an explicit construction of the $2$-form $\omega_{2}$. First we need to use two patches $U_{0}$ and $U_{1}$ to cover the real space $\mathbb{R}^{2}$ such that the singularity of the section $\boldsymbol{n}$ (at the origin) is contain entirely in $U_{0}$, and that $U_{1}$ covers $\mathbb{R}^{2} - D_{0}$, where $D_{0}$ is the disc covered by $U_{0}$. Let $\rho_{0}(r)$ and $\rho_{1}(r)$ be the partition of unity satisfying $\rho_{0}(r)+\rho_{1}(r)=1$ subordinate to $U_{0}$ and $U_{1}$. In the patch $U_{1}$, we use the configuration of $n_{\theta}$ discussed above and partition of unity to define a 1-form
\begin{equation}
\omega_{1}^{(1)} =  - \rho_{0} \frac{d n_{\theta}}{2\pi}  =  \rho_{0} \sum_{i=1}^{N} \frac{c_{i,i+1}}{N} \delta \left( \theta - \frac{2\pi c_{i,i+1}}{N} \right) d\theta,
\end{equation}
where we have used a short hand notation $c_{i,j}$ to denote $c(\Sigma_{1,ij})$ and $c_{N,N+1} = c_{N,1}$. In the patch $U_{0}$, we define the $1$-form $\omega_{1}^{(0)} = \rho_{1} dn_{\theta}/2\pi$.

Finally, we construct the two form 
\begin{equation}
\omega_{2} =  d \omega_{1}^{(1)} = - \frac{1}{2\pi} d \rho_{0} \wedge d n_{\theta}
\end{equation}
such that it has an integral period:
\begin{eqnarray}
\int_{M} \omega_{2} &=& 1.
\label{eqn:intw}
\end{eqnarray}
From the property of the partition of unity, $\omega_{2}$ has support near the intersection $U_{0} \cap U_{1}$. One can show that $d\omega_{1}^{(0)} = d\omega_{1}^{(1)}$ so that they piece together to a well-defined $2$-form $\omega_{2}$. We will then drop the superscript when it's not relevant to the context. Eq.~(\ref{eqn:intw}) can be checked explicitly as follows:
\begin{eqnarray}
\int_{M} \omega_{2} &=& \int_{M} d\omega_{1}^{(1)} 
\nonumber\\
&=& - \frac{1}{2\pi} \int_{M} d \rho_{0} \wedge d n_{\theta}
\nonumber\\
&=& - \frac{1}{2\pi} (\rho_{0}(\infty) - \rho_{0}(0)) \int d n_{\theta}
\nonumber\\
&=& \frac{1}{2\pi} \int \sum_{i=1}^{N} \frac{c_{i,i+1}}{N} \delta \left( \theta - \frac{2\pi c_{i,i+1}}{N} \right) d\theta
\nonumber\\
&=&1
\end{eqnarray}
Explicit form of the topological term is given by substituting this $2$-form $\omega_{2} = d \omega_{1}$ into Eq.~(\ref{eqn:Seff-cn}).

In this example, the coefficient in front of the topological term is $1$ by construction. In general, there is a coefficient $\kappa$ and the topological term takes the form:
\begin{equation}
\label{eq:2D_rotation_Seff_kappaAdomega1}
\kappa \int A \wedge d\omega_{1}.
\end{equation}
To show that the coefficient $\kappa$ is quantized, we integrate out the real space. By construction, we have
\begin{equation}
\label{eq:2D_rotation_Seff_kappa_quantization}
\kappa \int_{M \times S^{1}} A \wedge d\omega_{1} = \kappa \int_{S^{1}} A_{0} dt,
\end{equation}
which is precisely the effective action of a $0$d particle carrying charge-$\kappa$. Gauge invariance requires that $\kappa$ is quantized. Moreover, $\kappa$ is a mod $N$ integer as we now show. From the block equivalence relation in the topological crystal picture, a state with $0$ charge is equivalent to a state with $N$ charge at the rotational center. This can be understood as the following deformation process. Starting from a state with no charge, we bring in $N$ additional charge-$1$ particles to the rotational center while preserving the $C_{N}$ symmetry (and sending $N$ additional charge-$(-1)$ particles to infinite, which are not relevant to the bulk property). At the level of field theory, if there are $N$ charge-$1$ particles at the rotational center, the interface is described by a $2$-form $d\omega_{1}'$, in which $c'_{i,i+1} = N$. However, this $2$-form is trivial as one can check as follows:
\begin{eqnarray}
\int_{M} d\omega'_{1}{}^{(1)} &=& \frac{1}{2\pi} \sum_{i=1}^{N} \frac{c'_{i,i+1}}{N},
\nonumber\\
&=& \frac{1}{2\pi} \sum_{i=1}^{N} \frac{c'_{i,i+1} - dh_{i,i+1}}{N},
\nonumber\\
&=& 0
\end{eqnarray}
where, in the first equality, we have integrated over the $r$ and $\theta$ directions, and, in the second equality, we have shifted the cocycle $c'_{i,i+1}$ by a coboundary $dh_{i,i+1} = N$. This implies that $\kappa$ is a mod $N$ integer.

\subsubsection{Smooth limit}
Here we show that it's possible to deform the function $n_{\theta}$ to a smooth function such that the 2-form $\omega_{2}$ is smooth almost everywhere except having a singularity at the origin. We begin with the $1$-form $\omega_{1}$. To go to the smooth limit, the key is to consider the smooth $1$-form $d \tilde{n}_{\theta}$ of the angular variable $\tilde{n}_{\theta}$ such that
\begin{eqnarray}
\int_{\Sigma_{1}^{\vee}} d\tilde{n}_{\theta} &=& \int_{\theta_{0}}^{\theta_{0}+2\pi/N} d\tilde{n}_{\theta} \nonumber
\\
&=& \tilde{n}_{\theta}\left( \theta_{0}+\frac{2\pi}{N} \right) - \tilde{n}_{\theta}\left( \theta_{0} \right) \nonumber
\\
&=& \frac{2 \pi}{N} \tilde{c},
\label{eqn:ctilde}
\end{eqnarray}
where $\tilde{c} \in \mathbb{Z}$ is one-to-one correspondent to $c \in H^{1}(M,\mathbb{Z})$ in Eq.~(\ref{eqn:nt}), which is determined by the pullback $f^{*}\alpha_{1}$ with $\alpha_{1} \in H^{1}(BC_{N},\mathbb{Z}_{N})$. The smooth $2$-form $\tilde{\omega}_{2}$ is constructed in same way as in the discontinuous case:
\begin{equation}
\tilde{\omega}_{2} = d \tilde{\omega_{1}},
\end{equation}
where $\tilde{\omega}_{1} = -\rho_{0} d \tilde{n}_{\theta}/2\pi$ with $\tilde{n}_{\theta}$ satisfying Eq.~(\ref{eqn:ctilde}). Substituting this 2-form into Eq.~(\ref{eq:2D_rotation_Seff_kappaAdomega1}), we obtain the topological term in the smooth limit.

\subsection{3d topological crystalline insulators}
\label{sec:3dTCI}
Here we discuss the effective theories of the 3d topological crystalline insulators with $U(1) \times G_{pg}$, where $G_{pg}$ is a point group symmetry. The non-interacting classifications have been obtained in Ref.~\onlinecite{Shiozaki2019AHSS} by computing the Atiyah-Hirzebruch spectral sequence for $K$-homology, which is the rigorous mathematical framework of the topological crystal approach for free fermion systems. We will be focusing on one type of the 2nd order topological insulators~\cite{Benalcazar2017HOTI,Schindler2018HOTI,Varnava2018AI,Ahn2019C2T}, which break time-reversal symmetry and  host protected gapless chiral hinge modes in an open geometry. To illustrate the basic idea, we will consider $G_{pg} = C_{nv}$ for $n=1,2,3,4,6$ but our approach can be generalized to other point groups. Such 2nd order phases can be described by the topological crystal pictures of having some 2d integer quantum hall (IQH) states placing at some high symmetry planes. With proper open geometry, the protected chiral hinge modes are directly given by the gapless edge modes of the 2d IQH states~\cite{Schindler2018HOTI,Ahn2019C2T}.
It was known that such a type of 2nd order topological insolators can be well-described by the effective axion field proposed in \refcite{Qi2008TFT} (see also Ref.~\onlinecite{fu2021bulkhinge}).
It coincides with the fact that the 2nd order topological insulators that we considered can be well-described by the effective axion field, and the chiral hinge modes are the domain-wall modes between two gapped surface with opposite half quantum anomalous Hall effect~\cite{Schindler2018HOTI}.
In this section, we will show that the same topological terms can be reproduced by our approach.

\begin{figure}
\center
\includegraphics[width=1\columnwidth]{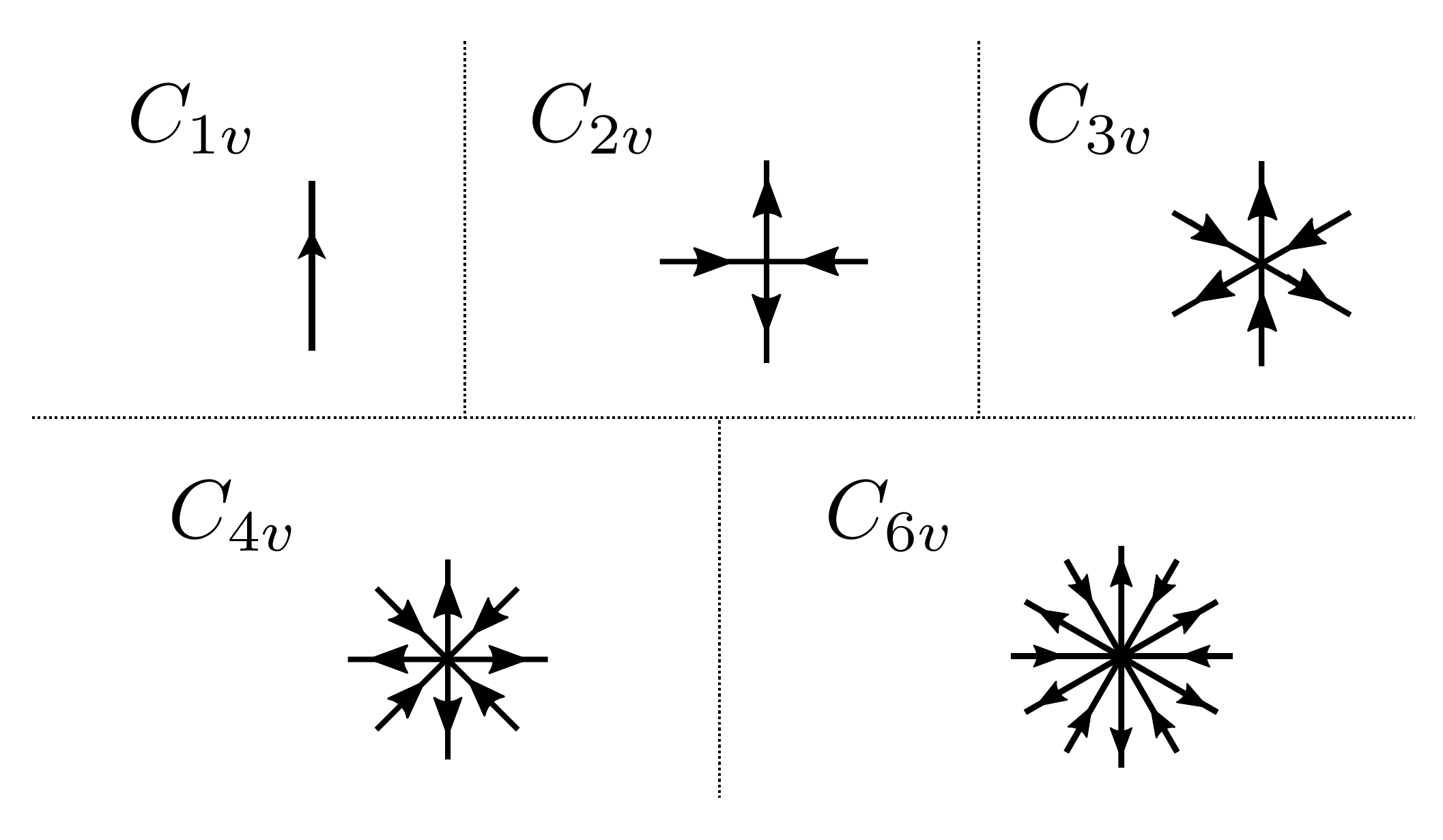}
\caption{The top surface view of the 3d topological crystals built from 2d IQH states for $C_{nv}$ point groups. Solid lines represent an IQH state on each mirror plane. The arrows represent the edge chiralities of the IQH states.
}
\label{fig:3dTCI}
\end{figure}

The topological crystal states that we are going to focus on are shown in Fig.~\ref{fig:3dTCI}. Those states are obtained by placing a IQH state at each reflection planes with the requirement that all the gapless modes at the rotational axis are gapped out while preserving the symmetry.

It has been shown in Ref.~\onlinecite{Shiozaki2019AHSS} that the bulk of the 2nd order phases can be described by the following $3+1$D massive Dirac theory:
\begin{equation}
\mathcal{L} = -i \bar{\psi} \gamma^{\mu} \partial_{\mu} \psi - i m_{0} \bar{\psi}\psi,
\end{equation}
where we use the following convention for the gamma matrices:
\begin{eqnarray}
\gamma^{0} &=&
\begin{pmatrix}
\mathbb{1} & 0  \\
0 & -\mathbb{1}  \\ 
\end{pmatrix}
= \tau^{3},
\nonumber\\
\gamma^{i} &=& 
\begin{pmatrix}
0 & \sigma^{i}  \\
-\sigma^{i} & 0  \\ 
\end{pmatrix}
=i\sigma^{i} \tau^{2},
\nonumber\\
\gamma^{5} &=&
\begin{pmatrix}
0 & \mathbb{1}  \\
\mathbb{1} & 0  \\ 
\end{pmatrix}
= \tau^{1}.
\end{eqnarray}
The reflection and the rotation symmetry in the $C_{nv}$ symmetry acts on the fermions by
\begin{eqnarray}
g_{r}^{y}: \psi(\boldsymbol{r}) &\rightarrow& i \sigma^{2} \tau^{3} \psi(g_{r}^{y} \boldsymbol{r}),
\nonumber\\
R: \psi(\boldsymbol{r}) &\rightarrow& e^{\frac{i}{2} \theta \sigma^{3}} \psi(R \boldsymbol{r}).
\end{eqnarray}
Other reflection symmetries in the $C_{nv}$ group can be generated by the combination of the $y$-reflection $g_{y}$ and the rotation $R$. 

To connect the massive Dirac theory to the topological crystal states, we add the following spatially dependent mass term:
\begin{equation}
\mathcal{L}_{m} = -i m \bar{\psi} e^{i\phi(\boldsymbol{r}) \gamma^{5}\tau^{2}} \psi.
\label{eqn:mass-3dcnv}
\end{equation}
The reflection and the rotational symmetry acts on the phase variable $\phi$ by
\begin{eqnarray}
g_{r}^{y} : \phi(\boldsymbol{r}) &\rightarrow& -\phi(g_{r}^{y}\boldsymbol{r}).
\nonumber\\
R : \phi(\boldsymbol{r}) &\rightarrow& \phi(R\boldsymbol{r}).
\end{eqnarray}
In order for Eq.~(\ref{eqn:mass-3dcnv}) to be invariant under the symmetry, we must have
\begin{equation}
 \phi(\boldsymbol{r}) = -\phi(g_{r}^{y}\boldsymbol{r})
\end{equation}
and 
\begin{equation}
 \phi(\boldsymbol{r}) = \phi(R\boldsymbol{r}).
\end{equation}

We are going to consider the interface configurations of $\phi$ such that there is an IQH state at each reflection plane. To simplified the discuss, let's first focus on $G_{pg} = C_{1v}$ in which there is only one reflection symmetry $g_{r}^{y}$. We are going to show that such interfaces of $\phi$ are classified by $H^{1}(BC_{1v},\mathbb{Z}^{r})$.

\begin{figure}
\center
\includegraphics[width=1\columnwidth]{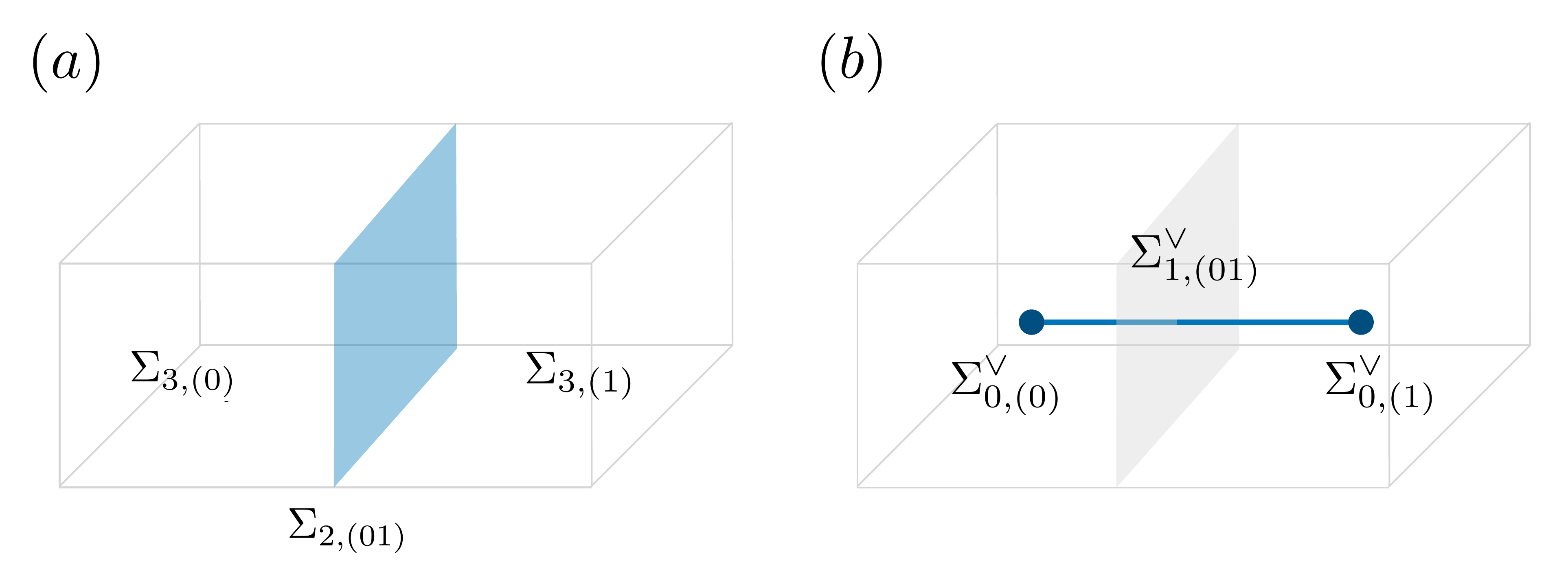}
\caption{(a) The cell-decomposition for 3d systems with reflection symmetry, and (b) the corresponding dual-cell decomposition.
}
\label{fig:r-cell-3d}
\end{figure}

The cell decomposition given by the fundamental domains is the shown in Fig.~\ref{fig:r-cell-3d}. There are two $3$-cells $\Sigma_{3,(0)}$,  $\Sigma_{3,(1)}$, and one $2$-cell $\Sigma_{2,(01)}$. Similar to the previous discussion, the map $f : M \rightarrow BC_{1v}$ maps the dual $0$-cells to the based point of $BC_{1v}$, and it maps the dual $1$-cell to the non-trivial loop in $BC_{1v}$ labeled by $g_{r}^{y}$.
 
\begin{figure}
\center
\includegraphics[width=0.8\columnwidth]{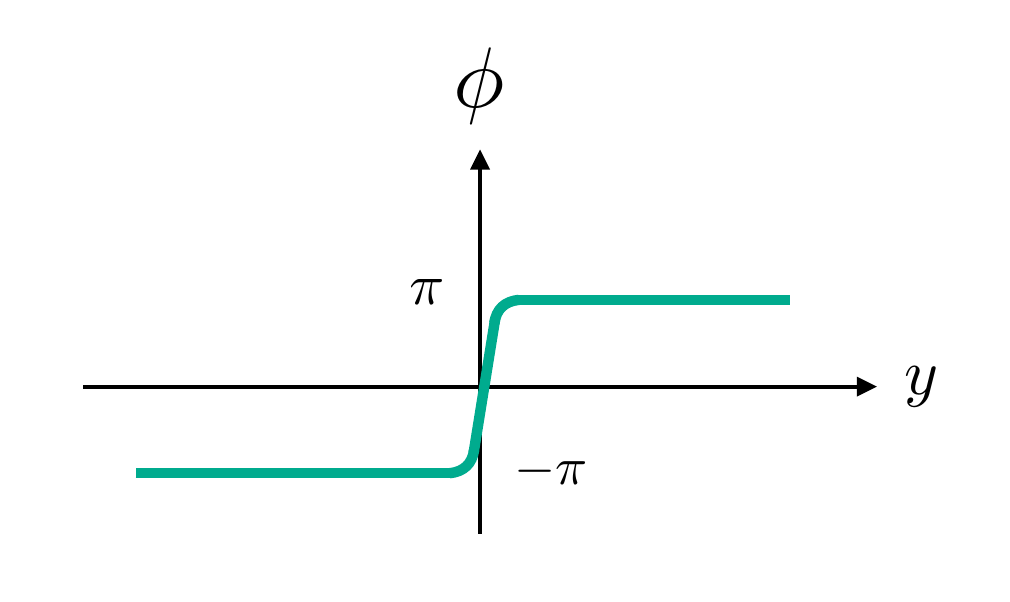}
\caption{A typical configuration of $\phi(\boldsymbol{r})$ as a function of $y$. $\phi(\boldsymbol{r})$ has no spatial variation along the $x$-, and $z$-directions.
}
\label{fig:phi-3d-r}
\end{figure}

We consider the configurations of $\phi(\boldsymbol{r})$ such that it's a constant function within the two $3$-cells $\Sigma_{3,(0)}$, $\Sigma_{3,(1)}$. At the intersecting $2$-cell $\Sigma_{2,(01)}$, we have the following relation:
\begin{equation}
\phi(\Sigma_{3,(1)}) = \phi(\Sigma_{3,(0)}) + 2\pi r(\Sigma_{2,(01)}),
\label{eqn:r-3d}
\end{equation}
where $r(\Sigma_{2,(01)}) \in \mathbb{Z}$. The reflection symmetry gives the following condition on $\phi$:
\begin{equation}
\phi(\Sigma_{3,(0)}) = - \phi(\Sigma_{3,(1)})
\label{eqn:prcond}
\end{equation}
There is also a non-trivial symmetry action on the integer-valued function $r$
\begin{equation}
g_{r}^{y} \cdot r = -r.
\label{eqn:rtwist-3d}
\end{equation}
However, there is a redundancy since, if we modify the configuration of $\phi(\boldsymbol{r})$ as
\begin{eqnarray}
\phi(\Sigma_{3,(i)}) &\rightarrow& \phi(\Sigma_{3,(i)}) + 2\pi h(\Sigma_{3,(i)}), 
\nonumber\\
r(\Sigma_{2,(01)}) &\rightarrow& r(\Sigma_{2,(01)}) + h(\Sigma_{3,(1)}) - h(\Sigma_{3,(0)}),
\end{eqnarray}
we obtain the same configuration. From these conditions, we see that $r$ is a $\mathbb{Z}$-valued cocycle in $H^{1}(M,\mathbb{Z}^{r})$ with a twisting coefficient due to the non-trivial action of the reflection Eq.~(\ref{eqn:rtwist-3d}). A typical configuration of $\phi$ is shown in Fig.~\ref{fig:phi-3d-r}. Similar to the discussion in Sec.~\ref{sec:1dr}, we identify $r$ to be the pull back of $c$: $r = f ^{*}c$, where $c \in H^{1}(BC_{1v},\mathbb{Z}^{r}) \cong \mathbb{Z}_{2}$. The deformation class of such $\phi(\boldsymbol{r})$ interfaces is classified by $\mathbb{Z}_{2}$.

With this systematic discussion on the interfaces, we are now ready to discussion the effective field theory. After coupling the fermions to the $U(1)$ gauge field and integrate out the massive fermions, the effective theory contains the following topological term
\begin{eqnarray}
S = \frac{1}{4\pi} \int AdA \wedge dP,
\end{eqnarray}
where we have defined $P = \phi/2\pi$. If we consider the non-trivial interface configuration of $\phi$ given by $r = 1$, it's easy to see that, at the interface, we have
\begin{equation}
    S_{\text{Interface}} = \frac{1}{4\pi} \int AdA,
\end{equation}
as we expected that there is an IQH state at the interface. In Ref.~\onlinecite{Shiozaki2019AHSS}, the explicit solution of the low-energy states that are localized at the interface is obtained, as well as an effective 2d massive Dirac Hamiltonian that describes an IQH state with a unit Chern number.  

\subsubsection{Smooth limit}
To go to the smooth limit, we simply replace the discontinuous 1-form $dP$ by the following smooth $1$-form $d\tilde{P}$ with integral periods satisfying
\begin{equation}
    \int_{\boldsymbol{r}_{0}}^{g_{r}^{y}\boldsymbol{r}_{0}} d\tilde{P} = r,
\label{eqn:P-int-3d}
\end{equation}
where $|\boldsymbol{r}_{0}|$ is much larger than the correlation length $\xi$.

The generalization to other $C_{nv}$ symmetry for $n=2,3,4,6$ is straightforward. In the smooth limit, we have a $1$-form $d\tilde{P}^{(n)}$ with integral periods satisfying Eq.~(\ref{eqn:P-int-3d}) and 
\begin{equation}
    \int_{\boldsymbol{r}_{0}}^{R\boldsymbol{r}_{0}} d\tilde{P}^{(n)} = 0.
\end{equation}
This comes from the requirement that the gapless modes at the rotational axis should be gapped out in order for the IQH states to be glued together. The resulting effective theory is given by
\begin{eqnarray}
S = \frac{1}{4\pi} \int AdA \wedge d\tilde{P}^{(n)}.
\label{eqn:S-TCI}
\end{eqnarray}

The above action is essentially the same as the action for effective axion field proposed in \refcite{Qi2008TFT}, where $\tilde{P}^{(n)}$ serves as the effective axion field.
Therefore, our method indeed reproduce the axion effective action for the 2nd order topological insulators. Moreover, our approach shows that the 1-form $d\tilde{P}^{(n)}$ constructed from the axion field $\tilde{P}^{(n)}$ should be classified by $H^{1}(BG_{pg},\mathbb{Z}^{r})$ for the 2nd order topological insulators.

\subsubsection{Physical responses}
For the sake of completeness, we briefly discuss the magnetoelectric responses for the 2nd order topological insulators governed by $\tilde{P}^{(n)}$.
When $\tilde{P}^{(n)}$ is a constant, the integrand of \eqnref{eqn:S-TCI} is a total derivative, and thus \eqnref{eqn:S-TCI} cannot have any response.
Therefore, we need certain spatial or temporal dependence in $\tilde{P}^{(n)}$ to get any responses.
%
Here the spatial dependence in $\tilde{P}^{(n)}$ come from
the mass interfaces in the bulk as discussed in the previous section.
In particular, we focus on the case where $\tilde{P}^{(n)}$ is smooth, and the space is infinite in order to avoid subtle issues that could appear on the boundary.
The responses we discuss below are essentially the same as in \refcite{Qi2008TFT}.

The general form of the magnetoelectric effect is given by the conserved current
\begin{equation}
    J^{\mu} = \frac{1}{2 \pi } \epsilon^{\mu\nu\lambda\delta} \partial_{\nu}\tilde{P}^{(n)} \partial_{\lambda} A_{\delta}.
\label{eqn:J-TCI}
\end{equation}
Specifically, when $\tilde{P}^{(n)}$ is static, the electric current can be induced by an applied electric field as
\begin{equation}
    J^{i} = \sigma^{ij} E_{j},
\end{equation}
with the $3$d Hall conductivity
\begin{equation}
    \sigma^{ij} = \frac{1}{2\pi} \epsilon^{ijk}\partial_{k}\tilde{P}^{(n)}.
\end{equation}
We note that the conductivity $\sigma^{ij}$ is not quantized by itself in general. The quantized quantity is given by
\begin{equation}
    \frac{1}{2} \int_{\boldsymbol{r}_{0}}^{g_{r}^{y}\boldsymbol{r}_{0}} \epsilon_{ijk} \sigma^{jk} dx^{i} = \frac{1}{2\pi}  \int_{\boldsymbol{r}_{0}}^{g_{r}^{y}\boldsymbol{r}_{0}} \partial_{i}\tilde{P}^{(n)} dx^{i} = \frac{1}{2\pi} r,
\end{equation}
which is quantized in the unit of $e^{2}/2 \pi h$ if we restore the proper unit. When $\tilde{P}^{(n)}$ is dynamic yet homogeneous, the electric current can be induced by an applied magnetic field as
\begin{equation}
    J^{i} = \frac{1}{2\pi} \partial_{t}\tilde{P}^{(n)} B^{i}\ .
\end{equation}


\section{Discussion and outlook}
\label{sec:summary}
In this work, we have proposed a general approach to characterize cSPT phases by its response to spatially dependent mass parameters with interfaces configurations. These mass interfaces implement the dimensional reduction procedure such that the bound states trapped at the interfaces are precisely the building blocks in the topological crystal picture. To illustrate the main idea, we have focused on the TCIs with both $U(1)$ charge conservation and the crystalline symmetry. We have shown that such mass interfaces with codimension $k$ are classified by $H^{k}(BG_{s},\mathbb{Z})$, and discussed the corresponding topological terms generated by integrating out the massive fermions. 

One physical correspondence of the spatially dependent mass terms in TCIs is non-homogeneous lattice distortions or strain~\cite{Ilan2020NatRevPseudoGauge,Yu2021PseudoGaugeDiracWeyl}. In the case when the couplings between electrons and the lattice distortions take the same form as the mass terms in this work, the topological terms will have a interpretations in the elasticity theory. This point of view provides a guiding principal on characterizing TCIs through lattice distortions.

The topological terms discussed in this paper is by no means an exhaustive list. In particular, these terms are incapable of describing the building blocks that transform non-trivially under the crystalline symmetry. It will be desirable to study this kind of terms in the future and matching with the formal classifications. 

When we consider more general topological crystalline phases, it might be the case that there is no local Lagrangian description for the building block of interest. A simple example is given by a weak topological crystalline superconductors protected by the translation symmetry. It can be think of as a stacking of 1d Kitaev chains. It's has been known that the Kitaev chain is characterized by the Arf invariant, which can not be expressed by a local differential form. One can nevertheless apply the dimensional reduction procedure by adding a mass term to a $2+1$D Dirac theory with the interfaces configurations with respective to the translation symmetry. We expect such mass interfaces are classified by $\tau \in H^{1}(B\mathbb{Z},\mathbb{Z})$, which can be bullback by the map $f:M \rightarrow B\mathbb{Z}$, giving $x = f^{*}\tau \in H^{1}(M,\mathbb{Z})$. The result is that there is a 1+1D Dirac theory describing the Kitaev chain at each mass interface. After integrating out the fermions, we can write the topological term in the following schematic form
\begin{equation}
    \int x \cup \text{Arf} \equiv \text{Arf}(\text{PD}(x)),
\end{equation}
where $\text{PD}(x)$ denotes a collection of codimension $1$ submanifolds, which is Poincar\'e dual to $x = f^{*}\tau \in H^{1}(M,\mathbb{Z})$. $\text{Arf}(\text{PD}(x))$ denotes the Arf invariant defined on the submanifolds $\text{PD}(x)$. The codimension $1$ submanifolds $\text{PD}(x)$ are precisely the location of the mass interfaces. These kinds of topological terms have been considered in Ref.~\onlinecite{Juven2019} from a more formal point of view. Applying our approach also leads to these kinds of topological terms naturally, and it will be interesting to study these kinds of terms more systematically in the future. Finally, we point out that our approach can be applied to non-invertible topological crystalline phases as well, which provides a way to study these phases in the continuous field theory framework while keeping a clear physical picture. 


\begin{acknowledgements}
S.-J.H. is grateful to Dominic Else, Abhinav Prem, and Andrey Gromov for related collaborations which inspire this work.
S.-J.H. acknowledges support from a JQI postdoctoral fellowship and the Laboratory for Physical Sciences.
C.-T.H. is supported by JSPS KAKENHI Grant No.19K14608.
J.~Y. is supported by the Laboratory for Physical Sciences.

\end{acknowledgements}


\appendix

\section{Review of the cellular cohomology}
\label{sec:cell-cohomology}
Let $X$ be a space with a cell decomposition in terms of CW-complexes. A cellular $k$-chain is a formal linear combination of oriented k-cells with integer coefficient $\mathbb{Z}$. These generate an abelian group $C_{k}(X,\mathbb{Z})$. The cellular $k$-cochain is defined to be a map
\begin{equation}
\alpha : C_{k}(X,\mathbb{Z}) \rightarrow A
\end{equation}
and these form a group denoted as $C^{k}(X,A)$. The pairing of a $k$-cochain $\alpha \in C^{k}(X,A)$ and a $k$-cycle $\Gamma \in C_{k}(X,\mathbb{Z})$ is a map $C^{k}(X,A) \otimes C_{k}(X,\mathbb{Z}) \rightarrow \mathbb{R}$, which we denote by
\begin{equation}
\int_{\Gamma} \alpha.
\end{equation}
The cellular coboundary map
\begin{equation}
\delta : C^{k}(X,A) \rightarrow C^{k+1}(X,A)
\end{equation}
is defined as
\begin{equation}
\int_{\Gamma} \delta \alpha = \int_{\partial \Gamma} \alpha.
\end{equation}
One can show that $\delta^{2} = 0$. We denote the kernel of $\delta$ as $Z^{k}(X,A)$, whose elements are cellular $k$-cocycles, and the image of $\delta$ in $Z^{k}(X,A)$ as $B^{k}(X,A)$, the group of exact cellular k-cocycles. The $k$th cellular cohomology of $X$ with coefficients in $A$ is defined as
\begin{equation}
H^{k}(X,A) = Z^{k}(X,A)/B^{k}(X,A).
\end{equation}


\section{Construction of the map $f : M \rightarrow BG_{s}$ for general space groups}
\label{sec:FD}

Here we give a construction of the map $f : M \rightarrow BG_{s}$. We begin by the following cell-decomposition of the real-space $M$, which is assumed to be the euclidean space $\mathbb{E}^{d}$.

Given a space group $G_{s}$, we can partition the euclidean space
$\mathbb{E}^{d}$ into fundamental domains accordingly. A \emph{fundamental domain} (FD), also know as an \emph{asymmetric unit} (AU) in crystallography, is a smallest simply connected closed
part of space from which, by application of all symmetry operations
of the space group, the whole of space is filled. Formally, the partition
is written as 
\begin{equation}
\mathbb{E}^{d}=\bigcup_{g\in G_{s}}g\mathcal{F},
\end{equation}
where $\mathcal{F}$ is a fundamental domain and $g\mathcal{F}$ its
image under the action of $g\in G_{s}$. If $g$ is not the identity of space group $G_{s}$,
then by definition $\mathcal{F}$ and $g\mathcal{F}$ only intersect
in their surfaces at most. The choice of fundamental domain is often not unique; a regular choice of fundamental domain for each wallpaper group and 3D space group is available in the International Tables for Crystallography.

This construction gives the euclidean space $\mathbb{E}^{d}$ a cell decomposition $\Sigma$. For example, in three dimensions, the 3-cells are the individual (non-overlapping) copies of FDs. The 2-cells lie on faces where two 3-cells meet, with the property that no two distinct points in the same 2-cell are related by symmetry. Similarly, 1-cells are edges where two or more faces meet, and 0-cells are points where edges meet. 

The construction of the map $f : M \rightarrow BG_{s}$ is based on the dual cell decomposition of the one given above. In particular, there is a one-to-one correspondence between $k$-cells $\Sigma_{k}$ and dual $(d-k)$-cells $\Sigma^{\vee}_{d-k}$ such that they intersect at a single point. Each fundamental domain $\mathcal{F}$ then corresponds to a dual $0$-cell $\Sigma^{\vee}_{0}$ and is labeled by a group element in $G_{s}$. Moreover, a dual 1-cell $\Sigma^{\vee}_{1}$ connecting a dual $0$-cell $\Sigma^{\vee}_{0}$ (associated to $\mathcal{F}$) to $g\Sigma^{\vee}_{0}$ (associated to $g\mathcal{F}$) is also labeled by a group element $g \in G_{s}$. The map $f$ is then constructed such that it maps these dual $0$-cells $\Sigma^{\vee}_{0}$ to the base point $\{*\}$ in $BG_{s}$, and maps a dual $1$-cell $\Sigma^{\vee}_{1}$ labeled by $g$ to a link in $BG_{s}$ labeled by the same $g \in \pi_{1}(BG_{s})$ and so on.


\section{Classifying space of space groups}
\label{sec:BG}
Let $\Gamma$ be the translation group in $\mathbb{R}^{d}$ and $P$ the point group. The $d$-dimensional space group $G_{s}$ fits into a short exact sequence,
 \begin{equation}
1 \rightarrow \Gamma \rightarrow G \rightarrow P \rightarrow 1.
\end{equation}
In general, $G_{s}$ is a subgroup of $\mathbb{R}^{d} \rtimes O(d)$. We can write an element of $G_{s}$ as $(v,P)$ with $v \in \Gamma$ and $p \in P$. 

Following Ref.~\onlinecite{Xiong2019thesis}, the classifying space of $G_{s}$ can be constructed as follows. First we note that the classifying space $B\Gamma$ of $\Gamma$ is the $d$-torus $T^{d} = \mathbb{R}^{d}/ \mathbb{Z}^{d}$. The point group $P$ has a non-trivial action on $B\Gamma$ by
\begin{equation}
p \cdot [v] = [a(p) + pv],
\end{equation}
where $[v]$ denotes an element in $B\Gamma$ with the representative $v \in \mathbb{R}^{d}$, and $a$ is a lift
\begin{equation}
a : P \rightarrow \mathbb{R}^{d}
\end{equation}
such that $(a(p),p) \in G_{s}$. For symmorphic space groups, $a$ can be chosen to be trivial. For nonsymmorphic space groups, $a$ has to be nontrivial. We have the usual point group action on the contractible universal cover $EP$ of the classifying space $BP$ of $P$. The classifying space $BG_{s}$ of $G_{s}$ can be construed as 
\begin{equation}
BG_{s} = B\Gamma \times_{P} EP,
\end{equation}
where $B\Gamma \times_{P} EP$ denotes the quotient space $(B\Gamma \times EP)/P$. One can show that this space is the same as 
\begin{equation}
BG_{s} = E\Gamma \times_{G_{s}} EP,
\end{equation}
where $G_{s}$ acts on $E\Gamma = \mathbb{R}^{d}$ according to the space-group action and on $EP$ by first projecting $G_{s}$ to $P$. The space $E\Gamma \times EP$ is the universal cover since it's contractable. 

In this paper, we only consider the symmetry group of the from $G = G_{s} \times G_{\text{int}}$, where $G_{\text{int}}$ is the internal symmetry group. Since $G_{\text{int}}$ has trivial action on $\mathbb{R}^{d}$, the classifying space of $G$ splits as
\begin{equation}
BG = BG_{s} \times BG_{\text{int}}.
\end{equation}


\section{The topological term of atomic insulators in two-dimensions}
\label{sec:2datomic}
Here we consider the topological terms of higher dimensional atomic insulators. We will illustrate the main idea in two dimensions, and generalization to higher-dimensions is straightforward. 

The symmetry group of a 2d atomic insulators is $U(1) \times \Gamma$, where $\Gamma = T_{x} \times T_{y} \cong \mathbb{Z}^{2}$, and the fermion parity is the $\mathbb{Z}_{2}$ subgroup of $U(1)$. It's known that the general classification of is $\mathbb{Z} \times \mathbb{Z}$. One of the $\z$ factor is the integer quantum hall state, which is not our focus. We will focus on the other $\z$ factor, which is the classification of 2d atomic insulators. The building block picture is having an atom carrying $U(1)$ charges per unit cell. We will focus on the charge-1 case. We assume that we have added the ancillas and performing the coarse-graining with respected to the ancillas lattice. The unit cell size is much larger than the lattice space of the ancillas lattice. The minimal low energy field theory is given by a two flavor massive Dirac theory:
\begin{equation}
\mathcal{L} = -i \bar{\Psi} \gamma^{\mu} \partial_{\mu} \Psi + i m_{0} \bar{\Psi} \sigma_{3} \Psi,
\label{eqn:2d-dirac}
\end{equation}
where $\Psi = (\psi_{1},\psi_{2})$, and $\sigma^{i}$ are Pauli matrices in the flavor space. The mass term here guarantees that there is no Chern number. 

To obtain the building block picture, we add the following spatially varying mass term:
\begin{equation}
\mathcal{L} = i m \bar{\Psi} (n^{1}\sigma_{1} + n^{2}\sigma_{2}) \Psi,
\end{equation}
where $n^{1}$, $n^{2}$ have spatial dependence. Translation symmetries $T_{x}$ and $T_{y}$ require that $n^{i}(\boldsymbol{r}) = n^{i}(\boldsymbol{r} + \boldsymbol{a})$ for $i = 1,2$. We would like to choose  configurations of $n^{1}$ and $n^{2}$ such that there is a charge-$1$ bound state at original lattice site. It turns out that it's enough to consider $n^{1}$ depends only on $x$ and $n^{2}$ on $y$. The conditions on $n^{1}$ and $n^{2}$ from the translation symmetries become 
\begin{equation}
n^{1}(x) = n^{1}(x+1), \; n^{2}(y) = n^{2}(y+1).
\label{eqn:cond-n1n2}
\end{equation}
The generic forms of $n^{1}$ and $n^{2}$ satisfying Eq.~(\ref{eqn:cond-n1n2}) are given by
\begin{eqnarray}
n^{1}(x) &=& x - \frac{1}{2\pi} \theta^{1}(x),
\nonumber\\
n^{2}(y) &=& y - \frac{1}{2\pi} \theta^{2}(y),
\end{eqnarray}
where $\theta^{1}(x)$, $\theta^{2}(y)$ are $\mathbb{R}/ 2\pi \mathbb{Z}$-valued functions and their values jump by $2\pi$ at the location of atoms. In other words, under that translations, we have
\begin{eqnarray}
\theta^{1}(x+1) &=& \theta^{1}(x) + 2\pi N^{1} , \ N^{1} \in \mathbb{Z} \ ,
\label{eqn:theta12-1}
\nonumber\\
\theta^{2}(y+1) &=& \theta^{2}(y) + 2\pi N^{2} , \ N^{2} \in \mathbb{Z} \ .
\label{eqn:theta12-2}
\end{eqnarray}
Eq,~(\ref{eqn:theta12-2}) implies that the parameter space is a 2-torus $T^{2} = S^{1} \times S^{1}$. If we consider the system with periodic boundary condition, we can think of the field $\theta$ as a map $\theta : S^{1} \times S^{1} \rightarrow  S^{1} \times S^{1}$.

We now discuss the general classification of the interface configurations of $\theta^{1}$ and $\theta^{2}$. Let $\Sigma_{2,(x,y)}$ and $\Sigma_{2,(x+1,y)}$ be two neighboring 2-cells that cover the two neighboring unit cell in real space, related by a translation in the $x$-direction, and similarly, $\Sigma_{2,(x,y)}$ and $\Sigma_{2,(x,y+1)}$ are two neighboring 2-cells related by a translation in the $y$-direction. According to Eq,~(\ref{eqn:theta12-2}), we have
\begin{eqnarray}
\theta^{1}(\Sigma_{2,(x+1,y)}) &=& \theta^{1}(\Sigma_{2,(x,y)}) + 2\pi N^{1}(\Sigma_{1,(x+1,x)}) ,
\nonumber\\
\theta^{2}(\Sigma_{2,(x,y+1)}) &=& \theta^{2}(\Sigma_{2,(x,y)}) + 2\pi N^{2}(\Sigma_{1,(y+1,y)}) ,
\label{eqn:theta12-cell}
\end{eqnarray}
where $N^{1} \in \mathbb{Z}$, $N^{2} \in \mathbb{Z}$, and $\Sigma_{1,(x+1,x)}$ is the 1-cell where the two neighboring 2-cells $\Sigma_{2,x+1}$ and $\Sigma_{2,x}$ meet, and similarly for $\Sigma_{1,(y+1,y)}$. It's easy to see that $N^{1}$ and $N^{2}$ satisfy
\begin{eqnarray}
N^{1}(\Sigma_{1,(x_{i},x_{j})}) + N^{1}(\Sigma_{1,(x_{j},x_{k})}) &=& N^{1}(\Sigma_{1,(x_{i},x_{k})}),
\nonumber\\
N^{2}(\Sigma_{1,(x_{i},x_{j})}) + N^{2}(\Sigma_{1,(x_{j},x_{k})}) &=& N^{2}(\Sigma_{1,(x_{i},x_{k})}).
\end{eqnarray}
There is a redundancy since, if we modify the configuration of $\theta^{1}$ as
\begin{widetext}
\begin{eqnarray}
\theta^{1}(\Sigma_{2,(x_{i},x_{j})}) &\rightarrow& \theta^{1}(\Sigma_{2,(x_{i},x_{j})}) + 2\pi h^{1}(\Sigma_{2,(x_{i},x_{j})}), \ h^{1}(\Sigma_{2,(x_{i},x_{j})}) \in \mathbb{Z}
\nonumber\\
N^{1}(\Sigma_{1,(x_{i}+1,x_{i})}) &\rightarrow& N^{1}(\Sigma_{1,(x_{i}+1,x_{i})}) + h^{1}(\Sigma_{2,(x_{i}+1,x_{j})}) - h^{1}(\Sigma_{2,(x_{i},x_{j})})
\end{eqnarray}
\end{widetext}
we obtain the same configuration of $\theta^{1}$. There is a similar redundancy for $N^{2}$ as well. Therefore, we see that $N^{1}$ and $N^{2}$ are $\mathbb{Z}$-valued cocycles in $H^{1}(M,\mathbb{Z})$. Moreover, $N^{1} = f ^{*}\alpha$ and $N^{2} = f ^{*}\beta$ are the pull back of the cocycle $\alpha \in H^{1}(BT_{x},\mathbb{Z})$ and $\beta \in H^{1}(BT_{y},\mathbb{Z})$. 

To obtain the effective field theory, we couple Eq,~(\ref{eqn:2d-dirac}) to a background $U(1)$ gauge field and integrate out the massive Dirac fermions. According to Ref.~\onlinecite{hsin2020berry}, we have the following topological term
\begin{eqnarray}
S_{\text{eff}} &=& \frac{1}{2} \int \epsilon^{\mu \nu \lambda} A_{\mu} \partial_{\nu} n^{I} \partial_{\lambda} n^{J} \tau_{2,IJ}(n) d^{3}x,
\nonumber\\
&=& \int A \wedge n^{*} \tau_{2}
\end{eqnarray}
where $\tau_{2} = \frac{1}{2} \tau_{2,IJ} dn^{I} \wedge dn^{J}$ is a $2$-form on the parameter space.
Here, since our parameter space is a 2-torus $T^{2}$ parametrized by $\theta^{I}$, the $2$-form $\tau_{2}$ should be proportional to the volume form of the $2$-torus: 
\begin{equation}
\tau_{2} = \frac{1}{8 \pi^{2}} \epsilon_{IJ} d\theta^{I} \wedge d\theta^{J}.
\end{equation}
The topological term becomes
\begin{equation}
S_{\text{eff}} = \frac{1}{8\pi^{2}} \int \epsilon^{\mu \nu \lambda} \epsilon_{IJ}   A_{\mu} \partial_{\nu} \theta^{I} \partial_{\lambda} \theta^{J}  d^{3}x.
\label{eqn:Seff-2d}
\end{equation}

Requiring that Eq.~(\ref{eqn:Seff-2d}) to be gauge invariant under $A_{\mu} \rightarrow A_{\mu} + \partial_{\mu} f$, we find that the following current has to be conserved
\begin{equation}
J^{\mu} = \frac{1}{2}\epsilon^{\mu \nu \lambda} \epsilon_{IJ} \partial_{\nu} \theta^{I} \partial_{\lambda} \theta^{J}.
\end{equation}

To gain more intuition, we focus on the charge density 
\begin{equation}
\rho = J^{0} = \frac{1}{2}\epsilon^{ k l} \epsilon_{IJ} \partial_{k} \theta^{I} \partial_{l} \theta^{J}.
\end{equation}
When the interface is discontinuous, we have
 \begin{equation}
\rho = \sum_{i} \delta(x - x_{i}) \delta(y - y_{i}).
\end{equation}
We thus recover the discrete nature of the charge density of an atomic insulator.

\subsection{Smooth limit}
Now we discuss how to take the smooth limit for the $\theta^{I}$ fields. To preserve the information of $H^{1}(B\Gamma,\mathbb{Z})$, we consider the following smooth 1-forms with integral periods:
\begin{eqnarray}
\int_{x_{0}}^{x_{0}+1} E^{1} dx &=& N^{1} \in \mathbb{Z},
\label{eqn:eicond-1}
\nonumber\\
\int_{y_{0}}^{y_{0}+1} E^{2} dy &=& N^{2} \in \mathbb{Z},
\label{eqn:eicond-2}
\end{eqnarray}
where we have defined $E^{I} = d\theta^{I}/2\pi$. The smooth limit of the two form $\tau_{2}$ then takes the form
\begin{equation}
\tau_{2} = \frac{1}{2} \epsilon_{IJ}  E^{I} \wedge E^{J}.
\end{equation}
Written in terms of the smooth differential forms, the effective action becomes
\begin{equation}
S_{\text{eff}} = \frac{1}{2} \int  \epsilon_{IJ} A \wedge   E^{I} \wedge E^{J}.
\label{eqn:2dtopo-atom}
\end{equation}

The simplest configurations of $\theta^{I}$ (and hence $E^{I}$) that satisfy Eq.~(\ref{eqn:eicond-1}) and (\ref{eqn:eicond-2}) are given by
\begin{eqnarray}
\theta^{1}(x) &=& 2\pi N^{1} x = b_{1} N^{1} x ,
\nonumber\\
\theta^{2}(y) &=& 2\pi N^{2} y = b_{2} N^{2} y ,
\end{eqnarray}
where $b_{1}$ and $b_{2}$ are the reciprocal lattice vector of $x$- and $y$-directions. This discussion can be straightforwardly generalized to higher dimensions.


\section{The topological term of rotation-invariant insulators in two dimensions}
\label{sec:2drotation}

In this section, we show how to derive the topological term of rotation-invariant insulators in two-dimensions by integrating out fermions.
We start from the following Lagrangian
\eq{
\mathcal{L} = i \bar{\psi} \gamma^{\mu} D_{\mu} \psi - m_{0} \bar{\psi} \psi - \bar{\psi} m \bsl{n}\cdot(\gamma^3,-\ii \gamma^5) \psi,
}
where $D_\mu=\partial_\mu+\ii A_\mu$, $\bsl{n}(\bsl{r})=(n^1,n^2) = n_r(\cos n_\theta, \sin n_\theta )$, $m_0$ is a constant, and the gamma matrices are
\eqa{
& \gamma^0=\tau_z\sigma_z\ ,\ \gamma^1=\ii \tau_y\sigma_z\ ,\ \gamma^2=-\ii \tau_x\sigma_z\ ,\ \gamma^3=\ii \tau_0\sigma_y\ ,\\
& \gamma^5=\ii \gamma^0\gamma^1\gamma^2\gamma^3\ .
}
Here the expressions of the gamma matrices differs from those in the main text by a factor $\ii$, but this difference has no influence on the resultant effective field theory.

\begin{figure}[t]
    \centering
    \includegraphics[width=\columnwidth]{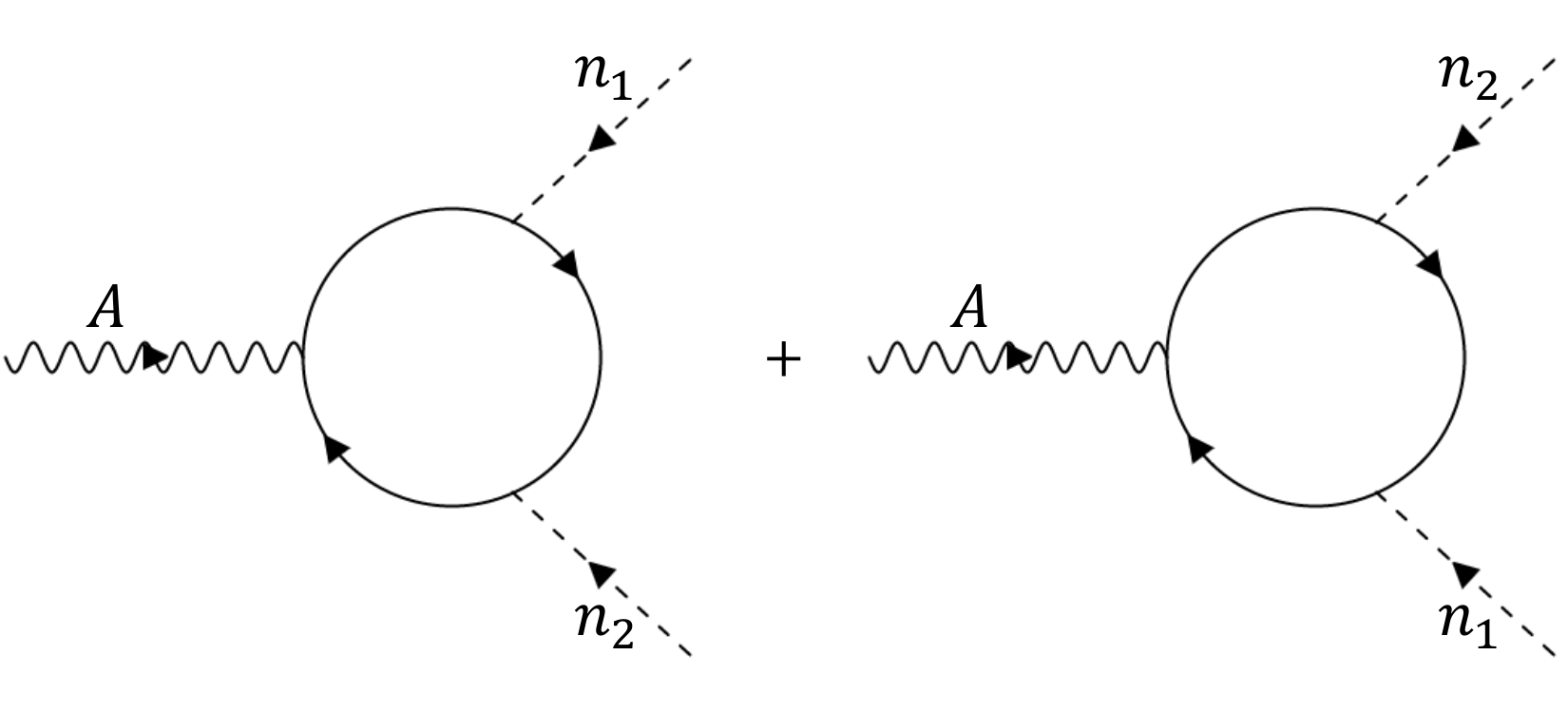}
    \caption{The Feynman diagrams for \eqnref{eq:2D_rotation_Seff}.
    The solid lines stand for the fermion field.}
    \label{fig:2D_rotation_FD}
\end{figure}

Performing the Fourier transformation gives the action in the momentum space as 
\begin{widetext}

\eq{
S=-\int_{k}\bar{\psi}_k G^{-1}(k) \psi_k - \int_{k}\int_{q} \bar{\psi}_{k+\frac{q}{2}} \left[ A^\mu (q) \gamma_\mu + m \bsl{n}(q)\cdot(\gamma^3,-\ii \gamma^5)\right] \psi_{k-\frac{q}{2}} \ ,
}
where $\mu=0,1,2 $, $k=(\omega,\bsl{k})$, $\int_{k}=\int d^3k/(2\pi)^3$, and $G^{-1}(k)=k_\mu\gamma^\mu+m_0$.
Here we use the same Fourier transformation rule for $\psi$, $A$, and $n$ as
\eq{
\psi_x=\int_k \ e^{\ii k x}\psi_k\ ,
}
and $k x=k_\mu x^\mu$.
Integrating out the fermions, the $n^1 n^2 A$ term is given by the two diagrams in \figref{fig:2D_rotation_FD}, and reads
\eqa{
\label{eq:2D_rotation_Seff}
& -\ii \int_{k,q_1,q_2} m^2 n^1 (q_1)  n^2 (-q_1-q_2) A_{\mu}(q_2) \left(\Tr[ G(k)\gamma^3 G(k-q_1) \gamma^5  G(k+q_2)  \gamma^\mu] + \Tr [G(k-q_2) \gamma^5 G(k+q_1) \gamma^3 G(k) \gamma^\mu] \right) \\
& = -\int_{k,q_1,q_2} m^2 n^1 (q_1)  n^2 (-q_1-q_2)  A_{\mu}(q_2) 8 m_0 \epsilon^{\mu\nu\rho} \frac{q_{1\nu} q_{2\rho}}{(k^2 +m_0^2)^3} +O (q^3) \\
& = \ii \frac{1}{4\pi} \frac{m^2}{m_0^2} \int d^3 x \epsilon^{\mu\nu\rho} \partial_\nu n_1 n_2 \partial_\rho  A_\mu +...\ .
}
Then, the leading order contribution to the corresponding effective action reads
\eq{
S_{eff} = -\frac{1}{4\pi} \frac{m^2}{m_0^2} \int d^3 x \epsilon^{\mu\nu\rho} \partial_\nu n_1 \partial_\rho n_2  A_\mu\ .
}
\end{widetext}
Substituting $\bsl{n}=n_r(\cos(n_\theta), \sin(n_\theta))$ and $\kappa=-m^2/(4 m_0^2)$ into the equation, we arrive at
\eq{
\frac{\kappa}{2\pi} \int d^3 x \epsilon^{\mu\nu\rho} A_\mu \partial_\nu n_r^2  \partial_\rho n_\theta\ .
}
The above expression is \eqnref{eq:2D_rotation_Seff_kappaAdomega1} with 
\eq{
d\omega_1 = \frac{1}{2\pi}\partial_\nu n_r^2  \partial_\rho n_\theta d x^\nu \wedge d x^\rho\ .
} 
The quantization of $\kappa$ can be derived from \eqnref{eq:2D_rotation_Seff_kappa_quantization}.


\bibliography{cSPT_EFT}

\begin{thebibliography}{84}%
\makeatletter
\providecommand \@ifxundefined [1]{%
 \@ifx{#1\undefined}
}%
\providecommand \@ifnum [1]{%
 \ifnum #1\expandafter \@firstoftwo
 \else \expandafter \@secondoftwo
 \fi
}%
\providecommand \@ifx [1]{%
 \ifx #1\expandafter \@firstoftwo
 \else \expandafter \@secondoftwo
 \fi
}%
\providecommand \natexlab [1]{#1}%
\providecommand \enquote  [1]{``#1''}%
\providecommand \bibnamefont  [1]{#1}%
\providecommand \bibfnamefont [1]{#1}%
\providecommand \citenamefont [1]{#1}%
\providecommand \href@noop [0]{\@secondoftwo}%
\providecommand \href [0]{\begingroup \@sanitize@url \@href}%
\providecommand \@href[1]{\@@startlink{#1}\@@href}%
\providecommand \@@href[1]{\endgroup#1\@@endlink}%
\providecommand \@sanitize@url [0]{\catcode `\\12\catcode `\$12\catcode
  `\&12\catcode `\#12\catcode `\^12\catcode `\_12\catcode `\%12\relax}%
\providecommand \@@startlink[1]{}%
\providecommand \@@endlink[0]{}%
\providecommand \url  [0]{\begingroup\@sanitize@url \@url }%
\providecommand \@url [1]{\endgroup\@href {#1}{\urlprefix }}%
\providecommand \urlprefix  [0]{URL }%
\providecommand \Eprint [0]{\href }%
\providecommand \doibase [0]{http://dx.doi.org/}%
\providecommand \selectlanguage [0]{\@gobble}%
\providecommand \bibinfo  [0]{\@secondoftwo}%
\providecommand \bibfield  [0]{\@secondoftwo}%
\providecommand \translation [1]{[#1]}%
\providecommand \BibitemOpen [0]{}%
\providecommand \bibitemStop [0]{}%
\providecommand \bibitemNoStop [0]{.\EOS\space}%
\providecommand \EOS [0]{\spacefactor3000\relax}%
\providecommand \BibitemShut  [1]{\csname bibitem#1\endcsname}%
\let\auto@bib@innerbib\@empty
\bibitem [{\citenamefont {Wen}(2019)}]{Wen2019review}%
  \BibitemOpen
  \bibfield  {author} {\bibinfo {author} {\bibfnamefont {Xiao-Gang}\
  \bibnamefont {Wen}},\ }\bibfield  {title} {\enquote {\bibinfo {title}
  {Choreographed entanglement dances: Topological states of quantum matter},}\
  }\href {\doibase 10.1126/science.aal3099} {\bibfield  {journal} {\bibinfo
  {journal} {Science}\ }\textbf {\bibinfo {volume} {363}} (\bibinfo {year}
  {2019}),\ 10.1126/science.aal3099}\BibitemShut {NoStop}%
\bibitem [{\citenamefont {Schnyder}\ \emph {et~al.}(2008)\citenamefont
  {Schnyder}, \citenamefont {Ryu}, \citenamefont {Furusaki},\ and\
  \citenamefont {Ludwig}}]{Schnyder2008}%
  \BibitemOpen
  \bibfield  {author} {\bibinfo {author} {\bibfnamefont {Andreas~P.}\
  \bibnamefont {Schnyder}}, \bibinfo {author} {\bibfnamefont {Shinsei}\
  \bibnamefont {Ryu}}, \bibinfo {author} {\bibfnamefont {Akira}\ \bibnamefont
  {Furusaki}}, \ and\ \bibinfo {author} {\bibfnamefont {Andreas W.~W.}\
  \bibnamefont {Ludwig}},\ }\bibfield  {title} {\enquote {\bibinfo {title}
  {Classification of topological insulators and superconductors in three
  spatial dimensions},}\ }\href {\doibase 10.1103/PhysRevB.78.195125}
  {\bibfield  {journal} {\bibinfo  {journal} {Phys. Rev. B}\ }\textbf {\bibinfo
  {volume} {78}},\ \bibinfo {pages} {195125} (\bibinfo {year}
  {2008})}\BibitemShut {NoStop}%
\bibitem [{\citenamefont {Kitaev}(2009)}]{Kitaev2009}%
  \BibitemOpen
  \bibfield  {author} {\bibinfo {author} {\bibfnamefont {Alexei}\ \bibnamefont
  {Kitaev}},\ }\bibfield  {title} {\enquote {\bibinfo {title} {Periodic table
  for topological insulators and superconductors},}\ }\href {\doibase
  10.1063/1.3149495} {\bibfield  {journal} {\bibinfo  {journal} {AIP Conference
  Proceedings}\ }\textbf {\bibinfo {volume} {1134}},\ \bibinfo {pages} {22--30}
  (\bibinfo {year} {2009})}\BibitemShut {NoStop}%
\bibitem [{\citenamefont {Ryu}\ \emph {et~al.}(2010)\citenamefont {Ryu},
  \citenamefont {Schnyder}, \citenamefont {Furusaki},\ and\ \citenamefont
  {Ludwig}}]{ryu2010}%
  \BibitemOpen
  \bibfield  {author} {\bibinfo {author} {\bibfnamefont {Shinsei}\ \bibnamefont
  {Ryu}}, \bibinfo {author} {\bibfnamefont {Andreas~P}\ \bibnamefont
  {Schnyder}}, \bibinfo {author} {\bibfnamefont {Akira}\ \bibnamefont
  {Furusaki}}, \ and\ \bibinfo {author} {\bibfnamefont {Andreas W~W}\
  \bibnamefont {Ludwig}},\ }\bibfield  {title} {\enquote {\bibinfo {title}
  {Topological insulators and superconductors: tenfold way and dimensional
  hierarchy},}\ }\href {http://stacks.iop.org/1367-2630/12/i=6/a=065010}
  {\bibfield  {journal} {\bibinfo  {journal} {New Journal of Physics}\ }\textbf
  {\bibinfo {volume} {12}},\ \bibinfo {pages} {065010} (\bibinfo {year}
  {2010})}\BibitemShut {NoStop}%
\bibitem [{\citenamefont {Gu}\ and\ \citenamefont {Wen}(2009)}]{gu2009}%
  \BibitemOpen
  \bibfield  {author} {\bibinfo {author} {\bibfnamefont {Zheng-Cheng}\
  \bibnamefont {Gu}}\ and\ \bibinfo {author} {\bibfnamefont {Xiao-Gang}\
  \bibnamefont {Wen}},\ }\bibfield  {title} {\enquote {\bibinfo {title}
  {Tensor-entanglement-filtering renormalization approach and
  symmetry-protected topological order},}\ }\href {\doibase
  10.1103/PhysRevB.80.155131} {\bibfield  {journal} {\bibinfo  {journal} {Phys.
  Rev. B}\ }\textbf {\bibinfo {volume} {80}},\ \bibinfo {pages} {155131}
  (\bibinfo {year} {2009})}\BibitemShut {NoStop}%
\bibitem [{\citenamefont {Pollmann}\ \emph {et~al.}(2010)\citenamefont
  {Pollmann}, \citenamefont {Turner}, \citenamefont {Berg},\ and\ \citenamefont
  {Oshikawa}}]{pollmann2010}%
  \BibitemOpen
  \bibfield  {author} {\bibinfo {author} {\bibfnamefont {Frank}\ \bibnamefont
  {Pollmann}}, \bibinfo {author} {\bibfnamefont {Ari~M.}\ \bibnamefont
  {Turner}}, \bibinfo {author} {\bibfnamefont {Erez}\ \bibnamefont {Berg}}, \
  and\ \bibinfo {author} {\bibfnamefont {Masaki}\ \bibnamefont {Oshikawa}},\
  }\bibfield  {title} {\enquote {\bibinfo {title} {Entanglement spectrum of a
  topological phase in one dimension},}\ }\href {\doibase
  10.1103/PhysRevB.81.064439} {\bibfield  {journal} {\bibinfo  {journal} {Phys.
  Rev. B}\ }\textbf {\bibinfo {volume} {81}},\ \bibinfo {pages} {064439}
  (\bibinfo {year} {2010})}\BibitemShut {NoStop}%
\bibitem [{\citenamefont {Fidkowski}\ and\ \citenamefont
  {Kitaev}(2011)}]{fidkowski2011}%
  \BibitemOpen
  \bibfield  {author} {\bibinfo {author} {\bibfnamefont {Lukasz}\ \bibnamefont
  {Fidkowski}}\ and\ \bibinfo {author} {\bibfnamefont {Alexei}\ \bibnamefont
  {Kitaev}},\ }\bibfield  {title} {\enquote {\bibinfo {title} {Topological
  phases of fermions in one dimension},}\ }\href {\doibase
  10.1103/PhysRevB.83.075103} {\bibfield  {journal} {\bibinfo  {journal} {Phys.
  Rev. B}\ }\textbf {\bibinfo {volume} {83}},\ \bibinfo {pages} {075103}
  (\bibinfo {year} {2011})}\BibitemShut {NoStop}%
\bibitem [{\citenamefont {Turner}\ \emph {et~al.}(2011)\citenamefont {Turner},
  \citenamefont {Pollmann},\ and\ \citenamefont {Berg}}]{turner2011}%
  \BibitemOpen
  \bibfield  {author} {\bibinfo {author} {\bibfnamefont {Ari~M.}\ \bibnamefont
  {Turner}}, \bibinfo {author} {\bibfnamefont {Frank}\ \bibnamefont
  {Pollmann}}, \ and\ \bibinfo {author} {\bibfnamefont {Erez}\ \bibnamefont
  {Berg}},\ }\bibfield  {title} {\enquote {\bibinfo {title} {Topological phases
  of one-dimensional fermions: An entanglement point of view},}\ }\href
  {\doibase 10.1103/PhysRevB.83.075102} {\bibfield  {journal} {\bibinfo
  {journal} {Phys. Rev. B}\ }\textbf {\bibinfo {volume} {83}},\ \bibinfo
  {pages} {075102} (\bibinfo {year} {2011})}\BibitemShut {NoStop}%
\bibitem [{\citenamefont {Chen}\ \emph
  {et~al.}(2011{\natexlab{a}})\citenamefont {Chen}, \citenamefont {Gu},\ and\
  \citenamefont {Wen}}]{chen2011_1dspt}%
  \BibitemOpen
  \bibfield  {author} {\bibinfo {author} {\bibfnamefont {Xie}\ \bibnamefont
  {Chen}}, \bibinfo {author} {\bibfnamefont {Zheng-Cheng}\ \bibnamefont {Gu}},
  \ and\ \bibinfo {author} {\bibfnamefont {Xiao-Gang}\ \bibnamefont {Wen}},\
  }\bibfield  {title} {\enquote {\bibinfo {title} {Classification of gapped
  symmetric phases in one-dimensional spin systems},}\ }\href {\doibase
  10.1103/PhysRevB.83.035107} {\bibfield  {journal} {\bibinfo  {journal} {Phys.
  Rev. B}\ }\textbf {\bibinfo {volume} {83}},\ \bibinfo {pages} {035107}
  (\bibinfo {year} {2011}{\natexlab{a}})}\BibitemShut {NoStop}%
\bibitem [{\citenamefont {Chen}\ \emph
  {et~al.}(2011{\natexlab{b}})\citenamefont {Chen}, \citenamefont {Gu},\ and\
  \citenamefont {Wen}}]{chen2011_1dcomplete}%
  \BibitemOpen
  \bibfield  {author} {\bibinfo {author} {\bibfnamefont {Xie}\ \bibnamefont
  {Chen}}, \bibinfo {author} {\bibfnamefont {Zheng-Cheng}\ \bibnamefont {Gu}},
  \ and\ \bibinfo {author} {\bibfnamefont {Xiao-Gang}\ \bibnamefont {Wen}},\
  }\bibfield  {title} {\enquote {\bibinfo {title} {Complete classification of
  one-dimensional gapped quantum phases in interacting spin systems},}\ }\href
  {\doibase 10.1103/PhysRevB.84.235128} {\bibfield  {journal} {\bibinfo
  {journal} {Phys. Rev. B}\ }\textbf {\bibinfo {volume} {84}},\ \bibinfo
  {pages} {235128} (\bibinfo {year} {2011}{\natexlab{b}})}\BibitemShut
  {NoStop}%
\bibitem [{\citenamefont {Schuch}\ \emph {et~al.}(2011)\citenamefont {Schuch},
  \citenamefont {P\'erez-Garc\'{\i}a},\ and\ \citenamefont
  {Cirac}}]{cirac2011}%
  \BibitemOpen
  \bibfield  {author} {\bibinfo {author} {\bibfnamefont {Norbert}\ \bibnamefont
  {Schuch}}, \bibinfo {author} {\bibfnamefont {David}\ \bibnamefont
  {P\'erez-Garc\'{\i}a}}, \ and\ \bibinfo {author} {\bibfnamefont {Ignacio}\
  \bibnamefont {Cirac}},\ }\bibfield  {title} {\enquote {\bibinfo {title}
  {Classifying quantum phases using matrix product states and projected
  entangled pair states},}\ }\href {\doibase 10.1103/PhysRevB.84.165139}
  {\bibfield  {journal} {\bibinfo  {journal} {Phys. Rev. B}\ }\textbf {\bibinfo
  {volume} {84}},\ \bibinfo {pages} {165139} (\bibinfo {year}
  {2011})}\BibitemShut {NoStop}%
\bibitem [{\citenamefont {Chen}\ \emph {et~al.}(2013)\citenamefont {Chen},
  \citenamefont {Gu}, \citenamefont {Liu},\ and\ \citenamefont
  {Wen}}]{chen2013cohomology}%
  \BibitemOpen
  \bibfield  {author} {\bibinfo {author} {\bibfnamefont {Xie}\ \bibnamefont
  {Chen}}, \bibinfo {author} {\bibfnamefont {Zheng-Cheng}\ \bibnamefont {Gu}},
  \bibinfo {author} {\bibfnamefont {Zheng-Xin}\ \bibnamefont {Liu}}, \ and\
  \bibinfo {author} {\bibfnamefont {Xiao-Gang}\ \bibnamefont {Wen}},\
  }\bibfield  {title} {\enquote {\bibinfo {title} {Symmetry protected
  topological orders and the group cohomology of their symmetry group},}\
  }\href {\doibase 10.1103/PhysRevB.87.155114} {\bibfield  {journal} {\bibinfo
  {journal} {Phys. Rev. B}\ }\textbf {\bibinfo {volume} {87}},\ \bibinfo
  {pages} {155114} (\bibinfo {year} {2013})}\BibitemShut {NoStop}%
\bibitem [{\citenamefont {Levin}\ and\ \citenamefont {Gu}(2012)}]{levin2012}%
  \BibitemOpen
  \bibfield  {author} {\bibinfo {author} {\bibfnamefont {Michael}\ \bibnamefont
  {Levin}}\ and\ \bibinfo {author} {\bibfnamefont {Zheng-Cheng}\ \bibnamefont
  {Gu}},\ }\bibfield  {title} {\enquote {\bibinfo {title} {Braiding statistics
  approach to symmetry-protected topological phases},}\ }\href {\doibase
  10.1103/PhysRevB.86.115109} {\bibfield  {journal} {\bibinfo  {journal} {Phys.
  Rev. B}\ }\textbf {\bibinfo {volume} {86}},\ \bibinfo {pages} {115109}
  (\bibinfo {year} {2012})}\BibitemShut {NoStop}%
\bibitem [{\citenamefont {Kapustin}(2014)}]{kapustin2014symmetry}%
  \BibitemOpen
  \bibfield  {author} {\bibinfo {author} {\bibfnamefont {Anton}\ \bibnamefont
  {Kapustin}},\ }\href@noop {} {\enquote {\bibinfo {title} {Symmetry protected
  topological phases, anomalies, and cobordisms: Beyond group cohomology},}\ }
  (\bibinfo {year} {2014}),\ \Eprint {http://arxiv.org/abs/1403.1467}
  {arXiv:1403.1467 [cond-mat.str-el]} \BibitemShut {NoStop}%
\bibitem [{\citenamefont {Else}\ and\ \citenamefont
  {Nayak}(2014)}]{Else2014spt}%
  \BibitemOpen
  \bibfield  {author} {\bibinfo {author} {\bibfnamefont {Dominic~V.}\
  \bibnamefont {Else}}\ and\ \bibinfo {author} {\bibfnamefont {Chetan}\
  \bibnamefont {Nayak}},\ }\bibfield  {title} {\enquote {\bibinfo {title}
  {Classifying symmetry-protected topological phases through the anomalous
  action of the symmetry on the edge},}\ }\href {\doibase
  10.1103/PhysRevB.90.235137} {\bibfield  {journal} {\bibinfo  {journal} {Phys.
  Rev. B}\ }\textbf {\bibinfo {volume} {90}},\ \bibinfo {pages} {235137}
  (\bibinfo {year} {2014})}\BibitemShut {NoStop}%
\bibitem [{\citenamefont {Freed}(2014)}]{freed2014shortrange}%
  \BibitemOpen
  \bibfield  {author} {\bibinfo {author} {\bibfnamefont {Daniel~S.}\
  \bibnamefont {Freed}},\ }\href@noop {} {\enquote {\bibinfo {title}
  {Short-range entanglement and invertible field theories},}\ } (\bibinfo
  {year} {2014}),\ \Eprint {http://arxiv.org/abs/1406.7278} {arXiv:1406.7278
  [cond-mat.str-el]} \BibitemShut {NoStop}%
\bibitem [{\citenamefont {Freed}\ and\ \citenamefont
  {Hopkins}(2019{\natexlab{a}})}]{freed2019reflection}%
  \BibitemOpen
  \bibfield  {author} {\bibinfo {author} {\bibfnamefont {Daniel~S.}\
  \bibnamefont {Freed}}\ and\ \bibinfo {author} {\bibfnamefont {Michael~J.}\
  \bibnamefont {Hopkins}},\ }\href@noop {} {\enquote {\bibinfo {title}
  {Reflection positivity and invertible topological phases},}\ } (\bibinfo
  {year} {2019}{\natexlab{a}}),\ \Eprint {http://arxiv.org/abs/1604.06527}
  {arXiv:1604.06527 [hep-th]} \BibitemShut {NoStop}%
\bibitem [{\citenamefont {Xiong}(2018)}]{Xiong_2018sptmin}%
  \BibitemOpen
  \bibfield  {author} {\bibinfo {author} {\bibfnamefont {Charles~Zhaoxi}\
  \bibnamefont {Xiong}},\ }\bibfield  {title} {\enquote {\bibinfo {title}
  {Minimalist approach to the classification of symmetry protected topological
  phases},}\ }\href {\doibase 10.1088/1751-8121/aae0b1} {\bibfield  {journal}
  {\bibinfo  {journal} {Journal of Physics A: Mathematical and Theoretical}\
  }\textbf {\bibinfo {volume} {51}},\ \bibinfo {pages} {445001} (\bibinfo
  {year} {2018})}\BibitemShut {NoStop}%
\bibitem [{\citenamefont {Gaiotto}\ and\ \citenamefont
  {Johnson-Freyd}(2019)}]{Gaiotto2019}%
  \BibitemOpen
  \bibfield  {author} {\bibinfo {author} {\bibfnamefont {Davide}\ \bibnamefont
  {Gaiotto}}\ and\ \bibinfo {author} {\bibfnamefont {Theo}\ \bibnamefont
  {Johnson-Freyd}},\ }\bibfield  {title} {\enquote {\bibinfo {title} {Symmetry
  protected topological phases and generalized cohomology},}\ }\href {\doibase
  10.1007/JHEP05(2019)007} {\bibfield  {journal} {\bibinfo  {journal} {Journal
  of High Energy Physics}\ }\textbf {\bibinfo {volume} {2019}},\ \bibinfo
  {pages} {7} (\bibinfo {year} {2019})}\BibitemShut {NoStop}%
\bibitem [{\citenamefont {Teo}\ and\ \citenamefont {Hughes}(2013)}]{Teo2013}%
  \BibitemOpen
  \bibfield  {author} {\bibinfo {author} {\bibfnamefont {Jeffrey C.~Y.}\
  \bibnamefont {Teo}}\ and\ \bibinfo {author} {\bibfnamefont {Taylor~L.}\
  \bibnamefont {Hughes}},\ }\bibfield  {title} {\enquote {\bibinfo {title}
  {Existence of majorana-fermion bound states on disclinations and the
  classification of topological crystalline superconductors in two
  dimensions},}\ }\href {\doibase 10.1103/PhysRevLett.111.047006} {\bibfield
  {journal} {\bibinfo  {journal} {Phys. Rev. Lett.}\ }\textbf {\bibinfo
  {volume} {111}},\ \bibinfo {pages} {047006} (\bibinfo {year}
  {2013})}\BibitemShut {NoStop}%
\bibitem [{\citenamefont {Shiozaki}\ and\ \citenamefont
  {Sato}(2014)}]{Shiozaki2014}%
  \BibitemOpen
  \bibfield  {author} {\bibinfo {author} {\bibfnamefont {Ken}\ \bibnamefont
  {Shiozaki}}\ and\ \bibinfo {author} {\bibfnamefont {Masatoshi}\ \bibnamefont
  {Sato}},\ }\bibfield  {title} {\enquote {\bibinfo {title} {Topology of
  crystalline insulators and superconductors},}\ }\href {\doibase
  10.1103/PhysRevB.90.165114} {\bibfield  {journal} {\bibinfo  {journal} {Phys.
  Rev. B}\ }\textbf {\bibinfo {volume} {90}},\ \bibinfo {pages} {165114}
  (\bibinfo {year} {2014})}\BibitemShut {NoStop}%
\bibitem [{\citenamefont {Isobe}\ and\ \citenamefont {Fu}(2015)}]{Isobe2015}%
  \BibitemOpen
  \bibfield  {author} {\bibinfo {author} {\bibfnamefont {Hiroki}\ \bibnamefont
  {Isobe}}\ and\ \bibinfo {author} {\bibfnamefont {Liang}\ \bibnamefont {Fu}},\
  }\bibfield  {title} {\enquote {\bibinfo {title} {Theory of interacting
  topological crystalline insulators},}\ }\href {\doibase
  10.1103/PhysRevB.92.081304} {\bibfield  {journal} {\bibinfo  {journal} {Phys.
  Rev. B}\ }\textbf {\bibinfo {volume} {92}},\ \bibinfo {pages} {081304}
  (\bibinfo {year} {2015})}\BibitemShut {NoStop}%
\bibitem [{\citenamefont {Shiozaki}\ \emph {et~al.}(2017)\citenamefont
  {Shiozaki}, \citenamefont {Sato},\ and\ \citenamefont {Gomi}}]{Shiozaki2017}%
  \BibitemOpen
  \bibfield  {author} {\bibinfo {author} {\bibfnamefont {Ken}\ \bibnamefont
  {Shiozaki}}, \bibinfo {author} {\bibfnamefont {Masatoshi}\ \bibnamefont
  {Sato}}, \ and\ \bibinfo {author} {\bibfnamefont {Kiyonori}\ \bibnamefont
  {Gomi}},\ }\bibfield  {title} {\enquote {\bibinfo {title} {Topological
  crystalline materials: General formulation, module structure, and wallpaper
  groups},}\ }\href {\doibase 10.1103/PhysRevB.95.235425} {\bibfield  {journal}
  {\bibinfo  {journal} {Phys. Rev. B}\ }\textbf {\bibinfo {volume} {95}},\
  \bibinfo {pages} {235425} (\bibinfo {year} {2017})}\BibitemShut {NoStop}%
\bibitem [{\citenamefont {Po}\ \emph {et~al.}(2017)\citenamefont {Po},
  \citenamefont {Vishwanath},\ and\ \citenamefont
  {Watanabe}}]{Po2017indicators}%
  \BibitemOpen
  \bibfield  {author} {\bibinfo {author} {\bibfnamefont {Hoi~Chun}\
  \bibnamefont {Po}}, \bibinfo {author} {\bibfnamefont {Ashvin}\ \bibnamefont
  {Vishwanath}}, \ and\ \bibinfo {author} {\bibfnamefont {Haruki}\ \bibnamefont
  {Watanabe}},\ }\bibfield  {title} {\enquote {\bibinfo {title} {Symmetry-based
  indicators of band topology in the 230 space groups},}\ }\href {\doibase
  10.1038/s41467-017-00133-2} {\bibfield  {journal} {\bibinfo  {journal}
  {Nature Communications}\ }\textbf {\bibinfo {volume} {8}},\ \bibinfo {pages}
  {50} (\bibinfo {year} {2017})}\BibitemShut {NoStop}%
\bibitem [{\citenamefont {Bradlyn}\ \emph {et~al.}(2017)\citenamefont
  {Bradlyn}, \citenamefont {Elcoro}, \citenamefont {Cano}, \citenamefont
  {Vergniory}, \citenamefont {Wang}, \citenamefont {Felser}, \citenamefont
  {Aroyo},\ and\ \citenamefont {Bernevig}}]{Bradlyn2017}%
  \BibitemOpen
  \bibfield  {author} {\bibinfo {author} {\bibfnamefont {Barry}\ \bibnamefont
  {Bradlyn}}, \bibinfo {author} {\bibfnamefont {L.}~\bibnamefont {Elcoro}},
  \bibinfo {author} {\bibfnamefont {Jennifer}\ \bibnamefont {Cano}}, \bibinfo
  {author} {\bibfnamefont {M.~G.}\ \bibnamefont {Vergniory}}, \bibinfo {author}
  {\bibfnamefont {Zhijun}\ \bibnamefont {Wang}}, \bibinfo {author}
  {\bibfnamefont {C.}~\bibnamefont {Felser}}, \bibinfo {author} {\bibfnamefont
  {M.~I.}\ \bibnamefont {Aroyo}}, \ and\ \bibinfo {author} {\bibfnamefont
  {B.~Andrei}\ \bibnamefont {Bernevig}},\ }\bibfield  {title} {\enquote
  {\bibinfo {title} {Topological quantum chemistry},}\ }\href {\doibase
  10.1038/nature23268} {\bibfield  {journal} {\bibinfo  {journal} {Nature}\
  }\textbf {\bibinfo {volume} {547}},\ \bibinfo {pages} {298--305} (\bibinfo
  {year} {2017})}\BibitemShut {NoStop}%
\bibitem [{\citenamefont {Kruthoff}\ \emph {et~al.}(2017)\citenamefont
  {Kruthoff}, \citenamefont {de~Boer}, \citenamefont {van Wezel}, \citenamefont
  {Kane},\ and\ \citenamefont {Slager}}]{Kruthoff2017}%
  \BibitemOpen
  \bibfield  {author} {\bibinfo {author} {\bibfnamefont {Jorrit}\ \bibnamefont
  {Kruthoff}}, \bibinfo {author} {\bibfnamefont {Jan}\ \bibnamefont {de~Boer}},
  \bibinfo {author} {\bibfnamefont {Jasper}\ \bibnamefont {van Wezel}},
  \bibinfo {author} {\bibfnamefont {Charles~L.}\ \bibnamefont {Kane}}, \ and\
  \bibinfo {author} {\bibfnamefont {Robert-Jan}\ \bibnamefont {Slager}},\
  }\bibfield  {title} {\enquote {\bibinfo {title} {Topological classification
  of crystalline insulators through band structure combinatorics},}\ }\href
  {\doibase 10.1103/PhysRevX.7.041069} {\bibfield  {journal} {\bibinfo
  {journal} {Phys. Rev. X}\ }\textbf {\bibinfo {volume} {7}},\ \bibinfo {pages}
  {041069} (\bibinfo {year} {2017})}\BibitemShut {NoStop}%
\bibitem [{\citenamefont {Fu}(2011)}]{Fu2011}%
  \BibitemOpen
  \bibfield  {author} {\bibinfo {author} {\bibfnamefont {Liang}\ \bibnamefont
  {Fu}},\ }\bibfield  {title} {\enquote {\bibinfo {title} {Topological
  crystalline insulators},}\ }\href {\doibase 10.1103/PhysRevLett.106.106802}
  {\bibfield  {journal} {\bibinfo  {journal} {Phys. Rev. Lett.}\ }\textbf
  {\bibinfo {volume} {106}},\ \bibinfo {pages} {106802} (\bibinfo {year}
  {2011})}\BibitemShut {NoStop}%
\bibitem [{\citenamefont {Ando}\ and\ \citenamefont
  {Fu}(2015)}]{Ando2015review}%
  \BibitemOpen
  \bibfield  {author} {\bibinfo {author} {\bibfnamefont {Yoichi}\ \bibnamefont
  {Ando}}\ and\ \bibinfo {author} {\bibfnamefont {Liang}\ \bibnamefont {Fu}},\
  }\bibfield  {title} {\enquote {\bibinfo {title} {Topological crystalline
  insulators and topological superconductors: From concepts to materials},}\
  }\bibfield  {booktitle} {\emph {\bibinfo {booktitle} {Annual Review of
  Condensed Matter Physics}},\ }\href {\doibase
  10.1146/annurev-conmatphys-031214-014501} {\bibfield  {journal} {\bibinfo
  {journal} {Annual Review of Condensed Matter Physics}\ }\textbf {\bibinfo
  {volume} {6}},\ \bibinfo {pages} {361--381} (\bibinfo {year}
  {2015})}\BibitemShut {NoStop}%
\bibitem [{\citenamefont {Chiu}\ \emph {et~al.}(2016)\citenamefont {Chiu},
  \citenamefont {Teo}, \citenamefont {Schnyder},\ and\ \citenamefont
  {Ryu}}]{Chiu2016review}%
  \BibitemOpen
  \bibfield  {author} {\bibinfo {author} {\bibfnamefont {Ching-Kai}\
  \bibnamefont {Chiu}}, \bibinfo {author} {\bibfnamefont {Jeffrey C.~Y.}\
  \bibnamefont {Teo}}, \bibinfo {author} {\bibfnamefont {Andreas~P.}\
  \bibnamefont {Schnyder}}, \ and\ \bibinfo {author} {\bibfnamefont {Shinsei}\
  \bibnamefont {Ryu}},\ }\bibfield  {title} {\enquote {\bibinfo {title}
  {Classification of topological quantum matter with symmetries},}\ }\href
  {\doibase 10.1103/RevModPhys.88.035005} {\bibfield  {journal} {\bibinfo
  {journal} {Rev. Mod. Phys.}\ }\textbf {\bibinfo {volume} {88}},\ \bibinfo
  {pages} {035005} (\bibinfo {year} {2016})}\BibitemShut {NoStop}%
\bibitem [{\citenamefont {Schindler}\ \emph {et~al.}(2018)\citenamefont
  {Schindler}, \citenamefont {Cook}, \citenamefont {Vergniory}, \citenamefont
  {Wang}, \citenamefont {Parkin}, \citenamefont {Bernevig},\ and\ \citenamefont
  {Neupert}}]{Schindler2018HOTI}%
  \BibitemOpen
  \bibfield  {author} {\bibinfo {author} {\bibfnamefont {Frank}\ \bibnamefont
  {Schindler}}, \bibinfo {author} {\bibfnamefont {Ashley~M.}\ \bibnamefont
  {Cook}}, \bibinfo {author} {\bibfnamefont {Maia~G.}\ \bibnamefont
  {Vergniory}}, \bibinfo {author} {\bibfnamefont {Zhijun}\ \bibnamefont
  {Wang}}, \bibinfo {author} {\bibfnamefont {Stuart S.~P.}\ \bibnamefont
  {Parkin}}, \bibinfo {author} {\bibfnamefont {B.~Andrei}\ \bibnamefont
  {Bernevig}}, \ and\ \bibinfo {author} {\bibfnamefont {Titus}\ \bibnamefont
  {Neupert}},\ }\bibfield  {title} {\enquote {\bibinfo {title} {Higher-order
  topological insulators},}\ }\href {\doibase 10.1126/sciadv.aat0346}
  {\bibfield  {journal} {\bibinfo  {journal} {Science Advances}\ }\textbf
  {\bibinfo {volume} {4}} (\bibinfo {year} {2018}),\
  10.1126/sciadv.aat0346}\BibitemShut {NoStop}%
\bibitem [{\citenamefont {Neupert}\ and\ \citenamefont
  {Schindler}(2018)}]{neupert2018rev}%
  \BibitemOpen
  \bibfield  {author} {\bibinfo {author} {\bibfnamefont {Titus}\ \bibnamefont
  {Neupert}}\ and\ \bibinfo {author} {\bibfnamefont {Frank}\ \bibnamefont
  {Schindler}},\ }\enquote {\bibinfo {title} {Topological crystalline
  insulators},}\ in\ \href {\doibase 10.1007/978-3-319-76388-0_2} {\emph
  {\bibinfo {booktitle} {Topological Matter: Lectures from the Topological
  Matter School 2017}}},\ \bibinfo {editor} {edited by\ \bibinfo {editor}
  {\bibfnamefont {Dario}\ \bibnamefont {Bercioux}}, \bibinfo {editor}
  {\bibfnamefont {J{\'e}r{\^o}me}\ \bibnamefont {Cayssol}}, \bibinfo {editor}
  {\bibfnamefont {Maia~G.}\ \bibnamefont {Vergniory}}, \ and\ \bibinfo {editor}
  {\bibfnamefont {M.}~\bibnamefont {Reyes~Calvo}}}\ (\bibinfo  {publisher}
  {Springer International Publishing},\ \bibinfo {address} {Cham},\ \bibinfo
  {year} {2018})\ pp.\ \bibinfo {pages} {31--61}\BibitemShut {NoStop}%
\bibitem [{\citenamefont {Song}\ \emph {et~al.}(2017)\citenamefont {Song},
  \citenamefont {Huang}, \citenamefont {Fu},\ and\ \citenamefont
  {Hermele}}]{Song2017dr}%
  \BibitemOpen
  \bibfield  {author} {\bibinfo {author} {\bibfnamefont {Hao}\ \bibnamefont
  {Song}}, \bibinfo {author} {\bibfnamefont {Sheng-Jie}\ \bibnamefont {Huang}},
  \bibinfo {author} {\bibfnamefont {Liang}\ \bibnamefont {Fu}}, \ and\ \bibinfo
  {author} {\bibfnamefont {Michael}\ \bibnamefont {Hermele}},\ }\bibfield
  {title} {\enquote {\bibinfo {title} {Topological phases protected by point
  group symmetry},}\ }\href {\doibase 10.1103/PhysRevX.7.011020} {\bibfield
  {journal} {\bibinfo  {journal} {Phys. Rev. X}\ }\textbf {\bibinfo {volume}
  {7}},\ \bibinfo {pages} {011020} (\bibinfo {year} {2017})}\BibitemShut
  {NoStop}%
\bibitem [{\citenamefont {Huang}\ \emph {et~al.}(2017)\citenamefont {Huang},
  \citenamefont {Song}, \citenamefont {Huang},\ and\ \citenamefont
  {Hermele}}]{Huang2017building}%
  \BibitemOpen
  \bibfield  {author} {\bibinfo {author} {\bibfnamefont {Sheng-Jie}\
  \bibnamefont {Huang}}, \bibinfo {author} {\bibfnamefont {Hao}\ \bibnamefont
  {Song}}, \bibinfo {author} {\bibfnamefont {Yi-Ping}\ \bibnamefont {Huang}}, \
  and\ \bibinfo {author} {\bibfnamefont {Michael}\ \bibnamefont {Hermele}},\
  }\bibfield  {title} {\enquote {\bibinfo {title} {Building crystalline
  topological phases from lower-dimensional states},}\ }\href {\doibase
  10.1103/PhysRevB.96.205106} {\bibfield  {journal} {\bibinfo  {journal} {Phys.
  Rev. B}\ }\textbf {\bibinfo {volume} {96}},\ \bibinfo {pages} {205106}
  (\bibinfo {year} {2017})}\BibitemShut {NoStop}%
\bibitem [{\citenamefont {Rasmussen}\ and\ \citenamefont
  {Lu}(2018)}]{rasmussen2018intrinsically}%
  \BibitemOpen
  \bibfield  {author} {\bibinfo {author} {\bibfnamefont {Alex}\ \bibnamefont
  {Rasmussen}}\ and\ \bibinfo {author} {\bibfnamefont {Yuan-Ming}\ \bibnamefont
  {Lu}},\ }\href@noop {} {\enquote {\bibinfo {title} {Intrinsically interacting
  topological crystalline insulators and superconductors},}\ } (\bibinfo {year}
  {2018}),\ \Eprint {http://arxiv.org/abs/1810.12317} {arXiv:1810.12317
  [cond-mat.str-el]} \BibitemShut {NoStop}%
\bibitem [{\citenamefont {Cheng}\ and\ \citenamefont
  {Wang}(2018)}]{cheng2018rotation}%
  \BibitemOpen
  \bibfield  {author} {\bibinfo {author} {\bibfnamefont {Meng}\ \bibnamefont
  {Cheng}}\ and\ \bibinfo {author} {\bibfnamefont {Chenjie}\ \bibnamefont
  {Wang}},\ }\href@noop {} {\enquote {\bibinfo {title} {Rotation
  symmetry-protected topological phases of fermions},}\ } (\bibinfo {year}
  {2018}),\ \Eprint {http://arxiv.org/abs/1810.12308} {arXiv:1810.12308
  [cond-mat.str-el]} \BibitemShut {NoStop}%
\bibitem [{\citenamefont {Shiozaki}\ \emph {et~al.}(2018)\citenamefont
  {Shiozaki}, \citenamefont {Xiong},\ and\ \citenamefont
  {Gomi}}]{shiozaki2018generalized}%
  \BibitemOpen
  \bibfield  {author} {\bibinfo {author} {\bibfnamefont {Ken}\ \bibnamefont
  {Shiozaki}}, \bibinfo {author} {\bibfnamefont {Charles~Zhaoxi}\ \bibnamefont
  {Xiong}}, \ and\ \bibinfo {author} {\bibfnamefont {Kiyonori}\ \bibnamefont
  {Gomi}},\ }\href@noop {} {\enquote {\bibinfo {title} {Generalized homology
  and atiyah-hirzebruch spectral sequence in crystalline symmetry protected
  topological phenomena},}\ } (\bibinfo {year} {2018}),\ \Eprint
  {http://arxiv.org/abs/1810.00801} {arXiv:1810.00801 [cond-mat.str-el]}
  \BibitemShut {NoStop}%
\bibitem [{\citenamefont {Xiong}\ and\ \citenamefont
  {Alexandradinata}(2018)}]{Xiong2018glide}%
  \BibitemOpen
  \bibfield  {author} {\bibinfo {author} {\bibfnamefont {Charles~Zhaoxi}\
  \bibnamefont {Xiong}}\ and\ \bibinfo {author} {\bibfnamefont
  {A.}~\bibnamefont {Alexandradinata}},\ }\bibfield  {title} {\enquote
  {\bibinfo {title} {Organizing symmetry-protected topological phases by
  layering and symmetry reduction: A minimalist perspective},}\ }\href
  {\doibase 10.1103/PhysRevB.97.115153} {\bibfield  {journal} {\bibinfo
  {journal} {Phys. Rev. B}\ }\textbf {\bibinfo {volume} {97}},\ \bibinfo
  {pages} {115153} (\bibinfo {year} {2018})}\BibitemShut {NoStop}%
\bibitem [{\citenamefont {Song}\ \emph {et~al.}(2018)\citenamefont {Song},
  \citenamefont {Zhang}, \citenamefont {Fang},\ and\ \citenamefont
  {Fang}}]{Song2018indicator}%
  \BibitemOpen
  \bibfield  {author} {\bibinfo {author} {\bibfnamefont {Zhida}\ \bibnamefont
  {Song}}, \bibinfo {author} {\bibfnamefont {Tiantian}\ \bibnamefont {Zhang}},
  \bibinfo {author} {\bibfnamefont {Zhong}\ \bibnamefont {Fang}}, \ and\
  \bibinfo {author} {\bibfnamefont {Chen}\ \bibnamefont {Fang}},\ }\bibfield
  {title} {\enquote {\bibinfo {title} {Quantitative mappings between symmetry
  and topology in solids},}\ }\href {\doibase 10.1038/s41467-018-06010-w}
  {\bibfield  {journal} {\bibinfo  {journal} {Nature Communications}\ }\textbf
  {\bibinfo {volume} {9}},\ \bibinfo {pages} {3530} (\bibinfo {year}
  {2018})}\BibitemShut {NoStop}%
\bibitem [{\citenamefont {Song}\ \emph {et~al.}(2019)\citenamefont {Song},
  \citenamefont {Huang}, \citenamefont {Qi}, \citenamefont {Fang},\ and\
  \citenamefont {Hermele}}]{Song2019tc}%
  \BibitemOpen
  \bibfield  {author} {\bibinfo {author} {\bibfnamefont {Zhida}\ \bibnamefont
  {Song}}, \bibinfo {author} {\bibfnamefont {Sheng-Jie}\ \bibnamefont {Huang}},
  \bibinfo {author} {\bibfnamefont {Yang}\ \bibnamefont {Qi}}, \bibinfo
  {author} {\bibfnamefont {Chen}\ \bibnamefont {Fang}}, \ and\ \bibinfo
  {author} {\bibfnamefont {Michael}\ \bibnamefont {Hermele}},\ }\bibfield
  {title} {\enquote {\bibinfo {title} {Topological states from topological
  crystals},}\ }\href {\doibase 10.1126/sciadv.aax2007} {\bibfield  {journal}
  {\bibinfo  {journal} {Science Advances}\ }\textbf {\bibinfo {volume} {5}}
  (\bibinfo {year} {2019}),\ 10.1126/sciadv.aax2007}\BibitemShut {NoStop}%
\bibitem [{\citenamefont {Song}\ \emph
  {et~al.}(2020{\natexlab{a}})\citenamefont {Song}, \citenamefont {Fang},\ and\
  \citenamefont {Qi}}]{Song2019real}%
  \BibitemOpen
  \bibfield  {author} {\bibinfo {author} {\bibfnamefont {Zhida}\ \bibnamefont
  {Song}}, \bibinfo {author} {\bibfnamefont {Chen}\ \bibnamefont {Fang}}, \
  and\ \bibinfo {author} {\bibfnamefont {Yang}\ \bibnamefont {Qi}},\ }\bibfield
   {title} {\enquote {\bibinfo {title} {Real-space recipes for general
  topological crystalline states},}\ }\href {\doibase
  10.1038/s41467-020-17685-5} {\bibfield  {journal} {\bibinfo  {journal}
  {Nature Communications}\ }\textbf {\bibinfo {volume} {11}},\ \bibinfo {pages}
  {4197} (\bibinfo {year} {2020}{\natexlab{a}})}\BibitemShut {NoStop}%
\bibitem [{\citenamefont {Okuma}\ \emph {et~al.}(2019)\citenamefont {Okuma},
  \citenamefont {Sato},\ and\ \citenamefont {Shiozaki}}]{Shiozaki2019AHSS}%
  \BibitemOpen
  \bibfield  {author} {\bibinfo {author} {\bibfnamefont {Nobuyuki}\
  \bibnamefont {Okuma}}, \bibinfo {author} {\bibfnamefont {Masatoshi}\
  \bibnamefont {Sato}}, \ and\ \bibinfo {author} {\bibfnamefont {Ken}\
  \bibnamefont {Shiozaki}},\ }\bibfield  {title} {\enquote {\bibinfo {title}
  {Topological classification under nonmagnetic and magnetic point group
  symmetry: Application of real-space atiyah-hirzebruch spectral sequence to
  higher-order topology},}\ }\href {\doibase 10.1103/PhysRevB.99.085127}
  {\bibfield  {journal} {\bibinfo  {journal} {Phys. Rev. B}\ }\textbf {\bibinfo
  {volume} {99}},\ \bibinfo {pages} {085127} (\bibinfo {year}
  {2019})}\BibitemShut {NoStop}%
\bibitem [{\citenamefont {Song}\ \emph
  {et~al.}(2020{\natexlab{b}})\citenamefont {Song}, \citenamefont {Xiong},\
  and\ \citenamefont {Huang}}]{Song2020}%
  \BibitemOpen
  \bibfield  {author} {\bibinfo {author} {\bibfnamefont {Hao}\ \bibnamefont
  {Song}}, \bibinfo {author} {\bibfnamefont {Charles~Zhaoxi}\ \bibnamefont
  {Xiong}}, \ and\ \bibinfo {author} {\bibfnamefont {Sheng-Jie}\ \bibnamefont
  {Huang}},\ }\bibfield  {title} {\enquote {\bibinfo {title} {Bosonic
  crystalline symmetry protected topological phases beyond the group cohomology
  proposal},}\ }\href {\doibase 10.1103/PhysRevB.101.165129} {\bibfield
  {journal} {\bibinfo  {journal} {Phys. Rev. B}\ }\textbf {\bibinfo {volume}
  {101}},\ \bibinfo {pages} {165129} (\bibinfo {year}
  {2020}{\natexlab{b}})}\BibitemShut {NoStop}%
\bibitem [{\citenamefont {Rasmussen}\ and\ \citenamefont
  {Lu}(2020)}]{Alex2020bhospt}%
  \BibitemOpen
  \bibfield  {author} {\bibinfo {author} {\bibfnamefont {Alex}\ \bibnamefont
  {Rasmussen}}\ and\ \bibinfo {author} {\bibfnamefont {Yuan-Ming}\ \bibnamefont
  {Lu}},\ }\bibfield  {title} {\enquote {\bibinfo {title} {Classification and
  construction of higher-order symmetry-protected topological phases of
  interacting bosons},}\ }\href {\doibase 10.1103/PhysRevB.101.085137}
  {\bibfield  {journal} {\bibinfo  {journal} {Phys. Rev. B}\ }\textbf {\bibinfo
  {volume} {101}},\ \bibinfo {pages} {085137} (\bibinfo {year}
  {2020})}\BibitemShut {NoStop}%
\bibitem [{\citenamefont {Huang}\ and\ \citenamefont
  {Hsu}(2021)}]{Huang2020tsc}%
  \BibitemOpen
  \bibfield  {author} {\bibinfo {author} {\bibfnamefont {Sheng-Jie}\
  \bibnamefont {Huang}}\ and\ \bibinfo {author} {\bibfnamefont {Yi-Ting}\
  \bibnamefont {Hsu}},\ }\bibfield  {title} {\enquote {\bibinfo {title}
  {Faithful derivation of symmetry indicators: A case study for topological
  superconductors with time-reversal and inversion symmetries},}\ }\href
  {\doibase 10.1103/PhysRevResearch.3.013243} {\bibfield  {journal} {\bibinfo
  {journal} {Phys. Rev. Research}\ }\textbf {\bibinfo {volume} {3}},\ \bibinfo
  {pages} {013243} (\bibinfo {year} {2021})}\BibitemShut {NoStop}%
\bibitem [{\citenamefont {Zhang}\ \emph {et~al.}(2020)\citenamefont {Zhang},
  \citenamefont {Yang}, \citenamefont {Qi},\ and\ \citenamefont
  {Gu}}]{Zhang2020}%
  \BibitemOpen
  \bibfield  {author} {\bibinfo {author} {\bibfnamefont {Jian-Hao}\
  \bibnamefont {Zhang}}, \bibinfo {author} {\bibfnamefont {Shuo}\ \bibnamefont
  {Yang}}, \bibinfo {author} {\bibfnamefont {Yang}\ \bibnamefont {Qi}}, \ and\
  \bibinfo {author} {\bibfnamefont {Zheng-Cheng}\ \bibnamefont {Gu}},\
  }\bibfield  {title} {\enquote {\bibinfo {title} {Real-space construction of
  crystalline topological superconductors and insulators in 2d interacting
  fermionic systems},}\ }\href@noop {} {\  (\bibinfo {year} {2020})},\ \Eprint
  {http://arxiv.org/abs/2012.15657} {arXiv:2012.15657 [cond-mat.str-el]}
  \BibitemShut {NoStop}%
\bibitem [{\citenamefont {Geier}\ \emph {et~al.}(2021)\citenamefont {Geier},
  \citenamefont {Fulga},\ and\ \citenamefont {Lau}}]{Geier2021defect}%
  \BibitemOpen
  \bibfield  {author} {\bibinfo {author} {\bibfnamefont {Max}\ \bibnamefont
  {Geier}}, \bibinfo {author} {\bibfnamefont {Ion~Cosma}\ \bibnamefont
  {Fulga}}, \ and\ \bibinfo {author} {\bibfnamefont {Alexander}\ \bibnamefont
  {Lau}},\ }\bibfield  {title} {\enquote {\bibinfo {title}
  {{Bulk-boundary-defect correspondence at disclinations in rotation-symmetric
  topological insulators and superconductors}},}\ }\href {\doibase
  10.21468/SciPostPhys.10.4.092} {\bibfield  {journal} {\bibinfo  {journal}
  {SciPost Phys.}\ }\textbf {\bibinfo {volume} {10}},\ \bibinfo {pages} {92}
  (\bibinfo {year} {2021})}\BibitemShut {NoStop}%
\bibitem [{\citenamefont {Huang}\ and\ \citenamefont
  {Hermele}(2018)}]{Huang2018surface}%
  \BibitemOpen
  \bibfield  {author} {\bibinfo {author} {\bibfnamefont {Sheng-Jie}\
  \bibnamefont {Huang}}\ and\ \bibinfo {author} {\bibfnamefont {Michael}\
  \bibnamefont {Hermele}},\ }\bibfield  {title} {\enquote {\bibinfo {title}
  {Surface field theories of point group symmetry protected topological
  phases},}\ }\href {\doibase 10.1103/PhysRevB.97.075145} {\bibfield  {journal}
  {\bibinfo  {journal} {Phys. Rev. B}\ }\textbf {\bibinfo {volume} {97}},\
  \bibinfo {pages} {075145} (\bibinfo {year} {2018})}\BibitemShut {NoStop}%
\bibitem [{\citenamefont {Huang}(2020)}]{Huang20204d}%
  \BibitemOpen
  \bibfield  {author} {\bibinfo {author} {\bibfnamefont {Sheng-Jie}\
  \bibnamefont {Huang}},\ }\bibfield  {title} {\enquote {\bibinfo {title} {4d
  beyond-cohomology topological phase protected by ${C}_{2}$ symmetry and its
  boundary theories},}\ }\href {\doibase 10.1103/PhysRevResearch.2.033236}
  {\bibfield  {journal} {\bibinfo  {journal} {Phys. Rev. Research}\ }\textbf
  {\bibinfo {volume} {2}},\ \bibinfo {pages} {033236} (\bibinfo {year}
  {2020})}\BibitemShut {NoStop}%
\bibitem [{\citenamefont {Thorngren}\ and\ \citenamefont
  {Else}(2018)}]{Else2018}%
  \BibitemOpen
  \bibfield  {author} {\bibinfo {author} {\bibfnamefont {Ryan}\ \bibnamefont
  {Thorngren}}\ and\ \bibinfo {author} {\bibfnamefont {Dominic~V.}\
  \bibnamefont {Else}},\ }\bibfield  {title} {\enquote {\bibinfo {title}
  {Gauging spatial symmetries and the classification of topological crystalline
  phases},}\ }\href {\doibase 10.1103/PhysRevX.8.011040} {\bibfield  {journal}
  {\bibinfo  {journal} {Phys. Rev. X}\ }\textbf {\bibinfo {volume} {8}},\
  \bibinfo {pages} {011040} (\bibinfo {year} {2018})}\BibitemShut {NoStop}%
\bibitem [{\citenamefont {Jiang}\ and\ \citenamefont {Ran}(2017)}]{Jiang2017}%
  \BibitemOpen
  \bibfield  {author} {\bibinfo {author} {\bibfnamefont {Shenghan}\
  \bibnamefont {Jiang}}\ and\ \bibinfo {author} {\bibfnamefont {Ying}\
  \bibnamefont {Ran}},\ }\bibfield  {title} {\enquote {\bibinfo {title} {Anyon
  condensation and a generic tensor-network construction for symmetry-protected
  topological phases},}\ }\href {\doibase 10.1103/PhysRevB.95.125107}
  {\bibfield  {journal} {\bibinfo  {journal} {Phys. Rev. B}\ }\textbf {\bibinfo
  {volume} {95}},\ \bibinfo {pages} {125107} (\bibinfo {year}
  {2017})}\BibitemShut {NoStop}%
\bibitem [{\citenamefont {Freed}\ and\ \citenamefont
  {Hopkins}(2019{\natexlab{b}})}]{freed2019invertiblecrystal}%
  \BibitemOpen
  \bibfield  {author} {\bibinfo {author} {\bibfnamefont {Daniel~S.}\
  \bibnamefont {Freed}}\ and\ \bibinfo {author} {\bibfnamefont {Michael~J.}\
  \bibnamefont {Hopkins}},\ }\href@noop {} {\enquote {\bibinfo {title}
  {Invertible phases of matter with spatial symmetry},}\ } (\bibinfo {year}
  {2019}{\natexlab{b}}),\ \Eprint {http://arxiv.org/abs/1901.06419}
  {arXiv:1901.06419 [math-ph]} \BibitemShut {NoStop}%
\bibitem [{\citenamefont {Debray}(2021)}]{debray2021invertible}%
  \BibitemOpen
  \bibfield  {author} {\bibinfo {author} {\bibfnamefont {Arun}\ \bibnamefont
  {Debray}},\ }\href@noop {} {\enquote {\bibinfo {title} {Invertible phases for
  mixed spatial symmetries and the fermionic crystalline equivalence
  principle},}\ } (\bibinfo {year} {2021}),\ \Eprint
  {http://arxiv.org/abs/2102.02941} {arXiv:2102.02941 [math-ph]} \BibitemShut
  {NoStop}%
\bibitem [{\citenamefont {Else}\ and\ \citenamefont
  {Thorngren}(2019)}]{Else2019defect}%
  \BibitemOpen
  \bibfield  {author} {\bibinfo {author} {\bibfnamefont {Dominic~V.}\
  \bibnamefont {Else}}\ and\ \bibinfo {author} {\bibfnamefont {Ryan}\
  \bibnamefont {Thorngren}},\ }\bibfield  {title} {\enquote {\bibinfo {title}
  {Crystalline topological phases as defect networks},}\ }\href {\doibase
  10.1103/PhysRevB.99.115116} {\bibfield  {journal} {\bibinfo  {journal} {Phys.
  Rev. B}\ }\textbf {\bibinfo {volume} {99}},\ \bibinfo {pages} {115116}
  (\bibinfo {year} {2019})}\BibitemShut {NoStop}%
\bibitem [{\citenamefont {Song}\ \emph {et~al.}(2021)\citenamefont {Song},
  \citenamefont {He}, \citenamefont {Vishwanath},\ and\ \citenamefont
  {Wang}}]{Song2021polarization}%
  \BibitemOpen
  \bibfield  {author} {\bibinfo {author} {\bibfnamefont {Xue-Yang}\
  \bibnamefont {Song}}, \bibinfo {author} {\bibfnamefont {Yin-Chen}\
  \bibnamefont {He}}, \bibinfo {author} {\bibfnamefont {Ashvin}\ \bibnamefont
  {Vishwanath}}, \ and\ \bibinfo {author} {\bibfnamefont {Chong}\ \bibnamefont
  {Wang}},\ }\bibfield  {title} {\enquote {\bibinfo {title} {Electric
  polarization as a nonquantized topological response and boundary luttinger
  theorem},}\ }\href {\doibase 10.1103/PhysRevResearch.3.023011} {\bibfield
  {journal} {\bibinfo  {journal} {Phys. Rev. Research}\ }\textbf {\bibinfo
  {volume} {3}},\ \bibinfo {pages} {023011} (\bibinfo {year}
  {2021})}\BibitemShut {NoStop}%
\bibitem [{\citenamefont {Gioia}\ \emph {et~al.}(2021)\citenamefont {Gioia},
  \citenamefont {Wang},\ and\ \citenamefont {Burkov}}]{gioia2021unquantized}%
  \BibitemOpen
  \bibfield  {author} {\bibinfo {author} {\bibfnamefont {Lei}\ \bibnamefont
  {Gioia}}, \bibinfo {author} {\bibfnamefont {Chong}\ \bibnamefont {Wang}}, \
  and\ \bibinfo {author} {\bibfnamefont {A.~A.}\ \bibnamefont {Burkov}},\
  }\href@noop {} {\enquote {\bibinfo {title} {Unquantized anomalies in
  topological semimetals},}\ } (\bibinfo {year} {2021}),\ \Eprint
  {http://arxiv.org/abs/2103.09841} {arXiv:2103.09841 [cond-mat.str-el]}
  \BibitemShut {NoStop}%
\bibitem [{\citenamefont {Manjunath}\ and\ \citenamefont
  {Barkeshli}(2021)}]{Naren2021}%
  \BibitemOpen
  \bibfield  {author} {\bibinfo {author} {\bibfnamefont {Naren}\ \bibnamefont
  {Manjunath}}\ and\ \bibinfo {author} {\bibfnamefont {Maissam}\ \bibnamefont
  {Barkeshli}},\ }\bibfield  {title} {\enquote {\bibinfo {title} {Crystalline
  gauge fields and quantized discrete geometric response for abelian
  topological phases with lattice symmetry},}\ }\href {\doibase
  10.1103/PhysRevResearch.3.013040} {\bibfield  {journal} {\bibinfo  {journal}
  {Phys. Rev. Research}\ }\textbf {\bibinfo {volume} {3}},\ \bibinfo {pages}
  {013040} (\bibinfo {year} {2021})}\BibitemShut {NoStop}%
\bibitem [{\citenamefont {Nissinen}\ and\ \citenamefont
  {Volovik}(2018)}]{Nissinen2018tetrads}%
  \BibitemOpen
  \bibfield  {author} {\bibinfo {author} {\bibfnamefont {J.}~\bibnamefont
  {Nissinen}}\ and\ \bibinfo {author} {\bibfnamefont {G.~E.}\ \bibnamefont
  {Volovik}},\ }\bibfield  {title} {\enquote {\bibinfo {title} {Tetrads in
  solids: from elasticity theory to topological quantum hall systems and weyl
  fermions},}\ }\href {\doibase 10.1134/S1063776118110080} {\bibfield
  {journal} {\bibinfo  {journal} {Journal of Experimental and Theoretical
  Physics}\ }\textbf {\bibinfo {volume} {127}},\ \bibinfo {pages} {948--957}
  (\bibinfo {year} {2018})}\BibitemShut {NoStop}%
\bibitem [{\citenamefont {Nissinen}\ and\ \citenamefont
  {Volovik}(2019)}]{Nissinen20193dqh}%
  \BibitemOpen
  \bibfield  {author} {\bibinfo {author} {\bibfnamefont {J.}~\bibnamefont
  {Nissinen}}\ and\ \bibinfo {author} {\bibfnamefont {G.~E.}\ \bibnamefont
  {Volovik}},\ }\bibfield  {title} {\enquote {\bibinfo {title} {Elasticity
  tetrads, mixed axial-gravitational anomalies, and ($3+1$)-d quantum hall
  effect},}\ }\href {\doibase 10.1103/PhysRevResearch.1.023007} {\bibfield
  {journal} {\bibinfo  {journal} {Phys. Rev. Research}\ }\textbf {\bibinfo
  {volume} {1}},\ \bibinfo {pages} {023007} (\bibinfo {year}
  {2019})}\BibitemShut {NoStop}%
\bibitem [{\citenamefont {{Nissinen}}(2020)}]{nissinen2020field}%
  \BibitemOpen
  \bibfield  {author} {\bibinfo {author} {\bibfnamefont {Jaakko}\ \bibnamefont
  {{Nissinen}}},\ }\bibfield  {title} {\enquote {\bibinfo {title} {{Field
  theory of higher-order topological crystalline response, generalized global
  symmetries and elasticity tetrads}},}\ }\href@noop {} {\bibfield  {journal}
  {\bibinfo  {journal} {arXiv e-prints}\ ,\ \bibinfo {eid} {arXiv:2009.14184}}
  (\bibinfo {year} {2020})},\ \Eprint {http://arxiv.org/abs/2009.14184}
  {arXiv:2009.14184 [cond-mat.str-el]} \BibitemShut {NoStop}%
\bibitem [{\citenamefont {Nissinen}\ \emph {et~al.}(2021)\citenamefont
  {Nissinen}, \citenamefont {Heikkil\"a},\ and\ \citenamefont
  {Volovik}}]{Nissinen2021}%
  \BibitemOpen
  \bibfield  {author} {\bibinfo {author} {\bibfnamefont {J.}~\bibnamefont
  {Nissinen}}, \bibinfo {author} {\bibfnamefont {T.~T.}\ \bibnamefont
  {Heikkil\"a}}, \ and\ \bibinfo {author} {\bibfnamefont {G.~E.}\ \bibnamefont
  {Volovik}},\ }\bibfield  {title} {\enquote {\bibinfo {title} {Topological
  polarization, dual invariants, and surface flat bands in crystalline
  insulators},}\ }\href {\doibase 10.1103/PhysRevB.103.245115} {\bibfield
  {journal} {\bibinfo  {journal} {Phys. Rev. B}\ }\textbf {\bibinfo {volume}
  {103}},\ \bibinfo {pages} {245115} (\bibinfo {year} {2021})}\BibitemShut
  {NoStop}%
\bibitem [{\citenamefont {Else}\ \emph {et~al.}(2021)\citenamefont {Else},
  \citenamefont {Huang}, \citenamefont {Prem},\ and\ \citenamefont
  {Gromov}}]{Else2021qc}%
  \BibitemOpen
  \bibfield  {author} {\bibinfo {author} {\bibfnamefont {Dominic~V.}\
  \bibnamefont {Else}}, \bibinfo {author} {\bibfnamefont {Sheng-Jie}\
  \bibnamefont {Huang}}, \bibinfo {author} {\bibfnamefont {Abhinav}\
  \bibnamefont {Prem}}, \ and\ \bibinfo {author} {\bibfnamefont {Andrey}\
  \bibnamefont {Gromov}},\ }\bibfield  {title} {\enquote {\bibinfo {title}
  {Quantum many-body topology of quasicrystals},}\ }\href {\doibase
  10.1103/PhysRevX.11.041051} {\bibfield  {journal} {\bibinfo  {journal} {Phys.
  Rev. X}\ }\textbf {\bibinfo {volume} {11}},\ \bibinfo {pages} {041051}
  (\bibinfo {year} {2021})}\BibitemShut {NoStop}%
\bibitem [{\citenamefont {Qi}\ \emph {et~al.}(2008)\citenamefont {Qi},
  \citenamefont {Hughes},\ and\ \citenamefont {Zhang}}]{Qi2008TFT}%
  \BibitemOpen
  \bibfield  {author} {\bibinfo {author} {\bibfnamefont {Xiao-Liang}\
  \bibnamefont {Qi}}, \bibinfo {author} {\bibfnamefont {Taylor~L.}\
  \bibnamefont {Hughes}}, \ and\ \bibinfo {author} {\bibfnamefont {Shou-Cheng}\
  \bibnamefont {Zhang}},\ }\bibfield  {title} {\enquote {\bibinfo {title}
  {Topological field theory of time-reversal invariant insulators},}\ }\href
  {\doibase 10.1103/PhysRevB.78.195424} {\bibfield  {journal} {\bibinfo
  {journal} {Phys. Rev. B}\ }\textbf {\bibinfo {volume} {78}},\ \bibinfo
  {pages} {195424} (\bibinfo {year} {2008})}\BibitemShut {NoStop}%
\bibitem [{Note1()}]{Note1}%
  \BibitemOpen
  \bibinfo {note} {To simplify the analysis, here we choose $a/l = 2m$ with $m
  \gg 1$ without loss of generality. One is free to choose $a/l$ to be an odd
  integers, which will not effect the results.}\BibitemShut {Stop}%
\bibitem [{Note2()}]{Note2}%
  \BibitemOpen
  \bibinfo {note} {More formally, we note that the function $\phi (x)$ defines
  a map $\phi : X \rightarrow \protect \mathfrak {M}_{1}$, where $\protect
  \mathfrak {M}_{1}$ is the space of $1+1$d fermionic short range entangled
  states with $U(1)$ symmetry. The configuration of $\phi (x)$ we choose gives
  a noncontractible loop in $\protect \mathfrak {M}_{1}$ every time we go
  through a unit cell and the bound state is associated to the winding number
  $\pi _{1}(\protect \mathfrak {M}_{1}) = \protect \mathbb {Z}$.}\BibitemShut
  {Stop}%
\bibitem [{\citenamefont {Thouless}(1983)}]{Thouless1983}%
  \BibitemOpen
  \bibfield  {author} {\bibinfo {author} {\bibfnamefont {D.~J.}\ \bibnamefont
  {Thouless}},\ }\bibfield  {title} {\enquote {\bibinfo {title} {Quantization
  of particle transport},}\ }\href {\doibase 10.1103/PhysRevB.27.6083}
  {\bibfield  {journal} {\bibinfo  {journal} {Phys. Rev. B}\ }\textbf {\bibinfo
  {volume} {27}},\ \bibinfo {pages} {6083--6087} (\bibinfo {year}
  {1983})}\BibitemShut {NoStop}%
\bibitem [{Note3()}]{Note3}%
  \BibitemOpen
  \bibinfo {note} {Recall that a closed $k$-form $\omega $ on M has integral
  periods if, for every smooth $k$-cycle $C$ in M, the integral $\DOTSI \intop
  \ilimits@ _{C} \omega $ is an integer. Moreover, a closed $k$-form $\omega $
  has integral periods if and only if the de Rham class of $\omega $ lies in
  the image of the change-of-coefficients map \begin {equation}
  H^{k}(M,\protect \mathbb {Z}) \rightarrow H^{k}(M,\protect \mathbb {R})
  \protect \cong H^{k}_{dR}(M), \end {equation} where $H^{k}_{dR}(M)$ denotes
  the de Rham cohomology of $M$ \cite {Simons2007}. Loosely speaking, a closed
  $k$-form with an integral period serves as a differential form representative
  of an element in $H^{k}(M,\protect \mathbb {Z})$.}\BibitemShut {Stop}%
\bibitem [{\citenamefont {Haldane}(1981)}]{Haldane1981}%
  \BibitemOpen
  \bibfield  {author} {\bibinfo {author} {\bibfnamefont {F.~D.~M.}\
  \bibnamefont {Haldane}},\ }\bibfield  {title} {\enquote {\bibinfo {title}
  {Effective harmonic-fluid approach to low-energy properties of
  one-dimensional quantum fluids},}\ }\href {\doibase
  10.1103/PhysRevLett.47.1840} {\bibfield  {journal} {\bibinfo  {journal}
  {Phys. Rev. Lett.}\ }\textbf {\bibinfo {volume} {47}},\ \bibinfo {pages}
  {1840--1843} (\bibinfo {year} {1981})}\BibitemShut {NoStop}%
\bibitem [{\citenamefont {{Kapustin}}\ and\ \citenamefont
  {{Spodyneiko}}(2020)}]{Kapustin2020thouless}%
  \BibitemOpen
  \bibfield  {author} {\bibinfo {author} {\bibfnamefont {Anton}\ \bibnamefont
  {{Kapustin}}}\ and\ \bibinfo {author} {\bibfnamefont {Lev}\ \bibnamefont
  {{Spodyneiko}}},\ }\bibfield  {title} {\enquote {\bibinfo {title}
  {{Higher-dimensional generalizations of the Thouless charge pump}},}\
  }\href@noop {} {\bibfield  {journal} {\bibinfo  {journal} {arXiv e-prints}\
  ,\ \bibinfo {eid} {arXiv:2003.09519}} (\bibinfo {year} {2020})},\ \Eprint
  {http://arxiv.org/abs/2003.09519} {arXiv:2003.09519 [cond-mat.str-el]}
  \BibitemShut {NoStop}%
\bibitem [{\citenamefont {Hsin}\ \emph {et~al.}(2020)\citenamefont {Hsin},
  \citenamefont {Kapustin},\ and\ \citenamefont {Thorngren}}]{hsin2020berry}%
  \BibitemOpen
  \bibfield  {author} {\bibinfo {author} {\bibfnamefont {Po-Shen}\ \bibnamefont
  {Hsin}}, \bibinfo {author} {\bibfnamefont {Anton}\ \bibnamefont {Kapustin}},
  \ and\ \bibinfo {author} {\bibfnamefont {Ryan}\ \bibnamefont {Thorngren}},\
  }\bibfield  {title} {\enquote {\bibinfo {title} {Berry phase in quantum field
  theory: Diabolical points and boundary phenomena},}\ }\href {\doibase
  10.1103/PhysRevB.102.245113} {\bibfield  {journal} {\bibinfo  {journal}
  {Phys. Rev. B}\ }\textbf {\bibinfo {volume} {102}},\ \bibinfo {pages}
  {245113} (\bibinfo {year} {2020})}\BibitemShut {NoStop}%
\bibitem [{\citenamefont {Ramamurthy}\ and\ \citenamefont
  {Hughes}(2015)}]{Ramamurthy2015}%
  \BibitemOpen
  \bibfield  {author} {\bibinfo {author} {\bibfnamefont {Srinidhi~T.}\
  \bibnamefont {Ramamurthy}}\ and\ \bibinfo {author} {\bibfnamefont
  {Taylor~L.}\ \bibnamefont {Hughes}},\ }\bibfield  {title} {\enquote {\bibinfo
  {title} {Patterns of electromagnetic response in topological semimetals},}\
  }\href {\doibase 10.1103/PhysRevB.92.085105} {\bibfield  {journal} {\bibinfo
  {journal} {Phys. Rev. B}\ }\textbf {\bibinfo {volume} {92}},\ \bibinfo
  {pages} {085105} (\bibinfo {year} {2015})}\BibitemShut {NoStop}%
\bibitem [{Note4()}]{Note4}%
  \BibitemOpen
  \bibinfo {note} {More precisely, the symmetry is $(U(1) \times
  C_{N})/\protect \mathbb {Z}_2$, as the $C_N$ rotation $U$ defined in
  Eq.~(\ref {eqn:2d-cn}) obeys $U^N = (-1)^{N_f}$, where $N_f$ is the total
  fermion number under $U(1)$.}\BibitemShut {Stop}%
\bibitem [{\citenamefont {{Cheng}}\ and\ \citenamefont
  {{Wang}}(2018)}]{Meng2018rot}%
  \BibitemOpen
  \bibfield  {author} {\bibinfo {author} {\bibfnamefont {Meng}\ \bibnamefont
  {{Cheng}}}\ and\ \bibinfo {author} {\bibfnamefont {Chenjie}\ \bibnamefont
  {{Wang}}},\ }\bibfield  {title} {\enquote {\bibinfo {title} {{Rotation
  Symmetry-Protected Topological Phases of Fermions}},}\ }\href@noop {}
  {\bibfield  {journal} {\bibinfo  {journal} {arXiv e-prints}\ ,\ \bibinfo
  {eid} {arXiv:1810.12308}} (\bibinfo {year} {2018})},\ \Eprint
  {http://arxiv.org/abs/1810.12308} {arXiv:1810.12308 [cond-mat.str-el]}
  \BibitemShut {NoStop}%
\bibitem [{\citenamefont {{Shiozaki}}(2019)}]{Shiozaki2019dirac}%
  \BibitemOpen
  \bibfield  {author} {\bibinfo {author} {\bibfnamefont {Ken}\ \bibnamefont
  {{Shiozaki}}},\ }\bibfield  {title} {\enquote {\bibinfo {title} {{The
  classification of surface states of topological insulators and
  superconductors with magnetic point group symmetry}},}\ }\href@noop {}
  {\bibfield  {journal} {\bibinfo  {journal} {arXiv e-prints}\ ,\ \bibinfo
  {eid} {arXiv:1907.09354}} (\bibinfo {year} {2019})},\ \Eprint
  {http://arxiv.org/abs/1907.09354} {arXiv:1907.09354 [cond-mat.mes-hall]}
  \BibitemShut {NoStop}%
\bibitem [{\citenamefont {Hason}\ \emph {et~al.}(2020)\citenamefont {Hason},
  \citenamefont {Komargodski},\ and\ \citenamefont
  {Thorngren}}]{Hason2020smith}%
  \BibitemOpen
  \bibfield  {author} {\bibinfo {author} {\bibfnamefont {Itamar}\ \bibnamefont
  {Hason}}, \bibinfo {author} {\bibfnamefont {Zohar}\ \bibnamefont
  {Komargodski}}, \ and\ \bibinfo {author} {\bibfnamefont {Ryan}\ \bibnamefont
  {Thorngren}},\ }\bibfield  {title} {\enquote {\bibinfo {title} {{Anomaly
  Matching in the Symmetry Broken Phase: Domain Walls, CPT, and the Smith
  Isomorphism}},}\ }\href {\doibase 10.21468/SciPostPhys.8.4.062} {\bibfield
  {journal} {\bibinfo  {journal} {SciPost Phys.}\ }\textbf {\bibinfo {volume}
  {8}},\ \bibinfo {pages} {62} (\bibinfo {year} {2020})}\BibitemShut {NoStop}%
\bibitem [{\citenamefont {Bott}\ and\ \citenamefont {Tu}(1982)}]{Bott1982}%
  \BibitemOpen
  \bibfield  {author} {\bibinfo {author} {\bibfnamefont {Raoul}\ \bibnamefont
  {Bott}}\ and\ \bibinfo {author} {\bibfnamefont {Loring~W.}\ \bibnamefont
  {Tu}},\ }\href@noop {} {\emph {\bibinfo {title} {Differential forms in
  algebraic topology}}},\ \bibinfo {series} {Graduate Texts in mathematics}\
  No.~\bibinfo {number} {82}\ (\bibinfo  {publisher} {Springer},\ \bibinfo
  {address} {New York},\ \bibinfo {year} {1982})\BibitemShut {NoStop}%
\bibitem [{\citenamefont {Benalcazar}\ \emph {et~al.}(2017)\citenamefont
  {Benalcazar}, \citenamefont {Bernevig},\ and\ \citenamefont
  {Hughes}}]{Benalcazar2017HOTI}%
  \BibitemOpen
  \bibfield  {author} {\bibinfo {author} {\bibfnamefont {Wladimir~A.}\
  \bibnamefont {Benalcazar}}, \bibinfo {author} {\bibfnamefont {B.~Andrei}\
  \bibnamefont {Bernevig}}, \ and\ \bibinfo {author} {\bibfnamefont
  {Taylor~L.}\ \bibnamefont {Hughes}},\ }\bibfield  {title} {\enquote {\bibinfo
  {title} {Quantized electric multipole insulators},}\ }\href {\doibase
  10.1126/science.aah6442} {\bibfield  {journal} {\bibinfo  {journal}
  {Science}\ }\textbf {\bibinfo {volume} {357}},\ \bibinfo {pages} {61--66}
  (\bibinfo {year} {2017})}\BibitemShut {NoStop}%
\bibitem [{\citenamefont {Varnava}\ and\ \citenamefont
  {Vanderbilt}(2018)}]{Varnava2018AI}%
  \BibitemOpen
  \bibfield  {author} {\bibinfo {author} {\bibfnamefont {Nicodemos}\
  \bibnamefont {Varnava}}\ and\ \bibinfo {author} {\bibfnamefont {David}\
  \bibnamefont {Vanderbilt}},\ }\bibfield  {title} {\enquote {\bibinfo {title}
  {Surfaces of axion insulators},}\ }\href {\doibase
  10.1103/PhysRevB.98.245117} {\bibfield  {journal} {\bibinfo  {journal} {Phys.
  Rev. B}\ }\textbf {\bibinfo {volume} {98}},\ \bibinfo {pages} {245117}
  (\bibinfo {year} {2018})}\BibitemShut {NoStop}%
\bibitem [{\citenamefont {Ahn}\ and\ \citenamefont {Yang}(2019)}]{Ahn2019C2T}%
  \BibitemOpen
  \bibfield  {author} {\bibinfo {author} {\bibfnamefont {Junyeong}\
  \bibnamefont {Ahn}}\ and\ \bibinfo {author} {\bibfnamefont {Bohm-Jung}\
  \bibnamefont {Yang}},\ }\bibfield  {title} {\enquote {\bibinfo {title}
  {Symmetry representation approach to topological invariants in
  ${C}_{2z}t$-symmetric systems},}\ }\href {\doibase
  10.1103/PhysRevB.99.235125} {\bibfield  {journal} {\bibinfo  {journal} {Phys.
  Rev. B}\ }\textbf {\bibinfo {volume} {99}},\ \bibinfo {pages} {235125}
  (\bibinfo {year} {2019})}\BibitemShut {NoStop}%
\bibitem [{\citenamefont {Fu}\ \emph {et~al.}(2021)\citenamefont {Fu},
  \citenamefont {Hu},\ and\ \citenamefont {Shen}}]{fu2021bulkhinge}%
  \BibitemOpen
  \bibfield  {author} {\bibinfo {author} {\bibfnamefont {Bo}~\bibnamefont
  {Fu}}, \bibinfo {author} {\bibfnamefont {Zi-Ang}\ \bibnamefont {Hu}}, \ and\
  \bibinfo {author} {\bibfnamefont {Shun-Qing}\ \bibnamefont {Shen}},\
  }\href@noop {} {\enquote {\bibinfo {title} {The bulk-hinge correspondence and
  three-dimensional quantum anomalous hall effect in second order topological
  insulators},}\ } (\bibinfo {year} {2021}),\ \Eprint
  {http://arxiv.org/abs/2102.12050} {arXiv:2102.12050 [cond-mat.mes-hall]}
  \BibitemShut {NoStop}%
\bibitem [{\citenamefont {Ilan}\ \emph {et~al.}(2020)\citenamefont {Ilan},
  \citenamefont {Grushin},\ and\ \citenamefont
  {Pikulin}}]{Ilan2020NatRevPseudoGauge}%
  \BibitemOpen
  \bibfield  {author} {\bibinfo {author} {\bibfnamefont {Roni}\ \bibnamefont
  {Ilan}}, \bibinfo {author} {\bibfnamefont {Adolfo~G.}\ \bibnamefont
  {Grushin}}, \ and\ \bibinfo {author} {\bibfnamefont {Dmitry~I.}\ \bibnamefont
  {Pikulin}},\ }\bibfield  {title} {\enquote {\bibinfo {title}
  {Pseudo-electromagnetic fields in 3d topological semimetals},}\ }\href
  {\doibase 10.1038/s42254-019-0121-8} {\bibfield  {journal} {\bibinfo
  {journal} {Nature Reviews Physics}\ }\textbf {\bibinfo {volume} {2}},\
  \bibinfo {pages} {29--41} (\bibinfo {year} {2020})}\BibitemShut {NoStop}%
\bibitem [{\citenamefont {Yu}\ and\ \citenamefont
  {Liu}(2021)}]{Yu2021PseudoGaugeDiracWeyl}%
  \BibitemOpen
  \bibfield  {author} {\bibinfo {author} {\bibfnamefont {Jiabin}\ \bibnamefont
  {Yu}}\ and\ \bibinfo {author} {\bibfnamefont {Chao-Xing}\ \bibnamefont
  {Liu}},\ }\bibfield  {title} {\enquote {\bibinfo {title} {Chapter five -
  pseudo-gauge fields in dirac and weyl materials},}\ }in\ \href {\doibase
  https://doi.org/10.1016/bs.semsem.2021.06.003} {\emph {\bibinfo {booktitle}
  {Topological Insulator and Related Topics}}},\ \bibinfo {series}
  {Semiconductors and Semimetals}, Vol.\ \bibinfo {volume} {108},\ \bibinfo
  {editor} {edited by\ \bibinfo {editor} {\bibfnamefont {Lu}~\bibnamefont
  {Li}}\ and\ \bibinfo {editor} {\bibfnamefont {Kai}\ \bibnamefont {Sun}}}\
  (\bibinfo  {publisher} {Elsevier},\ \bibinfo {year} {2021})\ pp.\ \bibinfo
  {pages} {195--224}\BibitemShut {NoStop}%
\bibitem [{\citenamefont {Guo}\ \emph {et~al.}(2020)\citenamefont {Guo},
  \citenamefont {Ohmori}, \citenamefont {Putrov}, \citenamefont {Wan},\ and\
  \citenamefont {Wang}}]{Juven2019}%
  \BibitemOpen
  \bibfield  {author} {\bibinfo {author} {\bibfnamefont {Meng}\ \bibnamefont
  {Guo}}, \bibinfo {author} {\bibfnamefont {Kantaro}\ \bibnamefont {Ohmori}},
  \bibinfo {author} {\bibfnamefont {Pavel}\ \bibnamefont {Putrov}}, \bibinfo
  {author} {\bibfnamefont {Zheyan}\ \bibnamefont {Wan}}, \ and\ \bibinfo
  {author} {\bibfnamefont {Juven}\ \bibnamefont {Wang}},\ }\bibfield  {title}
  {\enquote {\bibinfo {title} {Fermionic finite-group gauge theories and
  interacting symmetric/crystalline orders via cobordisms},}\ }\href {\doibase
  10.1007/s00220-019-03671-6} {\bibfield  {journal} {\bibinfo  {journal}
  {Communications in Mathematical Physics}\ }\textbf {\bibinfo {volume}
  {376}},\ \bibinfo {pages} {1073--1154} (\bibinfo {year} {2020})}\BibitemShut
  {NoStop}%
\bibitem [{\citenamefont {Xiong}(2019)}]{Xiong2019thesis}%
  \BibitemOpen
  \bibfield  {author} {\bibinfo {author} {\bibfnamefont {Zhaoxi}\ \bibnamefont
  {Xiong}},\ }\emph {\bibinfo {title} {Classification and Construction of
  Topological Phases of Quantum Matter}},\ \href@noop {} {Ph.D. thesis},\
  \bibinfo  {school} {Harvard University, Graduate School of Arts \& Sciences}
  (\bibinfo {year} {2019})\BibitemShut {NoStop}%
\bibitem [{\citenamefont {Simons}\ and\ \citenamefont
  {Sullivan}(2008)}]{Simons2007}%
  \BibitemOpen
  \bibfield  {author} {\bibinfo {author} {\bibfnamefont {James}\ \bibnamefont
  {Simons}}\ and\ \bibinfo {author} {\bibfnamefont {Dennis}\ \bibnamefont
  {Sullivan}},\ }\bibfield  {title} {\enquote {\bibinfo {title} {Axiomatic
  characterization of ordinary differential cohomology},}\ }\href {\doibase
  https://doi.org/10.1112/jtopol/jtm006} {\bibfield  {journal} {\bibinfo
  {journal} {Journal of Topology}\ }\textbf {\bibinfo {volume} {1}},\ \bibinfo
  {pages} {45--56} (\bibinfo {year} {2008})}\BibitemShut {NoStop}%
\end{thebibliography}%
\end{document}